# THE MAXWELL CLASS EXACT SOLUTIONS TO THE SCHRÖDINGER EQUATION AND CONTINUUM MECHANICS MODELS

E.E. Perepelkin[a,c,d,*], B.I. Sadovnikov[a], N.G. Inozemtseva[b], A.S. Medvedev[a]

[a] *Faculty of Physics, Lomonosov Moscow State University, Moscow, 119991 Russia*
[b] *Moscow Technical University of Communications and Informatics, Moscow, 123423 Russia*
[c] *Dubna State University, Moscow region, Dubna, 141980 Russia*
[d] *Joint Institute for Nuclear Research, Moscow region, Dubna, 141980 Russia*
[*] *Corresponding author: pevgeny@jinr.ru*

**Abstract**

By applying the nonlinear Legendre transform to the continuity equation, this paper derives exact solutions to the Schrödinger equation and the equations of continuum mechanics. A generalized Maxwell distribution has been used as the momentum density function. Explicit expressions for the vector fields of time independent flows, density distributions, quantum and classical potentials have been found, and a detailed mathematical and physical analysis of the results obtained has been carried out.

**Key words:** exact solution, Schrödinger equation, Legendre transform, nonlinear partial differential equation, continuum mechanics, generalized Maxwell distribution, rigorous result.

## Introduction

The continuity equation is a fundamental equation of mathematical physics. It is used in continuum mechanics [1–3], astrophysics and gravity [4–9], electrodynamics [10–12], statistical physics [13–15], plasma physics [16–18] and quantum mechanics [19]. Note that the continuity equation is the first equation in Vlasov's infinite self-linked chain of equations [20, 21]. Here are the first two equations of the chain

$$\partial_t f_1(\vec{r},t) + \text{div}_r \left[ f_1(\vec{r},t) \langle \vec{v} \rangle (\vec{r},t) \right] = 0, \qquad (\text{i.1})$$

$$\partial_t f_2(\vec{r},\vec{v},t) + \text{div}_r \left[ f_2(\vec{r},\vec{v},t) \vec{v} \right] + \text{div}_v \left[ f_2(\vec{r},\vec{v},t) \langle \dot{\vec{v}} \rangle (\vec{r},\vec{v},t) \right] = 0, \qquad (\text{i.2})$$

where

$$f_1(\vec{r},t) \stackrel{\text{def}}{=} \int_{\mathbb{R}^3} f_2(\vec{r},\vec{v},t) d^3v, \quad f_1(\vec{r},t) \langle \vec{v} \rangle (\vec{r},t) = \int_{\mathbb{R}^3} f_2(\vec{r},\vec{v},t) \vec{v} d^3v, \qquad (\text{i.3})$$

$$f_2(\vec{r},\vec{v},t) \langle \dot{\vec{v}} \rangle (\vec{r},\vec{v},t) = \int_{\mathbb{R}^3} f_3(\vec{r},\vec{v},\dot{\vec{v}},t) \dot{\vec{v}} d^3\dot{v}.$$

Depending on the problem under consideration, the functions $f_n$ may have various physical interpretations. In hydrodynamic and gas-dynamic problems, $f_n$ corresponds to mass density; in electrodynamics to charge density; in quantum mechanics to the probability density of a single particle; and in statistical physics, plasma physics, astrophysics and high-energy physics to a distribution function or probability density. The index «$n$» indicates the dimension of the generalized phase space of higher kinematic quantities [22]. For example, when $n=1$ the function $f_1(\vec{r},t)$ is defined in coordinate space, when $n=2$ the function $f_2(\vec{r},\vec{v},t)$ is considered in phase space, and when $n=3$ the function $f_3(\vec{r},\vec{v},\dot{\vec{v}},t)$ is defined on an extended (generalized) phase space that takes the acceleration $\dot{\vec{v}}$ into account. It should be noted that the power of electromagnetic radiation is related to the force of radiative friction, which is proportional to $\langle \ddot{\vec{v}} \rangle$. The quantity $\langle \ddot{\vec{v}} \rangle$ appears in the third Vlasov equation and in the Lorentz–Abraham–Dirac equation [23, 24].

The vector field $\langle \vec{v} \rangle (\vec{r},t)$ corresponds to the flux velocity in coordinate space. Its second averaging over coordinate space with the function $f_1(\vec{r},t)$ yields $\langle \langle \vec{v} \rangle \rangle (t)$ – the velocity of the system's centre of «mass» (here «mass» is reffered to as a general meaning and in particular as probability). The field $\langle \dot{\vec{v}} \rangle (\vec{r},\vec{v},t)$ corresponds to the acceleration specified in phase space. Repeated averaging over the function $f_2(\vec{r},\vec{v},t)$ yields the acceleration field $\langle \langle \dot{\vec{v}} \rangle \rangle (\vec{r},t)$ at each coordinate point in the medium. Triple averaging $\langle \langle \langle \dot{\vec{v}} \rangle \rangle \rangle (t)$ leads to the acceleration of the system's centre of «mass».

Integrating the second Vlasov equation (i.2) over the entire velocity space transforms it into the first Vlasov equation (i.1). If we multiply equation (i.2) by the velocity component $v_k$ and integrate it over the velocity space, we obtain the equation of motion for continuum mechanics

$$\frac{d}{dt}\langle v_k \rangle = \left(\partial_t + \langle \vec{v} \rangle \cdot \nabla_r \right)\langle v_k \rangle = -\frac{1}{f_1}\partial_\lambda P_{k\lambda} + \langle \langle \dot{v}_k \rangle \rangle, \tag{i.4}$$

$$P_{k\lambda} = \int_{\mathbb{R}^3} f_2 \left(v_k - \langle v_k \rangle\right)\left(v_\lambda - \langle v_\lambda \rangle\right) d^3v, \tag{i.5}$$

where the second-order moment $P_{k\lambda}$ corresponds to the pressure tensor. Thus, the left-hand side of equation (i.4) contains the total time derivative of the medium's flow velocity $\langle v_k \rangle$, whilst the right-hand side contains the force density, consisting of the pressure force $-\partial_\lambda P_{k\lambda}/f_1$ and the external force $\sim \langle \langle \dot{v}_k \rangle \rangle$. Multiplying the second Vlasov equation by $v^2$ and subsequently integrating it over the velocity space yields the law of energy conservation

$$\partial_t \left[\frac{f_1}{2}|\langle \vec{v} \rangle|^2 + \frac{1}{2}\mathrm{Tr}\,P_{kk}\right] + \partial_s\left[\frac{f_1}{2}|\langle \vec{v} \rangle|^2\langle v_s \rangle + \frac{1}{2}\langle v_s \rangle \mathrm{Tr}\,P_{kk} + \langle v_k \rangle P_{ks} + \frac{1}{2}\mathrm{Tr}\,P_{kks}\right] = \int_{\mathbb{R}^3} f_2 \langle \dot{v}_k \rangle v_k\, d^3v,$$

$$P_{kns} = \int_{\mathbb{R}^3} \left(v_k - \langle v_k \rangle\right)\left(v_n - \langle v_n \rangle\right)\left(v_s - \langle v_s \rangle\right) f_2 d^3v. \tag{i.6}$$

The first term on the left-hand side of (i.6) corresponds to the time derivative ($\partial_t$) of the energy density, which consists of the sum of the kinetic energy $f_1|\langle \vec{v} \rangle|^2/2$ and the internal energy $\mathrm{Tr}\,P_{kk}/2$. The repeated indices mean summation (the Einstien rule). The second term on the left-hand side of (i.6) corresponds to the divergence ($\partial_s$) of the energy flux density. The term $f_1|\langle \vec{v} \rangle|^2\langle v_s \rangle/2$ is the kinetic energy flux density, whilst $\langle v_s \rangle \mathrm{Tr}\,P_{kk}/2$ is the internal energy flux density. The quantity $\langle v_k \rangle P_{ks}$ is related to the work done by gravitational forces, and $\mathrm{Tr}\,P_{kks}/2$ corresponds to the heat flux [20]. The integral on the right-hand side of equation (i.6) accounts for the power work of external forces.

Another important property of equations (i.1)-(i.2) is their connection with quantum mechanics. Equation (i.1) reduces to the electromagnetic Schrödinger equation upon substituting $f_1 = |\psi|^2 \geq 0$, $\psi = |\psi|\exp(i\varphi) \in \mathbb{C}$ and using the Helmholtz decomposition into potential and vortex fields [25, 26]

$$\langle \vec{v} \rangle (\vec{r},t) = -\alpha \nabla_r \Phi(\vec{r},t) + \gamma \vec{A}(\vec{r},t) =$$
$$= i\alpha \nabla_r \left( \ln \left| \frac{\psi}{\psi^*} \right| + i \operatorname{Arg} \frac{\psi}{\psi^*} \right) + \gamma \vec{A} = i\alpha \nabla_r \operatorname{Ln} \frac{\psi}{\psi^*} + \gamma \vec{A}, \quad (i.7)$$

$$\Phi(\vec{r},t) \stackrel{\text{def}}{=} 2\varphi + 2\pi k, \; k \in \mathbb{Z}, \quad (i.8)$$

where $\alpha, \beta, \gamma$ are constants. Substituting (i.7)-(i.8) into (i.1) yields the equations

$$\frac{i}{\beta} \partial_t \psi = -\alpha\beta \left( \hat{p} - \frac{\gamma}{2\alpha\beta} \vec{A} \right)^2 \psi + U\psi, \quad (i.9)$$

$$-\frac{1}{\beta} \partial_t \varphi = -\frac{1}{2\alpha\beta} |\langle \vec{v} \rangle|^2 + V = H, \quad V = U + Q, \quad Q = \frac{\alpha}{\beta} \frac{\Delta_r |\psi|}{|\psi|}, \quad (i.10)$$

$$\frac{d}{dt} = (\partial_t + \langle \vec{v} \rangle \cdot \nabla_r)\langle \vec{v} \rangle = -\gamma(\vec{E} + \langle \vec{v} \rangle \times \vec{B}), \quad (i.11)$$

$$\vec{E} \stackrel{\text{def}}{=} -\partial_t \vec{A} - \frac{2\alpha\beta}{\gamma} \nabla_r V, \quad \vec{B} \stackrel{\text{def}}{=} \operatorname{curl}_r \vec{A}, \quad (i.12)$$

$$\operatorname{div}_r \vec{A} + \frac{2\alpha\beta}{\gamma} \frac{1}{c^2} \partial_t V = 0, \quad (i.13)$$

where $\hat{p} \stackrel{\text{def}}{=} \frac{i}{\beta} \nabla_r$, $U(\vec{r},t) \in \mathbb{R}$. When the constants $\alpha = -\hbar/2m$, $\beta = 1/\hbar$, $\gamma = -q/m$ are chosen, equation (i.9) reduces to the electromagnetic Schrödinger equation for the wave function $\psi$. Here, $\hbar$ is the Planck constant, $m$ is the mass, and $q$ is the charge. The operator $\hat{p}$ takes on the role of the momentum operator, and the function $U$ that of the potential energy. Equation (i.10) coincides with the Hamilton–Jacobi equation, in which the energy V is the sum of the potential energy $U$ and the quantum potential Q. The quantum potential Q appears in the de Broglie-Bohm «pilot-wave» theory [27–29]. Equation (i.11) corresponds to the equation of motion of a charged particle in electromagnetic fields (i.12). The vortex field $\vec{A}(\vec{r},t)$ from the Helmholtz decomposition corresponds to the vector potential of the magnetic field $\vec{B}$. Condition (i.13) is the Lorenz $\Psi$-gauge and breaks down into the standard Lorenz gauge in field theory and the gauge for the quantum potential [26].

From a comparison of the equations of motion (i.4) and (i.11) in the absence of magnetic fields it follows that the gradient of the quantum potential (force) can be interpreted as the quantum pressure force

$$\partial_k Q = \frac{m}{f_1} \partial_\lambda P_{k\lambda}. \quad (i.14)$$

We also note two important special cases of the second Vlasov equation (i.2). The first case is when there are no sources of dissipation $Q_2 = \operatorname{div}_v \langle \dot{\vec{v}} \rangle = 0$ [3]. In this case, equation (i.2) reduces to the well-known Liouville equation. The second case is when $\langle \dot{\vec{v}} \rangle$ admits the Vlasov-Moyal approximation [30]

$$f_2(\vec{r},\vec{v},t)\langle \dot{v}_k \rangle(\vec{r},\vec{v},t) = \sum_{n=0}^{+\infty} \frac{(-1)^{n+1}(\hbar/2)^{2n}}{m^{2n+1}(2n+1)!} \partial_k U(\vec{r},t)\left(\overleftarrow{\nabla}_r \cdot \vec{\nabla}_v\right)^{2n} f_2(\vec{r},\vec{v},t). \qquad (i.15)$$

Substituting (i.15) into the second Vlasov equation (i.2) transforms it into the Moyal equation [31] for the Wigner function [32–33] $f_2(\vec{r},\vec{v},t) = m^3 W(\vec{r},\vec{p},t)$ of a quantum system in phase space. Note that averaging (i.15) over the velocity space yields an expression for the right-hand side of the equation of motion (i.4)

$$\langle\langle \dot{v}_k \rangle\rangle(\vec{r},t) = -\frac{1}{m}\partial_k U(\vec{r},t). \qquad (i.16)$$

Thus, all quantum corrections (terms with coefficients $\hbar^{2n}$) disappear. In the classical limit, when $\hbar \ll 1$, only the first term remains in the approximation (i.15), that is, $m\langle \dot{v}_k \rangle = -\partial_k U$.

From the expressions (i.1)-(i.16) given above, a deep fundamental connection between classical and quantum physics follows within the framework of the so-called Wigner-Vlasov formalism, based on the first two equations of the infinite self-linked Vlasov chain equations.

Consequently, knowing the solution to equation (i.1) or (i.2), one can construct solutions to the Schrödinger equation (i.9), the Hamilton-Jacobi equation (i.10), the equations of motion (i.4), (i.11), find the quantum potential (i.10) and the pressure force (i.14), construct the vector field of the flow of a continuous medium $\langle \vec{v} \rangle(\vec{r},t)$ or the probability flux (i.7), and determine the space distribution of matter $f_1(\vec{r},t)$ or the probability density. All the above mentioned makes it possible to obtain all the information regarding both classical and quantum systems.

The existence of exact solutions to nonlinear problems in mathematical physics is of great importance for at least two reasons. The first is methodological. The theory of nonlinear partial differential equations is considerably more complex than that of linear equations [34–35]. It is impossible to introduce the concept of a general solution and it is only possible to have a limited number of particular solutions. Finding each particular solution to a nonlinear equation sometimes requires particularly sophisticated mathematical methods [36–43]. Knowledge of the exact solution to a nonlinear system allows one to analyse the nature of its behaviour without resorting to computationally intensive calculations. The second aspect is the use of exact solutions in the design and optimization of real physical installations [44, 45]. The fact is that the numerical methods used in software packages are not «perfect». Obtaining a correct calculated value requires the selection of a finite difference scheme, its order of accuracy, proof of the convergence of the iterative process, and proof of the stability of the numerical method. Such theorems in the field of computational mathematics have been proven for a sufficiently broad class of linear equations [46]. For nonlinear equations, this issue has been studied to a much lesser extent. Therefore, when writing program code that numerically solves a nonlinear problem, the question of the correctness of the result obtained always arises. Moreover, every numerical method has a multitude of free parameters that must be selected in some way. For example, the PIC (Particle In Cell) method is widely used in modelling plasma physics, astrophysics, accelerator physics and hydrodynamics. In this context, the algorithm for distributing charge density or mass requires the selection of an adaptive finite-element mesh and the type of its approximation. Next comes the solution of the field equations, followed by the transition from the Eulerian to the Lagrangian grid, and then the solution of the equations of motion. The procedure is multi-stage. How can one be sure that the result obtained is correct?

What should it be compared with? Therefore, the availability of exact solutions to a nonlinear problem allows one to test the correctness of the numerical algorithm and evaluate its characteristic parameters. In some cases, it is possible to embed the exact solution of a nonlinear equation into the finite-difference scheme itself, thereby increasing the order of accuracy of the numerical algorithm [47–49].

The aim of this work is to construct exact solutions to nonlinear problems in classical and quantum physics based on the Wigner-Vlasov formalism. The starting point is the first Vlasov equation (i.1), in which the density function $f_1$ has an explicit dependence on the flux modulus $\langle v \rangle = |\langle \vec{v} \rangle|$ and an implicit dependence on the coordinates

$$f_1(x,y) = F\left[|\langle \vec{v} \rangle(x,y)|\right], \tag{i.17}$$

where $F$ is a function to be chosen. When discussing distribution functions in momentum (velocity) space, one cannot overlook the well-known class of generalized Maxwell distributions

$$F_{n,\ell}(z) = \frac{N_{n,\ell} z^\ell}{\sigma_{\langle v \rangle,n,\ell}^\ell 2^{\ell/2}} \exp\left(-\frac{z^n}{\sigma_{\langle v \rangle,n,\ell}^n 2^{n/2}}\right), \tag{i.18}$$

where $n > 0$, $\sigma_{\langle v \rangle,n,\ell}$ is a characteristic length in the velocity space $\langle v \rangle$, and $N_{n,\ell}$ is the normalisation factor in coordinate space. In the special case where $n = \ell = 2$, the distribution (i.18) $F_{2,2}(z)$ formally coincides with the well-known Maxwell distribution. The generalized Maxwell distribution (i.18) ($n > 1$) is used in various branches of physics. At $n = 4$ it transitions into the Drüwestein distribution [50] and describes the energy distribution of electrons in a weakly ionized gas under conditions where the electron interaction cross-section begins to depend on their velocity. These conditions arise, for example, in certain astrophysical systems [51] and plasma physics [52]. In addition, the generalized Maxwell distribution finds application in statistical physics. In the case of general-form power-law Hamiltonians and an arbitrary number of degrees of freedom, the Tsallis distribution yields the single-particle generalized Maxwell-Tsallis distribution [53]. With a specific choice of parameters, distribution (i.18) reduces to the Weibull distribution [54], which is widely used in statistics [55], engineering [56] and medicine [57].

Note that relation (i.17) is not coincidental, it is used in field theory when solving the nonlinear problem of magnetostatics [48, 58–60]. Substituting (i.17) into (i.1) yields a time independent nonlinear second-order partial differential equation

$$\text{div}_r\left\{F\left[|\alpha \nabla_r \Phi(x,y)|\right]\nabla_r \Phi(x,y)\right\} = 0, \tag{i.19}$$

where, in the Helmholtz decomposition (i.7), only the potential component is considered. Note that for non-smooth potentials $\Phi$, the field $\nabla_r \Phi$ may be vortical, i.e. $\text{curl}_r \nabla_r \Phi \neq \vec{0}$. In this case, it can be assumed that $\vec{A} \sim \nabla_r \Phi$. Such an example will be considered in this paper.

The paper is structured as follows. In §1 using the nonlinear Legendre transform, equation (i.19) is reduced into a linear partial differential equation with variable coefficients. The coordinates in the new space are velocities (momenta). Depending on the type of momentum domain, the equation is of elliptic, parabolic or hyperbolic type. The characteristic equations are derived and their properties investigated. The canonical form of the equations in each domain is obtained. In §2 exact solutions are constructed in momentum space. The method of characteristics and the search for a solution in factored form (the product of the radial and angular parts) are considered. The radial part of the solution is represented by the Kummer confluent hypergeometric function, whilst the angular part is a superposition of trigonometric functions or a linear angular dependence. The case where the Kummer function reduces to the generalized Laguerre polynomials is analysed in detail. In §3 a solution to the original nonlinear equation (i.19) is derived by means of the inverse Legendre transform from momentum space to coordinate space. For the factorized solution in momentum space, an explicit expression is obtained for the solution $\Phi$ (i.19), the quantum Q and classical $U$ potentials (i.10), the vector velocity field $\langle \vec{v} \rangle$ (i.7) and the density $f_1$ in coordinate space. A detailed physical analysis is carried out of the system's dynamics, the direction of the flows and the forces acting upon them. For the linear angular part in momentum space, an exact solution to the Schrödinger equation in coordinate space with a vortex field of probability flux velocity has been found. This solution belongs to the class of solutions of the so-called $\Psi$ - model of micro- and macro-systems [61]. Relationships between the characteristic parameters of the distribution (i.18) and the standard deviations have been derived from the Heisenberg uncertainty principle. The main perspectives of the work are presented in the conclusion. Proofs of theorems, lemmas and intermediate calculations are given in Appendices A and B.

## §1 Nonlinear Legendre transformation

Consider a two $(n, \ell)$ parametric probability density function $f_{n,\ell}$ satisfying the time independent first Vlasov equation (i.1)

$$f_{n,\ell}(x,y) = F_{n,\ell}\left[\left|\langle \vec{v} \rangle(x,y)\right|\right]. \tag{1.1}$$

The integral of the function $F_{n,\ell}(z)$ over the space of its argument $z$ takes the form

$$\int_0^{+\infty} F_{n,\ell}(z)\,dz = \frac{N_{n,\ell}}{n} \Gamma\left(\frac{\ell+1}{n}\right) \sigma_{\langle v \rangle, n, \ell} \sqrt{2}, \tag{1.2}$$

where $\Gamma$ is the gamma function. We shall consider the vector field of the probability flux to be a potential

$$\langle \vec{v} \rangle (x,y) = -\alpha \nabla_r \Phi(x,y). \tag{1.3}$$

It should be noted that in the case of a non-smooth potential $\Phi$, the expansion (1.3) may correspond to a vortex field, for example, the $\Psi$-model [61]. Substituting expressions (1.1) and (1.3) into the first Vlasov equation (i.1), we obtain (i.19)

$$\left[1 + \Phi_x^2 h_{n,\ell}(|\alpha \nabla_r \Phi|)\right]\Phi_{xx} + 2 h_{n,\ell}(|\alpha \nabla_r \Phi|)\Phi_x \Phi_y \Phi_{xy} + \left[1 + \Phi_y^2 h_{n,\ell}(|\alpha \nabla_r \Phi|)\right]\Phi_{yy} = 0, \tag{1.4}$$

$$h_{n,\ell}(z) \stackrel{\text{def}}{=} \frac{\alpha^2 F'_{n,\ell}(z)}{z F_{n,\ell}(z)} = \frac{\alpha^2}{z^2}\left(\ell - \frac{n z^n}{\sigma^n_{\langle v \rangle, n, \ell} 2^{n/2}}\right). \tag{1.5}$$

Equation (1.4) is a nonlinear partial differential equation. In the special case where the average velocities $\langle v \rangle$ satisfy the condition

$$\frac{\langle v \rangle}{\sigma_{\langle v \rangle, n, \ell} \sqrt{2}} \sim \left(\frac{\ell}{n}\right)^{1/n} \Rightarrow \Delta_r \Phi \sim 0, \tag{1.6}$$

i.e. the phase of the wave function is close to a harmonic function. In the general case, according to [62], the nonlinear equation (1.5) admits linearisation using the nonlinear Legendre transformation [63]

$$\omega(\xi, \eta) + \Phi(x, y) = x\xi + y\eta,$$
$$\xi = \Phi_x, \ \eta = \Phi_y, \ x = \omega_\xi, \ y = \omega_\eta, \tag{1.7}$$
$$\Phi_{xx} = J\omega_{\eta\eta}, \ \Phi_{xy} = -J\omega_{\xi\eta}, \ \Phi_{yy} = J\omega_{\xi\xi}, \ J = \Phi_{xx}\Phi_{yy} - \Phi_{xy}^2 = \left(\omega_{\xi\xi}\omega_{\eta\eta} - \omega_{\xi\eta}^2\right)^{-1},$$

where $J$ is the Jacobian of the Legendre transformation. Applying transformation (1.7) to equation (1.5), we arrive at a linear equation with respect to the function $\omega$

$$\left[1 + \xi^2 \bar{h}_{n,\ell}(\rho)\right]\omega_{\eta\eta} - 2\xi\eta \bar{h}_{n,\ell}(\rho)\omega_{\xi\eta} + \left[1 + \eta^2 \bar{h}_{n,\ell}(\rho)\right]\omega_{\xi\xi} = 0, \tag{1.8}$$
$$\bar{h}_{n,\ell}(\rho) \stackrel{\text{def}}{=} h_{n,\ell}(\langle v \rangle) = \frac{\ell}{\rho^2} - \frac{n|\alpha|^n \rho^{n-2}}{\sigma_{\langle v \rangle, n, \ell}^n 2^{n/2}},$$

where $\rho^2 = \xi^2 + \eta^2$, $|\alpha|\rho = \langle v \rangle$. We determine the type of equation (1.8) by calculating its determinant $\Delta = a_{12}^2 - a_{11}a_{22}$, where $a_{12}, a_{11}, a_{22}$ are the coefficients of the equation

$$\Delta_{n,\ell}(\rho) = \xi^2\eta^2 \bar{h}_{n,\ell}^2(\rho) - \left[1 + \xi^2 \bar{h}_{n,\ell}(\rho)\right]\left[1 + \eta^2{}_{n,\ell}(\rho)\right] = -1 - \rho^2 \bar{h}_{n,\ell}(\rho),$$
$$\Delta_{n,\ell}(\rho) = \frac{n|\alpha|^n \rho^n}{\sigma_{\langle v \rangle, n, \ell}^n 2^{n/2}} - 1 - \ell. \tag{1.9}$$

According to (1.9), the determinant $\Delta_{n,\ell}$ depends only on the polar radius $\rho$. Consequently, the regions of ellipticity, parabolicity and hyperbolicity of equation (1.8) have radial symmetry (see Fig. 1, right)

$$\begin{cases} \rho < \rho_T, \ \Delta_{n,\ell} < 0, \ \textit{elliptical type} \\ \rho = \rho_T, \ \Delta_{n,\ell} = 0, \ \textit{parabolic type}, \\ \rho > \rho_T, \ \Delta_{n,\ell} > 0, \ \textit{hyperbolic type}, \end{cases} \quad \frac{|\alpha|\rho_T}{\sigma_{\langle v \rangle, n, \ell}\sqrt{2}} = \frac{\langle v_T \rangle}{\sigma_{\langle v \rangle, n, \ell}\sqrt{2}} = \left(\frac{\ell+1}{n}\right)^{1/n}. \tag{1.10}$$

Note that the expression for the quantity $\rho_T$ (1.10) is close to the expression (1.6). Thus, in the vicinity of the parabolic region $\rho \sim \rho_T$ (see Fig. 1, right), the phase function $\Phi$ of the wave function is close to a harmonic function. Since the parameter $\sigma_{\langle v \rangle, n, \ell}$ in the distribution (1.1) is free, without loss of generality, we define it as

$$\sigma_{\langle v \rangle, n, \ell} \stackrel{\text{def}}{=} \frac{\sigma_{\langle v \rangle}}{\sqrt{2}}\left(\frac{n}{\ell+1}\right)^{1/n} \Rightarrow \langle v_T \rangle = \sigma_{\langle v \rangle}, \ \rho_T = \frac{\sigma_{\langle v \rangle}}{|\alpha|}, \tag{1.11}$$

where $\sigma_{\langle v \rangle}$ is a constant that determines the region (1.10) of ellipticity, parabolicity and hyperbolicity of equation (1.8).

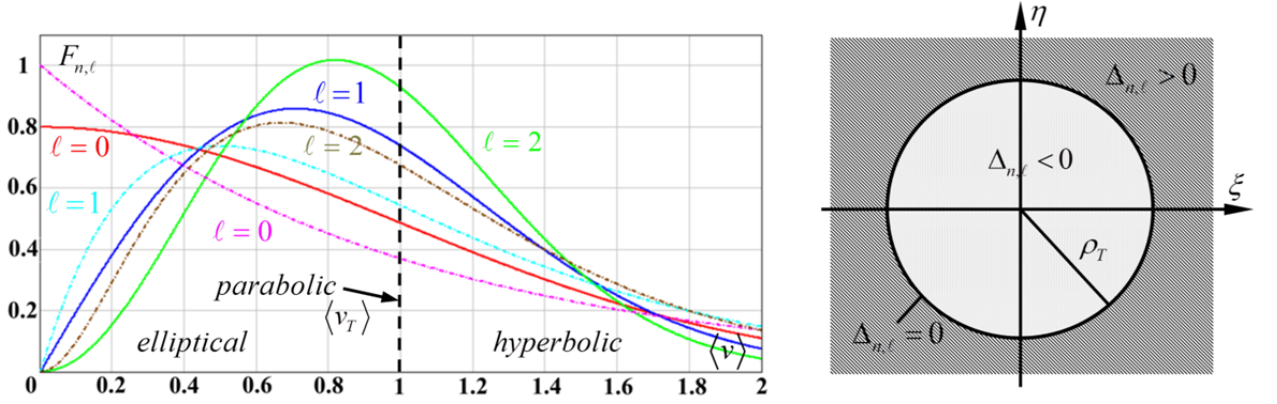

Fig. 1 Form of equation (1.14) depending on the region

For example, when $n = 2$ and $\ell = 0$, distribution (1.1) becomes a Gaussian distribution, and $\sigma_{\langle v \rangle}$ is its standard deviation. If we assume that the quantity $\sigma_{\langle v \rangle}$ formally satisfies the Heisenberg uncertainty principle $\sigma_{\langle v \rangle} \sigma_{\langle r \rangle} \geq \hbar/2$, then the region of ellipticity of equation (1.8) is bounded by $\rho_T \leq 1/\sigma_{\langle r \rangle}$. The representation (1.11) is convenient in that the quantity $\rho_T$ does not explicitly depend on the parameters $n, \ell$. Figure 1 (left) shows the distribution graphs of (i.18) for the parameters $n = 1, 2$ and $\ell = 0, 1, 2$. On the horizontal axis of Figure 1 (left), the quantity $\langle v \rangle$ is plotted in units of $\sigma_{\langle v \rangle}$. The solid line in the graph in Fig. 1 (left) corresponds to $n = 2$, whilst the dotted line corresponds to $n = 1$. According to (1.11), each distribution $F_{n,\ell}$ has its own value $\sigma_{\langle v \rangle, n, \ell}$, but $\langle v_T \rangle$ is the same for all of them (see Fig. 1, left).

Equation (1.8) takes a compact form in polar coordinates: $\xi = \rho \cos\theta$, $\eta = \rho \sin\theta$, $\omega(\xi, \eta) = u(\rho, \theta)$

$$u_{\rho\rho} + g_{n,\ell}(\rho)\left(\frac{1}{\rho}u_\rho + \frac{1}{\rho^2}u_{\theta\theta}\right) = 0, \qquad (1.12)$$

$$g_{n,\ell}(\rho) \stackrel{\text{def}}{=} 1 + \rho^2 \overline{h}_{n,\ell}(\rho) = \left(1 - \frac{\rho^n}{\rho_T^n}\right)(\ell + 1) = -\Delta_{n,\ell}(\rho) = -\rho^2 \overline{\Delta}_{n,\ell}(\rho), \qquad (1.13)$$

where $\overline{\Delta}_{n,\ell} = -g_{n,\ell}/\rho^2$ is the determinant of equation (1.12). Let us consider the question of reducing equation (1.12) to canonical form. The following statement holds.

**Theorem 1.** *If $\ell > -1$, then equation (1.12) has the characteristic*

$$\chi_{n,\ell}^{(h,\pm)}(\rho, \theta) = \frac{2}{n}\sqrt{\ell + 1}\left(\sqrt{\frac{\rho^n}{\rho_T^n} - 1} - \text{arctg}\sqrt{\frac{\rho^n}{\rho_T^n} - 1}\right) \pm \theta, \ \rho > \rho_T, \qquad (1.14)$$

$$\chi_{n,\ell}^{(e,\pm)}(\rho, \theta) = \frac{2}{n}\sqrt{\ell + 1}\left(\sqrt{1 - \frac{\rho^n}{\rho_T^n}} - \text{arcth}\sqrt{1 - \frac{\rho^n}{\rho_T^n}}\right) \pm \theta, \ 0 < \rho < \rho_T, \qquad (1.15)$$

*which reduce it to canonical form for the regions (1.10) of ellipticity and hyperbolicity, respectively*

$$\Omega_{\chi^{(+)}\chi^{(+)}} + \Omega_{\chi^{(-)}\chi^{(-)}} - \kappa_{n,\ell}^{(e)}\left[\Omega_{\chi^{(+)}} + \Omega_{\chi^{(-)}}\right] = 0, \qquad \Omega_{\chi^{(+)}\chi^{(-)}} + \kappa_{n,\ell}^{(h)}\left[\Omega_{\chi^{(+)}} + \Omega_{\chi^{(-)}}\right] = 0, \qquad (1.16)$$

$$\kappa_{n,\ell}^{(e)}(\rho) \stackrel{\text{def}}{=} \frac{n(\ell+1) + (n-2)\Delta_{n,\ell} + 2\Delta_{n,\ell}^2}{4(-\Delta_{n,\ell})^{3/2}}, \quad \kappa_{n,\ell}^{(h)}(\rho) \stackrel{\text{def}}{=} \frac{n(\ell+1) + (n-2)\Delta_{n,\ell} - 2\Delta_{n,\ell}^2}{8\Delta_{n,\ell}^{3/2}}, \qquad (1.17)$$

*where* $u(\rho,\theta) = \Omega\left[\chi_{n,\ell}^{(e/h,+)}(\rho,\theta), \chi_{n,\ell}^{(e/h,-)}(\rho,\theta)\right]$. *In the region of parabolicity,* $\rho = \rho_T$, *the solution* $u(\rho_T,\theta)$ *is an arbitrary smooth periodic function with respect to the argument* $\theta$, *that is* $u(\rho_T,\theta) = u(\rho_T,\theta+2\pi)$.

The proof of Theorem 1 is given in Appendix A.

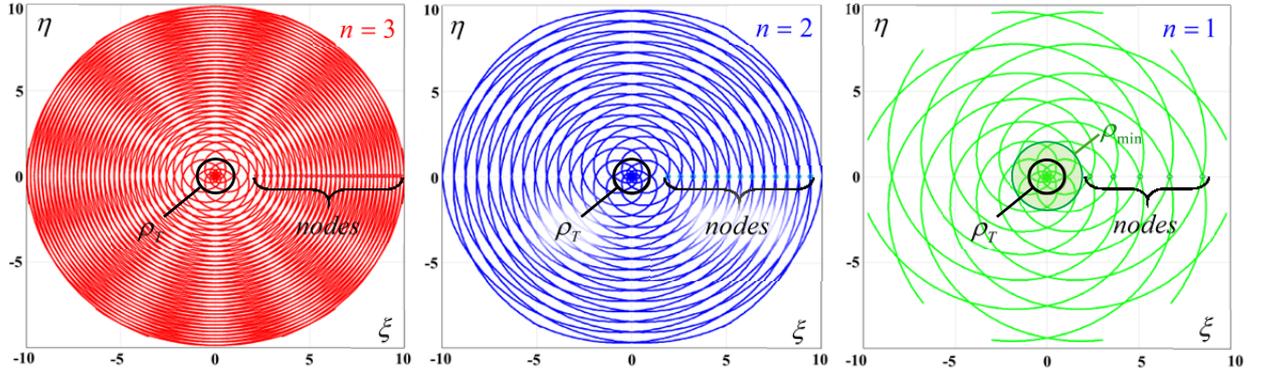

Fig. 2 Density of characteristic lines in the hyperbolicity region

Note that the implicit dependence of the functional coefficients (1.17) on the variables $\chi_{n,\ell}^{(e/h,\pm)}$ leads to difficulties in constructing exact solutions to the equations (1.16). Nevertheless, in certain special cases (see §2) it is possible to construct an exact solution to the hyperbolic equation (1.16) in explicit form. In this connection, let us examine in more detail the behaviour of the characteristic curves (1.14)-(1.15).

Figure 2 shows the graphs of the characteristic curves (1.14)-(1.15) for $\ell = 2$ and $n = 1, 2, 3$ in the three regions (1.10). The black line in Figure 2 indicates the boundary $\rho = \rho_T$ (a circle) on which equation (1.12) is parabolic. In Fig. 2, the main focus is on the region ($\rho > \rho_T$) of hyperbolicity of equation (1.12). In the elliptic region ($\rho < \rho_T$), the behaviour of the characteristics does not depend significantly on the parameter $n$. Therefore, as an example, Fig. 4 shows the characteristics for $n = \ell = 2$. The characteristics are plotted with a step size of $\pi/3$ along the angle $\chi_{n,\ell}^{(e/h,\pm)}$. The angle between the characteristics and the circle $\rho = \rho_T$ at their point of intersection is a straight line (see Fig. 4).

Fig. 2 shows that the density of the characteristic lines in the radial direction at $\rho > \rho_T$ (nodal points along the horizontal axis $\xi$) depends significantly on the parameter $n$. This dependence is determined by the graphs of the function $\rho'_\theta$ in the corresponding colours in Fig. 3. The values of the function $\rho'_\theta$ determine the tangent of the angle of inclination of the tangent to the graph of the characteristic (1.14)-(1.15)

$$\rho'_\theta(\rho) = \frac{\rho}{\sqrt{\pm\Delta_{n,\ell}(\rho)}}, \quad \begin{cases} +,\ \rho > \rho_T, \\ -,\ \rho < \rho_T, \end{cases} \qquad \lim_{\rho\to\infty}\rho'_\theta = \begin{cases} +\infty,\ 0 < n < 2, \\ \rho_T/\sqrt{\ell+1},\ n = 2, \\ 0,\ n > 2, \end{cases} \qquad (1.18)$$

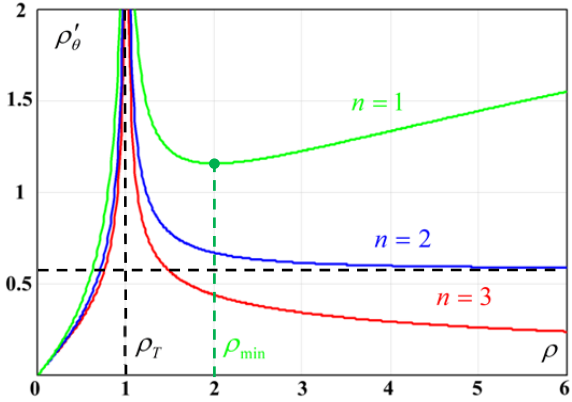

Fig. 3 Tangent of the angle of inclination of the tangent to the characteristic curve

where, for $\rho = \rho_T$, the determinant $\Delta_{n,\ell}(\rho_T) = 0$ and the function $\rho'_\theta$ tend to infinity, which corresponds to the angle of the tangent $\pi/2$ (see Fig. 4).

Figure 3 shows three graphs of the function (1.18) with the parameters $n = 1, 2, 3$ and $\ell = 2$. All three curves have a pole at the point $\rho = \rho_T$, which corresponds to the parabolic region of equation (1.12) and separates the elliptic region ($\rho < \rho_T$) from the hyperbolic region ($\rho > \rho_T$). The parameters $n = 1, 2, 3$ were not chosen at random. The fact is that, depending on the value of $n$, the behaviour of the tangent $\rho'_\theta$ (1.18) varies. When $n = 2$, the graph of the function (1.18) has a horizontal asymptote $\rho_T/\sqrt{\ell+1} \neq 0$ (see Fig. 3, blue line). Recall that the parameter $n = 2$ in distribution (i.18) corresponds to the Gaussian distribution (when $\ell = 0$) or the Maxwell distribution (when $\ell = 2$). In the case of $n = 2$ (see Fig. 2, blue curve), for large radii $\rho$ the characteristic curves have an almost constant slope towards the circle, which leads to the conservation of the distance between the characteristics (see Fig. 2, nodal points). For $n > 2$, the function $\rho'_\theta$ has a horizontal asymptote at zero (1.18) (see Fig. 3, red graph). As a result, the angle of inclination of the tangent to the characteristic curve tends to zero and the distance between the characteristics asymptotically tends to zero (see Fig. 2, red nodal points). When $n < 2$, the function $\rho'_\theta$ is non-monotonic. In Fig. 3 (green graph), the minimum of the function $\rho'_\theta$ is visible at the point $\rho_{\min} = \rho_T/\sqrt[n]{1-n/2}$. In the interval $\rho_T < \rho < \rho_{\min}$, a decrease in the angle of the tangent's slope is observed, and when $\rho > \rho_{\min}$, the angle of the slope begins to increase smoothly. Consequently, the distance between the characteristic curves increases indefinitely (see Fig. 2, green nodal points). In the region of ellipticity $0 < \rho < \rho_T$, the graphs of the characteristics (1.15) exhibit strong spiral curvature near the origin (see Fig. 4). Indeed, for $\rho \to 0$, the function (1.15) contains $\operatorname{arcth}\sqrt{1-\rho^n/\rho_T^n} \to \infty$.

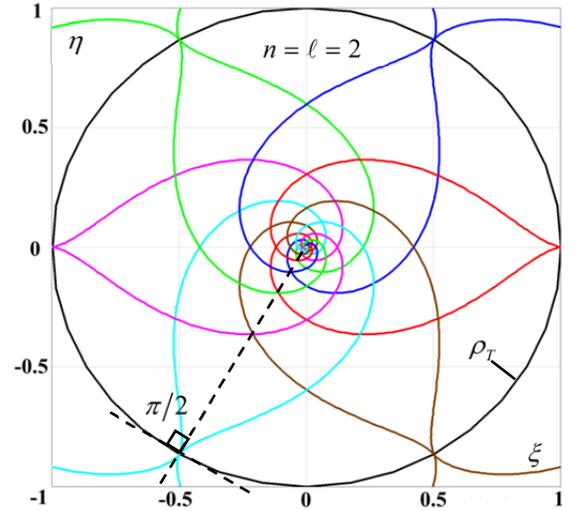

Fig. 4 Graphs of characteristics in the region ellipticity of equation (1.12)

Before proceeding to the direct derivation of exact solutions to equation (1.12), let us consider another form of its expression based on the method of separation of variables $u(\rho,\theta) \sim R(\rho)\Theta(\theta)$

$$R'' + g_{n,\ell}(\rho)\left(\frac{1}{\rho}R' - \frac{\lambda^2}{\rho^2}R\right) = 0, \quad \Theta'' + \lambda^2\Theta = 0, \qquad (1.19)$$

where $\lambda = const$. The function $\Theta$ takes the form

$$\Theta_\lambda(\theta) = \begin{cases} c_1\theta + c_2, & \lambda = 0, \\ c_1 \sin\lambda\theta + c_2 \cos\lambda\theta, & \lambda \neq 0, \end{cases} \qquad (1.20)$$

where $c_1, c_2$ are certain constants. In the case of periodic boundary conditions, $\Theta_\lambda(\theta + 2\pi) = \Theta_\lambda(\theta)$ takes the value $\lambda_k = k \in \mathbb{N}$. Note that satisfying the periodicity conditions is not mandatory. The reason is that, when performing the inverse Legendre transform back into coordinate space, to conserve the uniqueness of the transform, it is necessary to map only the angular sector $\theta_1 \leq \theta \leq \theta_2$. This issue will be discussed in more detail in §3.

Let us consider equation (1.19) in more detail for the function $R$.

**Theorem 2.** *Equation (1.19) for the function $R(\rho)$ reduces to the Hill equation for the function* $\bar{R}(\varsigma) = R[\rho(\varsigma)]$

$$\bar{R}''_{\varsigma\varsigma} + G_{n,\ell}(\bar{\rho})\bar{R} = 0, \qquad (1.21)$$

$$G_{n,\ell}(\bar{\rho}) = \begin{cases} \vartheta^2_{n,\ell}(\bar{\rho}), & \bar{\rho} \geq 1, \\ -\vartheta^2_{n,\ell}(\bar{\rho}), & \bar{\rho} < 1, \end{cases} \quad \vartheta_{n,\ell}(\bar{\rho}) = \frac{\lambda\sqrt{\ell+1}}{c_0 \rho_T} \bar{\rho}^\ell \sqrt{|\bar{\rho}^n - 1|} e^{-\frac{\ell+1}{n}\bar{\rho}^n}, \qquad (1.22)$$

*upon substitution*

$$\varsigma_{n,\ell}(\rho) = c_0 \rho_T \begin{cases} J_{n,k}(\bar{\rho}), & \ell = nk, \\ \displaystyle\sum_{k=0}^{+\infty} \frac{(\ell+1)^k \bar{\rho}^{nk-\ell}}{n^k(nk-\ell)k!}, & \ell \neq nk, \end{cases} \qquad (1.23)$$

$$J_{n,k}(x) \overset{def}{=} \begin{cases} \dfrac{1}{n}\mathrm{Ei}(\beta_{n,0}x^n), & k = 0, \\ \dfrac{\beta^k_{n,k}}{k\beta^{k-1}_{n,k-1}} J_{n,k-1}\!\left(\dfrac{\beta^{1/n}_{n,k}x}{\beta^{1/n}_{n,k-1}}\right) - \dfrac{e^{\beta_{n,k}x^n}}{knx^{kn}}, & k \in \mathbb{N}, \end{cases} \qquad (1.24)$$

$$\mathrm{Ei}(x) \overset{def}{=} \ln|x| + \sum_{k=1}^{+\infty}\frac{x^k}{k!\,k} + \gamma_e = \mathrm{v.p.}\int_{-\infty}^{x}\frac{e^t}{t}dt, \quad x \in \mathbb{R}, \qquad (1.25)$$

where $\bar{\rho} = \rho/\rho_T$, $\beta_{n,k} = k + 1/n$ and $\mathrm{Ei}(x)$ is the integral exponential function with the Euler–Mascheroni constant $\gamma_e$.

The proof of Theorem 2 is given in Appendix A.

**Remark 1.** For greater clarity when writing equation (1.21), the derivatives with respect to the variable $\varsigma_{n,\ell}$ are denoted by the derivatives with respect to the variable $\varsigma$. Thus, for each pair of parameters $n, \ell$ there is a corresponding variable substitution (1.23) and a function $G_{n,\ell}$. Note that the function $G_{n,\ell}$ explicitly depends on the variable $\rho$, whereas for the analysis of equation (1.21) a dependence on the variable $\varsigma$ is required. The explicit form of expressions (1.22)-(1.25) allows us to construct such a dependence. Fig. 5 (left) shows the distributions of $G_{n,\ell}$ with respect to $\varsigma_{n,\ell}$ for $n=2$ and $\ell>0$, $\ell_s^{(+)}=0.6+0.3s$, $s=0...4$, whilst Fig. 5 (right) shows the distributions for $n=2$ and $-1<\ell<0$, $\ell_s^{(-)}=-0.75+0.15s$.

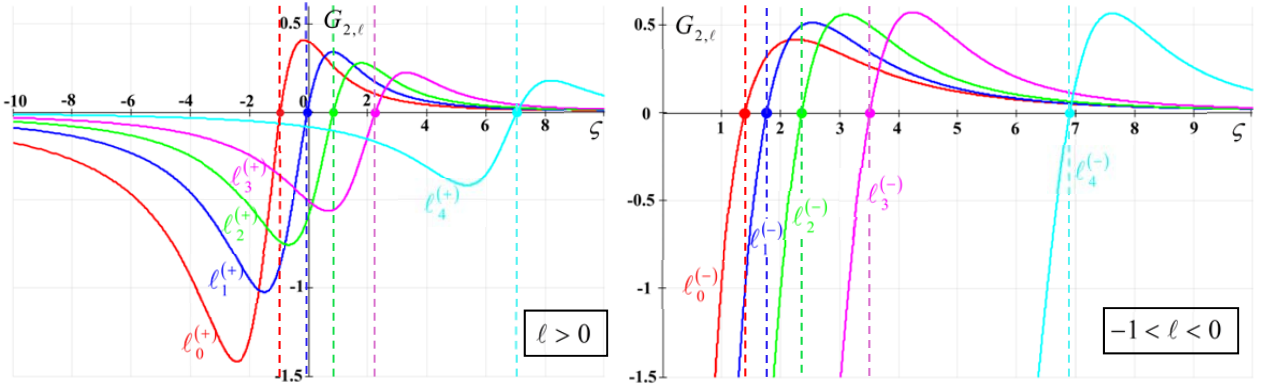

Fig. 5 Graphs of the frequency coefficient $G_{2,\ell}$ for equation (1.21)

The value $\varsigma_{n,\ell}(\rho_T)$ corresponds to the point where the graph of the function $G_{n,\ell}$ intersects the $\varsigma$ axis (see Fig. 5; for each graph, the point is marked with a different colour). At the point $\varsigma_{n,\ell}(\rho_T)$ equation (1.12) is of the parabolic type $G_{n,\ell}=0$. To the left of this point, at $\varsigma<\varsigma_{n,\ell}(\rho_T)$, equation (1.12) is of elliptic type and the value is $G_{n,\ell}<0$, whilst to the right, at $\varsigma>\varsigma_{n,\ell}(\rho_T)$, equation (1.12) is hyperbolic and $G_{n,\ell}>0$. The boundary between the regions of ellipticity and parabolicity in Fig. 5 is marked by a vertical dotted line. From a physical point of view, equation (1.21) is similar to the oscillation equation for a system with variable frequency $\sqrt{G_{n,\ell}}$. In the hyperbolic region ($G_{n,\ell}>0$), the frequency is a real function and is equal to $\vartheta_{n,\ell}$. Fig. 5 shows that near the point $\varsigma_{n,\ell}(\rho_T)$ the frequency $\vartheta_{n,\ell}$ changes sharply: a sharp rise, followed by a smooth asymptotic decline to zero. This behaviour is characteristic of both systems with $\ell>0$ (see Fig. 5, left) and systems with $-1<\ell<0$ (see Fig. 5, right). In the elliptic region $G_{n,\ell}<0$, so when $\varsigma<\varsigma_{n,\ell}(\rho_T)$, the frequency is a purely imaginary function $\pm i\vartheta_{n,\ell}$ and the oscillations degenerate. Furthermore, for positive values of the parameter $\ell>0$ (see Fig. 5, left), the function $\vartheta_{n,\ell}\to 0$ when $\varsigma_{n,\ell}\to -\infty$ ($\rho\to 0$). In this case, the solutions to equation (1.21) are close to linear functions, since $\bar{R}''_{\varsigma\varsigma}\to 0$. For negative values of the parameter $-1<\ell<0$ (see Fig. 5, right), the function $\vartheta_{n,\ell}\sim\rho^{\ell}\to\infty$ at $\rho\to 0$ ($\varsigma_{n,\ell}\to -\infty$), which may formally correspond to solutions of equation (1.21) with a sharp decline or rise.

We present another method for constructing a solution to the differential equation (1.19) for the component $R(\rho)$. From the theory of differential equations, it is known that the solution to equation (1.19) for the function $R(\rho)$ can be represented as the product of an analytic function and a singularity function at the pole $\rho=0$

$$R_{\lambda}^{(1)}(\rho)=\rho^{\nu_1}\sum_{k=0}^{+\infty}c_k^{(1)}\rho^k, \qquad (1.26)$$

where $\upsilon_1$ is the root with the largest real part of the characteristic equation

$$\upsilon^2 + \upsilon\left[g_{n,\ell}(0)-1\right] - \lambda^2 g_{n,\ell}(0) = 0. \tag{1.27}$$

If $\upsilon_2$ differs from $\upsilon_1$ by a non-integer amount and $\upsilon_2 \neq \upsilon_1$, then the second linearly independent solution is given by

$$R_\lambda^{(2)}(\rho) = \rho^{\upsilon_2} \sum_{k=0}^{+\infty} c_k^{(2)} \rho^k, \tag{1.28}$$

otherwise

$$R_\lambda^{(2)}(\rho) = c_0 R_\lambda^{(1)}(\rho) \ln\rho + \rho^{\upsilon_2} \sum_{k=0}^{+\infty} c_k^{(3)} \rho^k, \tag{1.29}$$

where $c_0$ is a constant, and $c_k^{(1)}$, $c_k^{(2)}$, $c_k^{(3)}$ are certain expansion coefficients to be determined, for example, by the Fubini method. It follows from expression (1.13) that $g_{n,\ell}(0) = \ell+1$, therefore, the roots of equation (1.27)

$$\upsilon_{1,2} = -\ell/2 \pm \sqrt{\ell^2/4 + \lambda^2(\ell+1)}, \qquad \upsilon_1 - \upsilon_2 = \sqrt{\ell^2 + 4\lambda^2(\ell+1)}. \tag{1.30}$$

From the condition $-1 < \ell$ it follows that the roots are $\upsilon_{1,2} \in \mathbb{R}$. Depending on the values of $\ell$ and $\lambda$, different variants of solutions (1.28) and (1.29) are possible. For example, for periodic conditions (1.20) $\lambda_k = k \in \mathbb{N}$ and for $\ell = 0$ (Gaussian distribution) according to (1.30) $\upsilon_1 - \upsilon_2 = 2k$, which leads to solution (1.29).

### §2 Exact solutions in the momentum representation

Let us proceed to the direct construction of exact solutions to equation (1.8) in momentum space. We begin with the hyperbolic domain (see Fig. 1) $\Delta_{n,\ell} > 0$ (1.10).

*Hyperbolic type*

**Theorem 3.** *In the hyperbolic domain $\Delta_{n,\ell} > 0$, for $\ell > -1$, equation (1.16) has a partial solution of the form*

$$\Omega\left[\chi_{n,\ell}^{(h,+)}, \chi_{n,\ell}^{(h,-)}\right] = \bar\Omega\left[\mu_{n,\ell}^{(+)}\right] \stackrel{\text{def}}{=} \frac{c_1\sqrt{\ell+1}}{n} e^{-\frac{\ell+1}{n}} \begin{cases} I_{n,k}(\varepsilon_n+1)+c_2, & \ell = nk, \\ \sum_{k=0}^{+\infty} \frac{(\ell+1)^k (\varepsilon_n+1)^{k-\ell/n}}{n^{k-1}(kn-\ell)k!} + c_2, & \ell \neq nk, \end{cases} \tag{2.1}$$

$$I_{n,k}(x) \stackrel{\text{def}}{=} \begin{cases} \text{Ei}(\beta_{n,0}x), & k = 0, \\ \dfrac{\beta_{n,k}^k}{k\beta_{n,k-1}^{k-1}} I_{n,k-1}\left(\dfrac{\beta_{n,k}x}{\beta_{n,k-1}}\right) - \dfrac{e^{\beta_{n,k}x}}{kx^k}, & k \in \mathbb{N}, \end{cases} \tag{2.2}$$

$$\mu_{n,\ell}^{(+)} \stackrel{\text{def}}{=} \frac{1}{2}\left[\chi_{n,\ell}^{(h,+)} + \chi_{n,\ell}^{(h,-)}\right] = \frac{2\sqrt{\ell+1}}{n}\left(\varepsilon_n^{1/2} - \text{arctg}\,\varepsilon_n^{1/2}\right), \tag{2.3}$$

*where $c_1, c_2$ are constant values and $\varepsilon_n = \bar{\rho}^n - 1$.*

The proof of Theorem 3 is given in Appendix A.

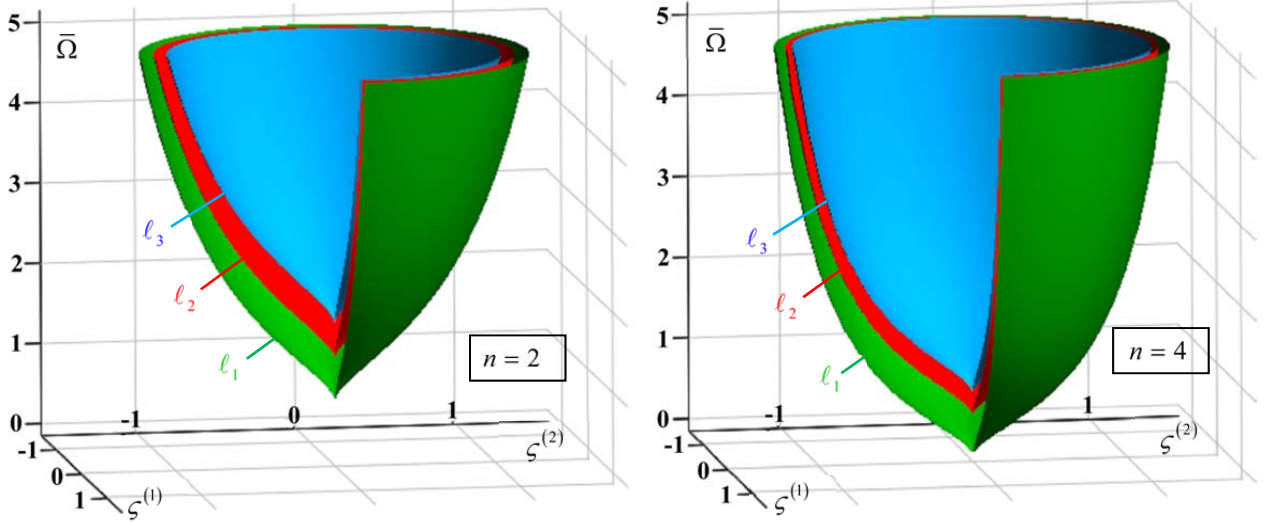

Fig. 6 Graphs of partial solutions to the hyperbolic equation (1.16)

**Remark 2.** Note that equation (1.16) has a coefficient $\kappa_{n,\ell}^{(h)}$, which depends implicitly on the variables $\chi_{n,\ell}^{(h,\pm)}$ and explicitly on the radius $\rho$. Nevertheless, the specified dependence (1.17) allows us to find a partial solution (2.1) to equation (1.16). This result is linked to the transition from the canonical coordinates $\chi_{n,\ell}^{(h,\pm)}$ (1.14) to the coordinate $\mu_{n,\ell}^{(+)}$, which depends only on $\rho$. The transition from $\chi_{n,\ell}^{(h,\pm)}$ to $\mu_{n,\ell}^{(\pm)}$ leads to a separation of the radial $\rho$ and angular $\theta$ dependencies. Indeed, according to (1.14) and (2.3), the variable $\mu_{n,\ell}^{(+)}$ is expressed solely in terms of $\rho$, whilst $\mu_{n,\ell}^{(-)} = \theta$. Thus, the solution (2.1) obtained possesses angular symmetry and does not depend on the angle $\theta$.

Fig. 6 shows graphs of the function (2.1) in the coordinate axes $\varsigma_{n,\ell}^{(1)} = \mu_{n,\ell}^{(+)} \cos \mu_{n,\ell}^{(-)}$, $\varsigma_{n,\ell}^{(2)} = \mu_{n,\ell}^{(+)} \sin \mu_{n,\ell}^{(-)}$. The transition from the coordinate system $\chi_{n,\ell}^{(h,\pm)}$ to the system $\mu_{n,\ell}^{(\pm)}$ leads to the «collapse» of the regions of ellipticity and parabolicity (see Fig. 1, right) into a single point – the origin (see Fig. 6). On the left-hand side of Fig. 6, the solution $\bar{\Omega}$ corresponds to the distribution parameter (i.18) $n = 2$, whilst on the right-hand side of Fig. 6 it corresponds to the parameter $n = 4$. In Fig. 6, for each value of $n$, three graphs have been plotted, corresponding to different values of the parameter $\ell_j = 0.3(j+1)$, $j = 1...3$. To illustrate the relative positions of the surfaces, their graphs in Fig. 6 have been plotted only for the range $0 \leq \mu_{n,\ell}^{(-)} \leq 3\pi/2$. A comparison of the graphs on the left and right in Fig. 6 shows that an increase in the parameter $n$ in the vicinity of the origin (for small $\mu_{n,\ell}^{(+)}$) leads to a gentler rise in $\bar{\Omega}$.

*Parabolic type*

The region of parabolicity of equation (1.12) is a circle of radius $\rho_T$ (see Fig. 1 on the right), on which, according to (1.13), $g_{n,\ell}(\rho_T) = 0$. Consequently, $u(\rho_T, \theta) = \bar{u}(\theta)$ is an arbitrary smooth function, for example, the expression (1.20).

*Elliptic type*

Let us map the region of ellipticity onto the unit circle using the change of variables employed in the previous theorems

$$R(\rho) = R(\bar{\rho}\rho_T) \stackrel{\text{def}}{=} \bar{R}(\bar{\rho}), \quad g_{n,\ell}(\rho) = g_{n,\ell}(\bar{\rho}\rho_T) \stackrel{\text{def}}{=} \bar{g}_{n,\ell}(\bar{\rho}) = (\ell+1)(1-\bar{\rho}^n), \quad (2.4)$$

$$\bar{R}''_{\bar{\rho}\bar{\rho}} + \bar{g}_{n,\ell}(\bar{\rho})\left(\frac{1}{\bar{\rho}}\bar{R}'_{\bar{\rho}} - \frac{\lambda^2}{\bar{\rho}^2}\bar{R}\right) = 0, \quad (2.5)$$

where $0 < \bar{\rho} < 1$. Knowing the explicit form of the coefficient $\bar{g}_{n,\ell}(\bar{\rho})$, equation (2.5) can be reduced to a well-known form.

**Theorem 4.** *Equation (2.5) for the function $\bar{R}(\bar{\rho})$ reduces to the Kummer equation for the function $T(\tau)$*

$$\tau T''_{\tau\tau} + \left[b^{(\pm)}_{n,\ell,\lambda} - \tau\right]T'_\tau - a^{(\pm)}_{n,\ell,\lambda}T = 0, \quad (2.6)$$

*where*

$$a^{(\pm)}_{n,\ell,\lambda} = \frac{v^{(\pm)}_{\ell,\lambda} - \lambda^2}{n}, \quad b^{(\pm)}_{n,\ell,\lambda} = \frac{2v^{(\pm)}_{\ell,\lambda} + n + \ell}{n}, \quad v^{(\pm)}_{\ell,\lambda} = -\ell/2 \pm \sqrt{(\ell/2)^2 + \lambda^2(\ell+1)}, \quad (2.7)$$

$$\bar{R}^{(\pm)}_{n,\ell,\lambda}(\bar{\rho}) = \bar{\rho}^{v^{(\pm)}_{\ell,\lambda}} T^{(\pm)}_{n,\ell,\lambda}(\tau), \quad \tau = \frac{\ell+1}{n}\bar{\rho}^n, \quad (2.8)$$

*whose solution $T^{(\pm)}_{n,\ell,\lambda}(\tau)$ can be expressed as a superposition of the Kummer function $M$ and the Tricomi function $\Psi$*

$$T^{(\pm)}_{n,\ell,\lambda}(\tau) = c_1 M\left(a^{(\pm)}_{n,\ell,\lambda}, b^{(\pm)}_{n,\ell,\lambda}, \tau\right) + c_2 \Psi\left(a^{(\pm)}_{n,\ell,\lambda}, b^{(\pm)}_{n,\ell,\lambda}, \tau\right). \quad (2.9)$$

The proof of Theorem 4 is given in Appendix B.

**Remark 3.** It follows from expression (2.7) that $v^{(+)}_{\ell,\lambda} > 0$, and $v^{(-)}_{\ell,\lambda} < 0$. The values of the numbers $v^{(\pm)}_{\ell,\lambda}$ coincide exactly with the previously given values of the roots $v_1, v_2$ (1.30). Therefore, by default, we shall assume the equivalence of the notations $v_1 = v^{(+)}_{\ell,\lambda}$ and $v_2 = v^{(-)}_{\ell,\lambda}$. Even the form of the solutions (1.26) and (1.28) is similar to the representation (2.8), with the sole difference that $\bar{R}^{(+)}_{n,\ell,\lambda}$ corresponds to the value $v^{(-)}_{\ell,\lambda}$ and $\bar{R}^{(-)}_{n,\ell,\lambda}$ to the value $v^{(+)}_{\ell,\lambda}$. Indeed, the roots $\bar{v}_{1,2}$ of the characteristic equation (1.27) for the Kummer equation (2.8) take the form

$$\begin{cases} \overline{\upsilon}_1 = 0, \ \overline{\upsilon}_2 = -\frac{1}{n}\sqrt{\ell^2 + 4\lambda^2(\ell+1)}, \ T_{n,\ell,\lambda}^{(+)}(\tau) \sim \tau^{\overline{\upsilon}_2}, \ "+", \\ \overline{\upsilon}_1 = \frac{1}{n}\sqrt{\ell^2 + 4\lambda^2(\ell+1)}, \ \overline{\upsilon}_2 = 0, \ T_{n,\ell,\lambda}^{(-)}(\tau) \sim \tau^{\overline{\upsilon}_1}, \ "-". \end{cases} \quad (2.10)$$

The behaviour of the function $T_{n,\ell,\lambda}^{(\pm)}(\tau)$ in (2.10) corresponds to the properties of the Kummer and Tricomi functions. It is known that the Kummer function $M$ is unbounded at infinity (the case $T_{n,\ell,\lambda}^{(-)}$), whilst the Tricomi function $\Psi$ is bounded at infinity (the case $T_{n,\ell,\lambda}^{(+)}$). Substituting the expressions (2.10) into the representation (2.8) yields the singularity functions $\overline{R}_{n,\ell,\lambda}^{(+)}(\overline{\rho}) \sim \overline{\rho}^{\upsilon_{\ell,\lambda}^{(-)}}$ (pole) and $\overline{R}_{n,\ell,\lambda}^{(-)}(\overline{\rho}) \sim \overline{\rho}^{\upsilon_{\ell,\lambda}^{(+)}}$ (zero).

The functions $M$ and $\Psi$ are also known as the Kummer functions of the first and second kind, and they are confluent hypergeometric functions. For example, $M(a,b,z) = {}_1F_1(a;b;z)$, whilst $\Psi$ can be expressed via $M$

$$\Psi(a,b,z) = \frac{\Gamma(1-b)}{\Gamma(a+1-b)} M(a,b,z) + \frac{\Gamma(b-1)}{\Gamma(a)} z^{1-b} M(a+1-b, 2-b, z). \quad (2.11)$$

Depending on the values of the parameters $a_{n,\ell,\lambda}^{(\pm)}$ and $b_{n,\ell,\lambda}^{(\pm)}$, the Kummer functions transform into a wide range of well-known orthogonal polynomials and special functions. As an example, consider the generalized Laguerre polynomials $L_k^{(\overline{\alpha})}(z)$ (see Appendix B).

**Lemma 1.** *Let $\ell > -1$ and $n > 0$, $k \in \mathbb{N}_0$ and $\lambda \geq 1$ then for the solution $M\left(a_{n,\ell,\lambda}^{(+)}, b_{n,\ell,\lambda}^{(+)}, z\right)$ of equation (2.6) to map onto the generalized Laguerre polynomials $L_k^{(\overline{\alpha})}(z)$ it is necessary that the parameters $(n,\ell)$ of the distribution (i.18) satisfy the conditions*

$$\begin{cases} a_{\ell,\lambda}^{(+)} = -k, \\ b_{\ell,\lambda}^{(+)} = 1+\overline{\alpha}, \end{cases} \Rightarrow \begin{cases} kn\ell = (\lambda^2 - kn)^2 - \lambda^2, \\ kn \leq \lambda\sqrt{\lambda^2 - 1}, \end{cases} \Rightarrow \begin{cases} M(-k, 1+\overline{\alpha}, z) = c_0 L_k^{(\overline{\alpha})}(z), \\ \overline{\alpha} \neq 0, \end{cases} \quad (2.12)$$

*where $c_0 \stackrel{\text{def}}{=} \Gamma(1+k)\Gamma(1+\overline{\alpha})/\Gamma(1+\overline{\alpha}+k)$ and $\mathbb{N}_0 = \mathbb{N} \cup \{0\}$.*

The proof of Lemma 1 is given in Appendix B.

Using equations (2.12), one can determine the parameters $(n,\ell)$ of the distribution (i.18) to which the solution of equation (2.6) in the form of generalized Laguerre polynomials corresponds. For example, for the distribution (i.18) with parameters $n = \ell = 2$, the following generalized Laguerre polynomials and their corresponding numbers are obtained $\lambda$

$$\begin{aligned} &\lambda = 1, L_0^{(2)}(z); &&\lambda = \sqrt{5}, L_1^{(4)}(z); &&\lambda = 2\sqrt{2}, L_2^{(5)}(z); \\ &\lambda = 4, L_5^{(7)}(z); &&\lambda = \sqrt{21}, L_7^{(8)}(z); &&\lambda = \sqrt{33}, L_{12}^{(10)}(z). \end{aligned} \quad (2.13)$$

In the case where $n = 2$ and the values of $\lambda \in \mathbb{N}$ from expressions (2.12) imply that

$$\lambda = 2 : \{\ell = 0,\ L_1^{(2)}(z)\};$$

$$\lambda = 3 : \{L_1^{(17)}(z),\ \ell = 20\};\ \{L_2^{(7)}(z),\ \ell = 4\};\ \{L_3^{(3)}(z),\ \ell = 0\}; \qquad (2.14)$$

$$\lambda = 4 : \begin{array}{l} \{\ell = 90,\ L_1^{(59)}(z)\};\ \{\ell = 32,\ L_2^{(28)}(z)\};\ \{\ell = 14,\ L_3^{(17)}(z)\};\ \{\ell = 6,\ L_4^{(11)}(z)\}; \\ \{\ell = 2,\ L_5^{(7)}(z)\};\ \{\ell = 0,\ L_6^{(4)}(z)\};\ \{\ell = -6/7 > -1,\ L_7^{(11/7)}(z)\}. \end{array}$$

The number of possible variants of (2.14) for each value of $\lambda$ is determined by the inequality (2.12) over the range of values $k$. From the condition $\bar{\alpha} \neq 0$ it follows that the Laguerre polynomials $L_k(z) = L_k^{(0)}(z)$ at $\ell > -1$ for the distribution (i.18) will be absent from the solution to equation (2.6).

Thus, for the case (2.12), the solution to equation (1.12) is constructed from a superposition of functions of the form

$$u_{n,\ell,\lambda}(\rho,\theta) = (-1)^{k(n,\ell,\lambda)} \left(\frac{\rho}{\rho_T}\right)^{\upsilon_{n,\ell}^{(+)}} L_{k(n,\ell,\lambda)}^{(b_{\ell,\lambda}^{(+)}-1)}\left(\frac{\ell+1}{n}\frac{\rho^n}{\rho_T^n}\right)\Theta_\lambda(\theta). \qquad (2.15)$$

Note that the solutions (2.15) hold not only in the elliptic region of equation (1.12), but also in the parabolic and hyperbolic regions. Fig. 7 shows the graphs of the function (2.15) for $n = 2$ and $\lambda = 3$. The values $\ell$ and $k$ are related by the relations (2.12) and were found earlier (2.14).

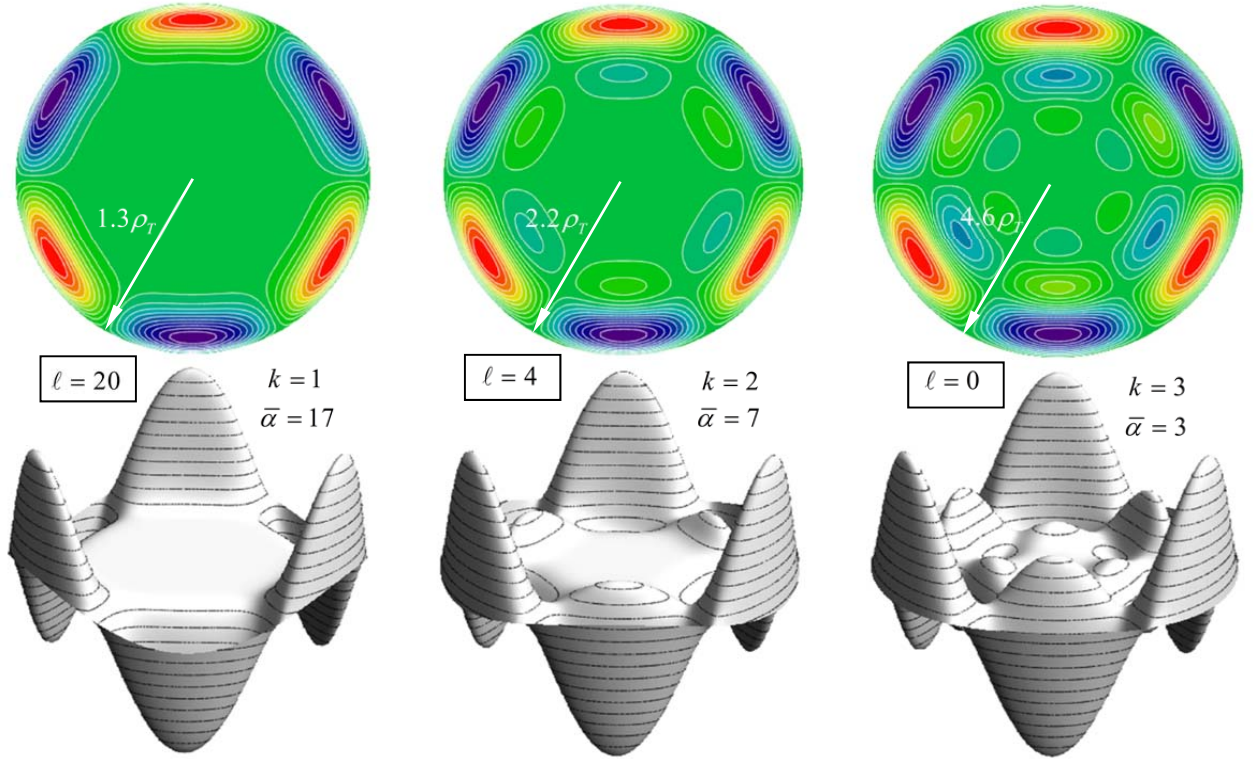

Fig. 7 Graphs of the solutions (2.15) to equation (1.12) for $n = 2$ and $\lambda = 3$

The lower part of Fig. 7 corresponds to three isometric projections of the graphs of the function (2.15). The upper part of Fig. 7 shows three top views illustrating the contour lines of the function (2.15). Red corresponds to the maximum values, and blue to the minimum values. Since the generalized Laguerre polynomials grow substantially (in magnitude) for large values of the argument, for clarity in Fig. 7 the values of the functions (2.15) are multiplied by the weighting function $\bar{\rho}^{\bar{\alpha}/2} e^{-\bar{\rho}/2}$. This weighting function coincides with the weighting function of the generalized Laguerre polynomials. In the region of ellipticity ($\rho < \rho_T$), the generalized Laguerre polynomials are monotonic functions. The appearance of radial oscillations (see Fig. 7) occurs at values of $\rho > \rho_T$. The number of such maxima and minima is determined by the order $k$ of the polynomial $L_k^{(\bar{\alpha})}$. Thus, in Fig. 7 on the left, there is only one radial maximum/minimum for $L_1^{(17)}$, whilst in Fig. 7 on the right there are three for $L_3^{(3)}$. The azimuthal oscillations in Fig. 7 are determined by function (1.20), which in this case is $\Theta_3(\theta) = \sin 3\theta$.

Thus, by specifying the parameters $(n, \ell)$ of distribution (1.1), one can obtain various explicit forms of solutions to equation (1.12) in the momentum space $(m\xi, m\eta)$. The original nonlinear equation (1.4) for the phase of the wave function $\varphi = \Phi/2$ is defined in coordinate space $(x, y)$. Consequently, it is necessary to construct the inverse Legendre transform (1.7) of the solutions to equation (1.12) in momentum space to the solutions to equation (1.4) in coordinate space.

## §3 Solution in coordinate representation

The original first Vlasov equation (i.19) is defined in coordinate space. In the preceding sections, the corresponding solutions to the first Vlasov equation in momentum space have been constructed. Thus, it is necessary to perform the inverse Legendre transform (1.7) from momentum space $-\alpha(m\xi, m\eta)$ to coordinate space $(x, y)$. Due to the nonlinear nature of the Legendre transform, the solutions of the linear equation with variable coefficients (1.8) will become particular solutions of the original nonlinear partial differential equation (i.19). As an example, let us consider the factorised solution $u \sim R(\rho)\Theta(\theta)$ of equation (1.8), described in detail in the previous paragraph.

**Theorem 5.** *Let $u_{n,\ell,\lambda}(\rho, \theta)$ be a factorised solution of equation (1.12), restricted to the origin, such that $\lambda \neq 0$ and $\lambda \neq 1$ then the corresponding particular solution $\Phi_{n,\ell,\lambda}$ of the nonlinear equation (1.4) takes the form*

$$\Phi_{n,\ell,\lambda}(x,y) = u_{n,\ell,\lambda}(\rho,\theta)\left[\mathcal{R}_{n,\ell,\lambda}(\rho) - 1\right], \quad \begin{pmatrix} x \\ y \end{pmatrix} = \frac{u_{n,\ell,\lambda}(\rho,\theta)}{\rho} \mathcal{Q}(\theta) \begin{pmatrix} \mathcal{R}_{n,\ell,\lambda}(\rho) \\ \Upsilon_\lambda(\theta) \end{pmatrix}, \quad (3.1)$$

$$\mathcal{R}_{n,\ell,\lambda}(\rho) \stackrel{\text{def}}{=} \upsilon_{\ell,\lambda}^{(+)} + (\ell+1)\frac{\rho^n}{\rho_T^n}\mathcal{M}\left(a_{n,\ell,\lambda}^{(+)}, b_{n,\ell,\lambda}^{(+)}, \frac{\ell+1}{n}\frac{\rho^n}{\rho_T^n}\right), \quad \mathcal{Q}(\theta) \stackrel{\text{def}}{=} \begin{pmatrix} \cos\theta & -\sin\theta \\ \sin\theta & \cos\theta \end{pmatrix}, \quad (3.2)$$

$$\mathcal{M}(a,b,\tau) \stackrel{\text{def}}{=} \frac{d}{d\tau}\ln M(a,b,\tau) = \frac{a}{b}\frac{M(a+1,b+1,\tau)}{M(a,b,\tau)}, \quad \Upsilon_\lambda(\theta) \stackrel{\text{def}}{=} \frac{d}{d\theta}\ln\Theta_\lambda = \lambda\frac{c_1 - c_2\,\text{tg}\,\lambda\theta}{c_1\,\text{tg}\,\lambda\theta + c_2}, \quad (3.3)$$

*where $c_1, c_2$ are constant values from the representation (1.20) of the function $\Theta_\lambda$. The Jacobian $J_{n,\ell,\lambda}$ of the Legendre transformation (1.7) takes the form*

$$J_{n,\ell,\lambda}^{-1}(\rho,\theta) = -\frac{u_{n,\ell,\lambda}^2(\rho,\theta)}{\rho^4}\{\mathcal{R}_{n,\ell,\lambda}^2(\rho)[g_{n,\ell}(\rho)+\Upsilon_\lambda^2(\theta)] - 2\mathcal{R}_{n,\ell,\lambda}(\rho)[\lambda^2 g_{n,\ell}(\rho)+\Upsilon_\lambda^2(\theta)] +$$
$$+ \lambda^4 g_{n,\ell}(\rho) + \Upsilon_\lambda^2(\theta)\}, \tag{3.4}$$
$$J_{n,\ell,\lambda}^{-1}(\rho_T,\theta_e) = 0, \tag{3.5}$$

*where the angle $\theta_e$ is determined by the condition $\Theta_\lambda'(\theta_e)=0$. If $\lambda=1$, then $J_{n,\ell,1}^{-1}(\rho,\theta)=0$ and the inverse Legendre transform is not possible.*

The proof of Theorem 5 is given in Appendix B.

**Remark 4.** The result of Theorem 5 extends to the case of the second linearly independent solution (2.9), expressed in terms of the Tricomi function $\Psi(a,b,\tau)$. In this case, the function $\mathcal{M}(a,b,\tau)$ takes the form $\Psi_\tau'(a,b,\tau)/\Psi(a,b,\tau)$. Since the function $\Psi(a,b,\tau)$, according to (2.11), can be explicitly expressed in terms of the Kummer function $M(a,b,\tau)$, the expression (3.3) for the updated function $\mathcal{M}(a,b,\tau)$ is known.

The angle $\theta_e$ corresponds to an extremum of the function $\Theta_\lambda$ and can be expressed as $\theta_e = (\pi/2 + \pi k - \theta_0)/\lambda$, $k \in \mathbb{Z}$, where $\mathrm{tg}\,\theta_0 = c_2/c_1$. The point $(\rho_T, \theta_e)$ lies in the parabolic region, in which the solution has only angular dependence (see §2). The Legendre transformation is a tangent transformation, thus (3.5) for $\Theta_\lambda'(\theta_e)=0$ is natural. In the case where $\lambda=0$, the results (3.1)-(3.3) of Theorem 5 remain valid for $\Upsilon_0(\theta) = c_1/(c_1\theta + c_2)$. For $\lambda=1$ the solution is $u_{n,\ell,\lambda}(\rho,\theta) \sim A\xi + B\eta + C$, i.e. it is the equation of a plane for which the uniqueness of the Legendre tangent transformation is lost.

Note that the function $u_{n,\ell,\lambda}(\rho,\theta)$ in the condition of Theorem 5 does not, in the general case, coincide with the function (2.15). The expression (2.15) is one of the possible forms of the function $u_{n,\ell,\lambda}(\rho,\theta)$. When considering the boundary value problem for equation (1.12) in a circular region of radius $\rho_0$, one can formulate the Dirichlet problem with the boundary condition $u(\rho_0,\theta) = u_0(\theta)$. In this case, the function $u_0(\theta)$ will yield the Fourier coefficients $c_1^{(k)}, c_2^{(k)}$ for the function (1.20) $\Theta_\lambda(\theta)$, where $\lambda_k = k \in \mathbb{N}$. The solution $u_{n,\ell}(\rho,\theta)$ to equation (1.12) will be expressed as a superposition of harmonics $u_{n,\ell,\lambda}(\rho,\theta)$ and the inverse Legendre transform can be applied to it.

In the inverse Legendre transform, the one-to-one correspondence may be violated in the form of a multivalent effect of the function. For example, a circular region in momentum space $0 \le \theta < 2\pi$ maps to a region with polar angle $0 \le \phi < \phi_0$, where $\phi_0 > 2\pi$. In such cases, in momentum coordinates, one may consider a region in the form of an angular sector.

As an example, we shall construct solutions in coordinate space for cases (2.14)-(2.15). We shall begin by mapping the solution from the momentum domain of ellipticity ($\rho < \rho_T$) to the coordinate domain. Fig. 8 shows the distributions of the flow velocity $\langle \vec{v} \rangle(x,y)$ (top three figures) and the density $f_{n,\ell}(x,y)$ (bottom three figures) for different values of $\lambda$ and $\ell$. The parameter $n=2$ is the same for all distributions.

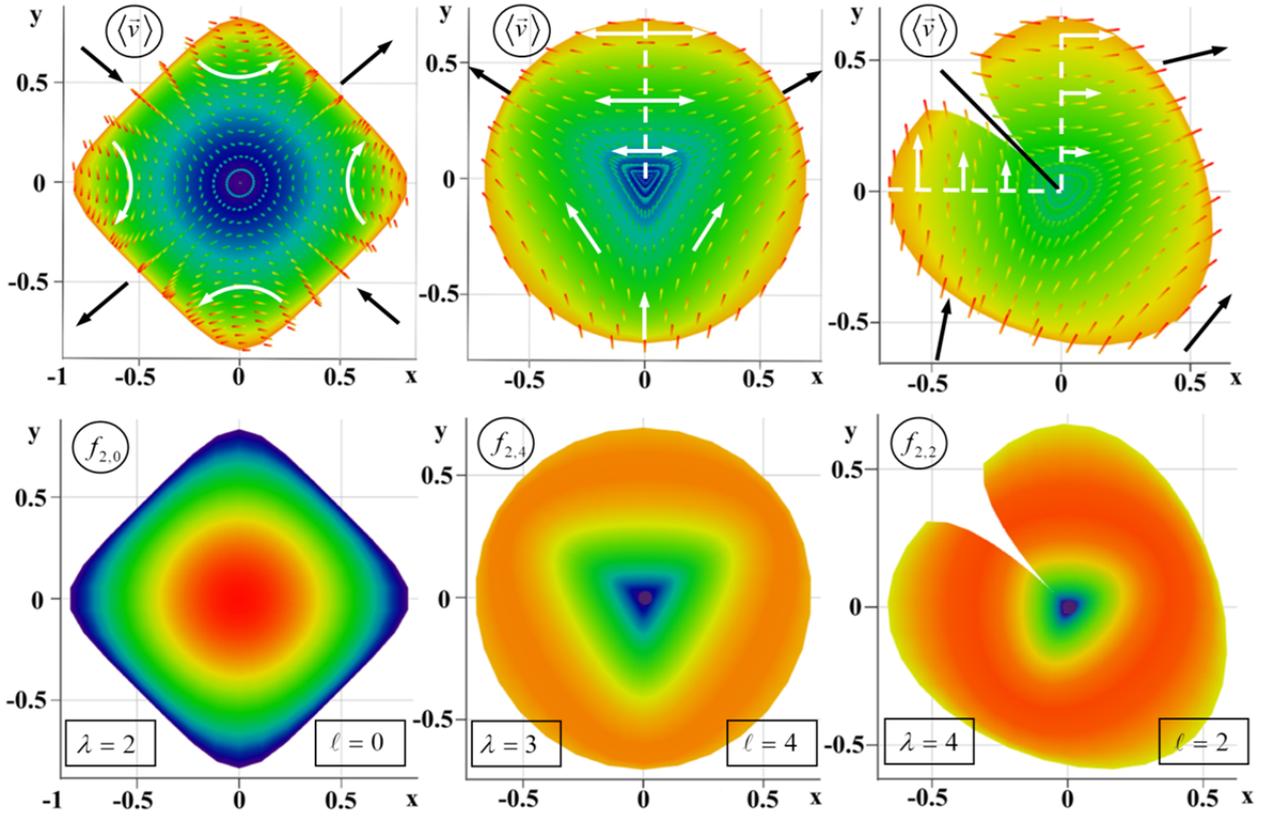

Fig. 8 Graphs of the velocity $\langle \vec{v} \rangle$ and density $f$ in a coordinate system for the elliptical case

For $\lambda = 2$, $\ell = 0$, the initial momentum region was a circle of radius $\rho_T$, which was mapped onto the coordinate region in the form of a «diamond» (see Fig. 8, top left and bottom left plots). In the case of $\lambda = 3$, $\ell = 4$, the momentum region was a semicircle $0 \leq \theta \leq \pi$, and the coordinate region was close to the shape of a «circle» (see Fig. 8, middle top and middle bottom graphs) with a vertical cut at $\phi = \pi/2$ (see Fig. 8, white dotted line). For the values $\lambda = 4$, $\ell = 2$, the momentum region is a circular sector of two types. The first type has an angle of $0 \leq \theta \leq \pi/2$ and the second $-12° \leq \theta \leq 102°$. The momentum region of the first type is mapped onto a circular sector with an angle of $-\pi \leq \phi \leq \pi/2$. In Fig. 8 (top right) the vertical and horizontal white dotted lines indicate the boundaries of the first-type region. The momentum region of the second type is mapped onto the coordinate space in the form of a «heart» or a «leaf» of a tree with an oblique cut along the black solid line (see Fig. 8, top right).

The red colour in Fig. 8 corresponds to the maximum values of the velocity distribution modulus $\langle v \rangle (x, y)$ and density $f_{2,\ell}(x, y)$, whilst the blue colour corresponds to the minimum values. The direction of the velocity vector field $\langle \vec{v} \rangle (x, y)$ in Fig. 8 (top three plots) is shown as small coloured dashes, as well as large white and black arrows.

Let us consider each of the three cases $\lambda = 2, 3, 4$ separately. We begin with the vector field $\langle \vec{v} \rangle (x, y)$ where $\lambda = 2$ and $\ell = 0$ (see Fig. 8, top left). A flow comes from two directions (top left and bottom right) (black arrows) and then goes out in two directions (top right and bottom left) (black arrows). Inside this region the diagonal incoming flows collide and after scattering come out in diagonal directions. It can be seen that in the vicinity of the centre of the flow collision, the velocity is close to zero (blue), whilst the maximum flow velocity is at the boundary of the region (red). At the corners of the region, white arrows indicate strong flow vortices $\langle \vec{v} \rangle$ caused by the proximity of the incoming and outgoing flows. Knowing the distribution $\langle v \rangle(x, y)$ one can find the corresponding density distribution $f_{n,\ell}(x, y)$ using formula (1.1), shown at the bottom of Fig. 8. Since $\ell = 0$, the distribution $f_{2,0}$ is the Gaussian distribution with a maximum at zero velocities (see Fig. 1, left). Indeed, in Fig. 8 (bottom left), the maximum density is at the centre, where there is almost no motion due to the collision of the flows.

In the case of $\lambda = 3$ and $\ell = 4$ (see Fig. 8, centre), the flow enters from below and exits from above on the right and left sides (black arrows) of the vertical section (white dotted line). The large white arrows show the characteristic directions of flow within the region, whilst the small coloured lines illustrate the flow in detail. The length of the white arrows corresponds to the magnitude of the flow velocity. Note that the vertical section (white dotted line) is the source of flow in the horizontal directions (to the right and left). The magnitude of the flow from the section increases with distance from the centre (the length of the horizontal white arrows increases). As in the previous case, the minimum velocity is in the central region. Since $\ell = 4$, the maximum of the density function $f_{2,4}$, unlike in the previous case, is attained at non-zero velocities. Indeed, in Fig. 8 (centre and bottom), large values (red) of the density $f_{2,4}$ are located at the periphery of the coordinate domain, whilst in the central region the density is minimal (blue).

As noted above, for the values $\lambda = 4$, $\ell = 2$, two types of coordinate domains were obtained. The domain of the first type is a special case of the domain of the second type. A notable feature, which is the reason for its separate consideration, is that the direction of the flow $\langle \vec{v} \rangle$ at its boundary (white dotted line) is normal (orthogonal). In Fig. 8 on the right, the white arrows indicate normal inflow through the vertical boundary and normal outflow through the horizontal boundary. As the angle of the solution domain increases, the problem of multi-leafness of the Legendre transformation for the coordinate domain may arise. In this case, a second-type domain with a wedge-shaped section is selected as an extended continuation of the first-type domain. As can be seen in Fig. 8 (right), the flow direction $\langle \vec{v} \rangle$ continues into the extended coordinate part of the second-type domain. There is flow at the boundary of the wedge-shaped section. The flow enters through the lower part of the section boundary and exits through the upper part of the section boundary. The flow velocity is minimal in the central region (blue and green) and increases with distance from it (orange and red). The length of the white arrows also indicates this. When $n = 2$ and $\ell = 2$, the density distribution (i.18) is effectively the Maxwell distribution (see Fig. 1, left) therefore, as in the previous case ($\lambda = 3$), the density $f_{2,2}$ (see Fig. 8, bottom right) is minimal at the centre of the region (minimum velocities) and increases in a certain neighbourhood (maximum velocity).

Let us note one common feature of the distributions shown in Fig. 8. All velocity distributions $\langle \vec{v} \rangle(x, y)$ contain a coordinate region in which the velocity is zero. This characteristic of the distributions $\langle \vec{v} \rangle(x, y)$ is determined by the shape of the initial momentum domain. Indeed, all the momentum domains shown were either a circle or a sector centred at the origin of the momentum space $-\alpha(m\xi, m\eta) = (0,0)$, i.e. with zero velocity. The shape of the momentum region influences the presence of different velocity directions. For example, in the first case for $\lambda = 2$ and $\ell = 0$, the momentum region was a circle, i.e. it contained all possible velocity directions $\langle \vec{v} \rangle$ ranging from 0 to $2\pi$. Looking at Fig. 8 (top left), it can be seen that the distribution $\langle \vec{v} \rangle(x, y)$ contains all directions from 0 to $2\pi$. In the second case, for $\lambda = 3$ and $\ell = 4$ the impulse region is a semicircle with velocity directions $\langle \vec{v} \rangle$ from 0 and $\pi$. Consequently, in Fig. 8 (centre, top), the flow moves from bottom to top and diverges to the right and left, corresponding to angles from 0 to $\pi$. A similar situation applies to the final case $\lambda = 4$

, $\ell = 2$. In the first type of momentum region, the direction of the velocities $\langle \vec{v} \rangle$ ranges from 0 to $\pi/2$, which is fully consistent with the distribution $\langle \vec{v} \rangle(x,y)$ in Fig. 8 (top right). The same situation applies to the second type of momentum region.

The second general feature of the distributions in Fig. 8 is the boundedness of the velocity magnitudes $\langle v \rangle < |\alpha| \rho_T$, caused by the size of the elliptic momentum domain ($\rho < \rho_T$) of equation (1.8). A greater variation in the magnitude of the velocity $\langle v \rangle$ can be obtained by mapping the hyperbolic momentum domain with $\rho > \rho_T$. Thus, the variety of directions of the velocity vector fields $\langle \vec{v} \rangle$ is determined by the shape of the momentum domain, whilst the range of velocity magnitudes $\langle v \rangle$ is determined by its dimensions.

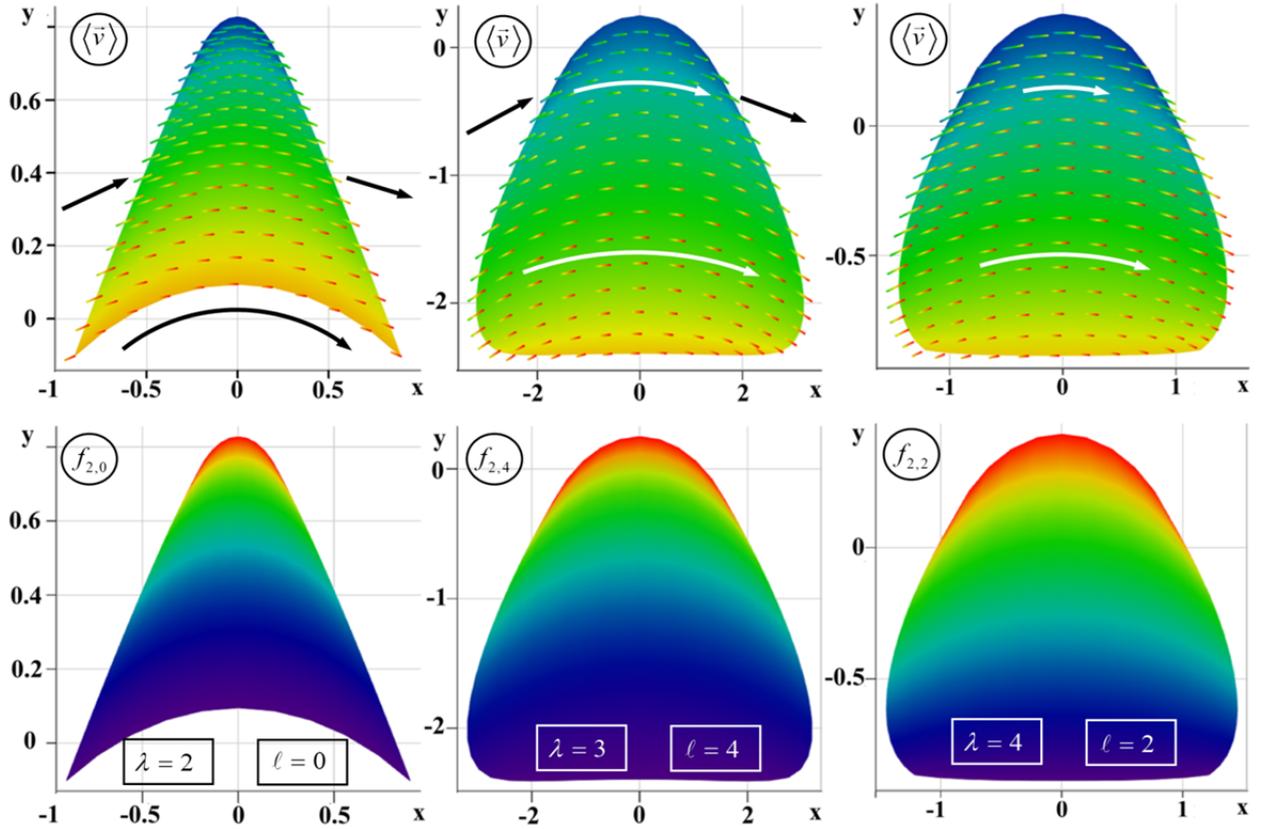

Fig. 9 Graphs of the velocity $\langle \vec{v} \rangle$ and density $f$ for the hyperbolic case

Figure 9 shows the distributions of velocities $\langle \vec{v} \rangle$ and densities $f_{2,\ell}$ for the momentum region in which equation (1.8) is of the hyperbolic type. The general structure of Fig. 9 and the notation used therein are analogous to the structure and notation introduced in Fig. 8. The same distribution parameters $(n,\ell)$ and values $\lambda$ are used. The momentum region is a radial sector

$$(\rho,\theta): \quad \rho_T < \rho_1 \leq \rho \leq \rho_2, \quad |\theta| \leq \theta_{max}, \qquad (3.6)$$

where the values $\rho_1, \rho_2$ and $\theta_{max}$ are chosen from the condition of univalent of the coordinate domain in the inverse Legendre transform. From the form of the domain (3.6), it follows that the spread in the velocity directions $\langle \vec{v} \rangle$ ranges from $-\theta_{max}$ to $\theta_{max}$ and in the magnitude $\langle v \rangle$ ranges from $|\alpha|\rho_1$ to $|\alpha|\rho_2$. The following parameters of the momentum domains were used in constructing the distributions in Fig. 9

$$\lambda = 2, \ \ell = 0: \ \rho_1 = 1.8\rho_T, \ \rho_2 = 2.4\rho_T, \ \theta_{\max} = 12°, \quad (3.7)$$

$$\lambda = 3, \ \ell = 4: \ \rho_1 = 1.5\rho_T, \ \rho_2 = 1.89\rho_T, \ \theta_{\max} = 15°, \quad (3.8)$$

$$\lambda = 4, \ \ell = 2: \ \rho_1 = 1.45\rho_T, \ \rho_2 = 1.75\rho_T, \ \theta_{\max} = 12°. \quad (3.9)$$

In all three upper graphs in Fig. 9, the velocity distributions are consistent with the form of the momentum domain (3.6), i.e. $\langle v_x \rangle > 0$, whilst $\langle v_y \rangle$ may be positive, negative or zero. It follows from expressions (3.7)-(3.9) that when $\lambda = 2,3,4$ the maximum $|\alpha|\rho_2$ and minimum $|\alpha|\rho_1$ values of the velocity decrease (the length of the white arrows in Fig. 9). This behaviour of the velocity $\langle v \rangle$ also determines the nature of the density distribution $f_{2,\ell}$ (the bottom three graphs in Fig. 9). Since the region of hyperbolicity is under consideration (see Fig. 1, left), the velocity values lie at the tail (tail) of the density distributions $f_{2,\ell}$. Consequently, the minimum velocity value $\langle v \rangle$ leads to the maximum possible density value $f_{2,\ell}$, whilst the maximum velocity $\langle v \rangle$ yields the minimum density $f_{2,\ell}$. Consequently, in Fig. 9 (density plots $f_{2,\ell}$), the upper part of the region has a red hue, whilst the lower part is blue. Since the minimum value $\langle v \rangle$ decreases as $\lambda = 2,3,4$, the size of the region with a red hue increases (see Fig. 9).

When considering the hyperbolicity region of equation (1.8), recall the solution $\bar{\Omega}\left[\mu_{n,\ell}^{(+)}\right]$ (2.1), which corresponds to the case $\lambda = 0$. Let us construct the inverse Legendre transform for it.

**Lemma 2.** *The solution* $\bar{\Omega}\left[\mu_{n,\ell}^{(+)}\right]$ *(2.1) of equation (1.16) for* $\Delta_{n,\ell} > 0$ *and* $\ell > -1$ *in momentum space corresponds to the solution* $\Phi_{n,\ell}(x,y)$ *of the nonlinear equation (1.4) in coordinate space of the form*

$$\Phi_{n,\ell}(x,y) = \rho \bar{\varsigma}_{n,\ell}(\rho) - R_{n,\ell}(\rho), \quad \begin{pmatrix} x \\ y \end{pmatrix} = \frac{\bar{\varsigma}_{n,\ell}(\rho)}{\rho} \begin{pmatrix} \xi \\ \eta \end{pmatrix}, \quad (3.10)$$

$$\bar{\varsigma}_{n,\ell}(\rho) = c_0 \left(\frac{\rho_T}{\rho}\right)^{\ell+1} \exp\left(\frac{\ell+1}{n} \frac{\rho^n}{\rho_T^n}\right), \quad \bar{\Omega}\left[\mu_{n,\ell}^{(+)}\right] = R_{n,\ell}(\rho), \quad (3.11)$$

*where* $R_{n,\ell}(\rho)$ *is the solution to equation (1.19) for* $\lambda = 0$, $c_0$ *is a constant and in accordance with (2.3), it is taken into account that* $\mu_{n,\ell}^{(+)} = \mu_{n,\ell}^{(+)}(\varepsilon_n)$ *and* $\varepsilon_n = \varepsilon_n(\rho)$.

The proof of Lemma 2 is given in Appendix B.

**Remark 5.** It follows from transformation (3.10) that $r = \bar{\varsigma}_{n,\ell}(\rho)$ and $\phi = \theta$, where $x = r\cos\phi$, $y = r\sin\phi$. Thus, there is an explicit relationship between the coordinate radius-vector $\vec{r}$ and the velocity $\langle \vec{v} \rangle(x,y) = |\alpha|\vec{r}$. Since there exists a minimum value $\rho_{\min} = \rho_T$ for the region of momentum hyperbolicity, there also exists a minimum radius (3.11) $r_{\min} = \bar{\varsigma}_{n,\ell}(\rho_T) = \exp\left[(\ell+1)/n\right]$ in coordinate space, where $c_0 = 1$.

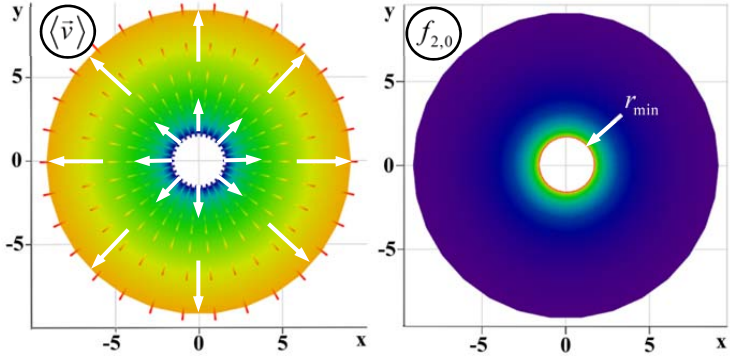

Fig. 10 Solution of (2.1) in the coordinate representation (3.10)

As an example, Fig. 10 (left) shows the velocity vector field $\langle \vec{v} \rangle (x,y)$ for the solution (3.10)-(3.11), whilst Fig. 10 (right) shows the corresponding density function $f_{2,0}(x,y)$ for $n=2$ and $\ell=0$. In constructing Fig. 10, an momentum domain in the form of a closed ring ( $0 \leq \theta < 2\pi$ ) with an inner radius of $\rho_T$ and an outer radius of $2.5\rho_T$ was used, which corresponds to the hyperbolic type of equation (1.16). As mentioned previously, when mapping the momentum domain onto the coordinate domain, there exists an «empty» region in the form of a circle of radius $r_{\min}$ (see Fig. 10, right). The flux is directed along the radial lines and increases with the radius (Fig. 10, left). According to Fig. 1 (left), the hyperbolic region lies at the «tail» of the distribution $f_{2,0}$, so the maximum density is attained in the neighbourhood of the minimum radius $r_{\min}$ (see Fig. 10, right).

**Theorem 6.** *Let $u_{n,\ell,\lambda}(\rho,\theta)$ be the factorised solution to equation (1.12), $\lambda \neq 0$, $\lambda \neq 1$ and $\ell > -1$ then the expression for the quantum potentia $lQ(r,\phi) = \bar{Q}(\rho,\theta)$ (i.10) takes the form*

$$\bar{Q}_{n,\ell,\lambda}(\rho,\theta) = \frac{\alpha \rho^2}{2\beta u_{n,\ell,\lambda}^2(\rho,\theta)} \mathcal{C}_{n,\ell,\lambda}\left[\mathcal{R}_{n,\ell,\lambda}(\rho) - 1, \mathcal{R}_{n,\ell,\lambda}(\rho) - \lambda^2, g_{n,\ell}(\rho), \Upsilon_\lambda(\theta)\right], \quad (3.11)$$

$$\mathcal{C}_{n,\ell,\lambda}(z_1, z_2, z_3, z_4) \stackrel{\text{def}}{=} \mathcal{A}_{n,\ell,\lambda}(z_1, z_2, z_3, z_4) + \mathcal{B}_{n,\ell}(z_1, z_2, z_3, z_4),$$

$$\mathcal{A}_{n,\ell,\lambda}(z_1, z_2, z_3, z_4) \stackrel{\text{def}}{=} \frac{z_3 - 1}{(z_1^2 z_4^2 + z_3 z_2^2)^3} \sum_{k=0}^{2} A_{n,\ell,\lambda}^{(k)}(z_1, z_2, z_3) z_4^{2k}, \quad (3.12)$$

$$\mathcal{B}_{n,\ell}(z_1, z_2, z_3, z_4) \stackrel{\text{def}}{=} \frac{z_1^2 z_4^2 + z_2^2}{(z_1^2 z_4^2 + z_3 z_2^2)^2}\left[\frac{1}{2}(z_3 - 1)^2 + (z_3 - 1)(n-1) - n\ell\right], \quad (3.13)$$

$$A_{n,\ell,\lambda}^{(k)}(z_1, z_2, z_3) \stackrel{\text{def}}{=} \begin{cases} z_2^3 \{\lambda^2 z_1 (1 - z_3) z_3 + z_2[(1-n)z_3 + n(\ell+1)]\}, & k = 0, \\ z_1 z_2 \{z_1 z_2 [2z_3^2 + (3+n)(1-z_3) + n\ell] + 3(1-z_3)(z_2^2 z_3 - \lambda^2 z_1^2)\}, & k = 1, \\ z_1^3 [z_1 - z_2(1 - z_3)], & k = 2, \end{cases} \quad (3.14)$$

*where the relationship between the coordinate $(x,y)$ and momentum $(\rho,\theta)$ representations is determined by the transformation (3.1).*

The proof of Theorem 6 is given in Appendix B.

Since the problem under consideration is time independent, the potential $U$ can be found from the Hamilton–Jacobi equation (i.10). Indeed, the phase $\varphi$ of the wave function $\psi$ takes the form

$$\varphi(\vec{r},t) = \bar{\varphi}(r,\phi) - \beta E t, \qquad (3.15)$$

where the constant $E$ corresponds to the energy of the system and the phase $2\bar{\varphi}(r,\phi) = \Phi(r,\phi)$ is the solution to the original nonlinear equation (1.4). Consequently, equation (i.10) takes the form

$$U_{n,\ell,\lambda}(r,\phi) = \frac{1}{4\alpha\beta}\left|\langle\vec{v}\rangle(r,\phi)\right|^2 - Q_{n,\ell,\lambda}(r,\phi) + E, \qquad (3.16)$$

where the quantum potential $Q_{n,\ell,\lambda}$ is given by expressions (3.11)-(3.14) and the kinetic energy is determined via the flux velocity $\langle v \rangle = |\alpha|\rho$.

**Remark 6.** When calculating the quantum potential $Q$, an expression of the form $Q(\vec{r}) = \tilde{Q}(\vec{r}) + C$ is obtained, where $C = const$ and $\tilde{Q}(\vec{r})$ is some function without a free constant. From a physical point of view, in the eigen coordinate system, the value is $C = E$. In this case, the potential (3.16) does not depend on an arbitrary constant. This situation is standard for quantum systems. A quantum system with potential $U$ may have a set of wave functions $\psi_s$ corresponding to various energy levels $E_s$. Each wave function $\psi_s$ corresponds to its own quantum potential

$$Q_s(\vec{r}) = -\frac{\hbar^2}{2m}\frac{\Delta_r|\Psi_s|}{|\Psi_s|} = \tilde{Q}(\vec{r}) + E_s. \qquad (3.17)$$

When substituting (3.17) into equation (3.16) the constant $E_s$ cancels out and the potential $U$ is the same for all quantum states with number $s$.

Fig. 11 shows the distributions of the potential energy $U_{n,\ell,\lambda}$ (3.16). For the sake of comparative analysis, the values of $n, \ell, \lambda$ correspond to those considered earlier in Fig. 8. The top three graphs in Fig. 11 show the isometric projection of the potential function $U_{n,\ell,\lambda}$ whilst the corresponding bottom three graphs are contour lines (top view) of the potential function. Since the potential $U_{n,\ell,\lambda}$ takes on large negative and positive values in the central region, the values $U_{n,\ell,\lambda}$ have been clipped to improve image contrast. Consequently, in Fig. 11 (bottom right and bottom centre), there are white areas where the values $U_{2,4,3}$ and $U_{2,2,4}$ exceed the set limits.

Let us consider the physical interpretation of the graphs in Fig. 11. We shall begin with the cases $\ell = 0$ and $\lambda = 2$ (the top-left and bottom graphs in Fig. 11). As can be seen in Fig. 11 (top-left), the potential energy surface $U_{2,0,2}$ is flat in the central region, but drops sharply as one moves away from it. Consequently, the external force $\vec{F} = -\nabla_r U$ will be small in the central region and large at its boundary. Since the potential energy decreases sharply along the radius, the force is directed away from the centre and is denoted in Fig. 11 (bottom) by the vector $\vec{F}_+$. This repulsive force counteracts the external, inflowing flows $\langle\vec{v}\rangle$ (see Fig. 8, top left). Upon encountering a sharp potential barrier, the flows $\langle\vec{v}\rangle$ reduce their velocity and enter the central region almost at rest (see the blue central region in Fig. 8, top left). As a result, the density in the central region increases $f_{2,0}$ (the red central region in Fig. 8, bottom left). The accumulated flow in the central region is pushed outwards by the force $\vec{F}_+$ (in Fig. 11, bottom left) along two diagonal directions (black arrows in Fig. 8, top left).

For $\ell = 4$ and $\lambda = 3$ (the central top and bottom graphs in Fig. 11), the potential energy surface $U_{2,4,3}$ has a complex shape. In the central region, there are three «petals» with an infinite potential barrier and three «petals» with an infinitely deep well. These potential barriers and wells in Fig. 11 (bottom centre) correspond to the white regions in the centre. In the vicinity of the potential barriers, there is a repulsive force $\vec{F}_+$, and in the vicinity of the potential wells there is an attractive force $\vec{F}_-$ (see Fig. 11, bottom centre). As shown in Fig. 8 (top centre), there is an incoming flow from below $\langle \vec{v} \rangle$, which encounters a potential barrier in the form of a lower vertical lobe (see Fig. 11, bottom centre). As a result, a repulsive force $\vec{F}_+$ acts on the incoming flow $\langle \vec{v} \rangle$, slowing it down. The potential barrier itself, in the form of a thin lobe, resembles a «knife» cutting the incoming flow in two (see Fig. 8, centre top). Each of the two resulting flows is attracted towards the central region by the force $\vec{F}_-$ from the potential well (the blue region in Fig. 11, centre bottom). Next comes another lobe of the potential barrier, which impedes rectilinear motion and directs the flows in diagonal directions towards the periphery of the region. A similar situation occurs with horizontal flows emerging from the vertical slit (see Fig. 8, top centre). Each horizontal flow formed in such a way comes to the potential barrier petal and exits the region in the corresponding diagonal direction. According to the density distribution $f_{2,4}$ (see Fig. 8, bottom centre), the main flow $\langle \vec{v} \rangle$ flows along the perimeter of the region and is scarcely present in its central part.

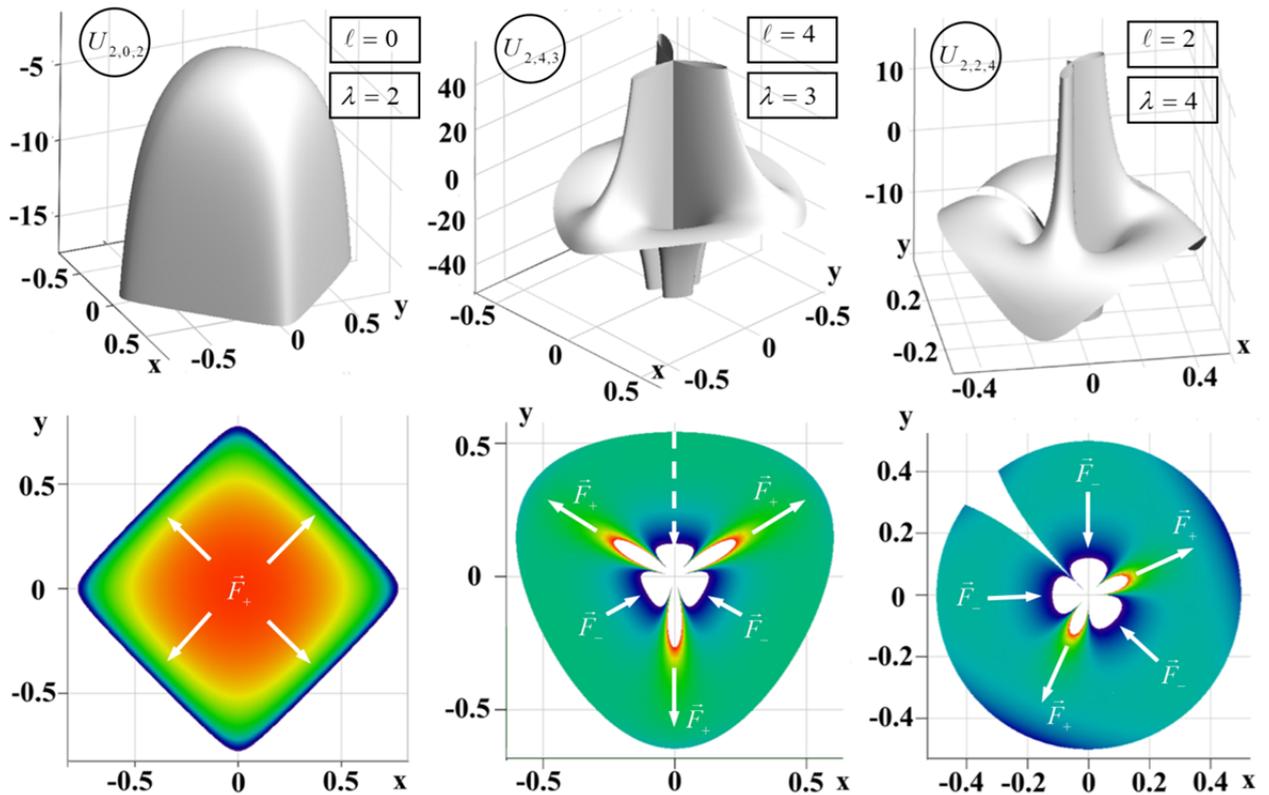

Fig. 11 Potential energy (3.16) and characteristic direction of external forces

In the latter case, for $\ell = 2$ and $\lambda = 4$ (see Fig. 11, top right and bottom), the potential energy surface $U_{2,2,4}$ has a complex shape, partly resembling the previous distribution $U_{2,4,3}$. In the region under consideration $U_{2,2,4}$ has two potential barriers and three potential wells in the shape of petals (see Fig. 11, bottom right). This potential energy configuration creates a characteristic distribution of forces $\vec{F}_{\pm}$ shown in Fig. 11 (bottom right). According to Fig. 8 (top right, black arrow), the flow $\langle \vec{v} \rangle$ enters the region from the bottom left diagonally. Next, the incoming flow is «split» by the potential barrier petal (see Fig. 11, bottom right) into two flows. One flow moves diagonally to the right, whilst the other moves vertically upwards towards the cut in the region (see Fig. 8, top right). The flow entering the upper part of the region from the split is attracted towards the centre by the potential well, and then encounters the potential barrier, which directs it diagonally towards the exit of the region. As in the previous case, the principal density $f_{2,2}$ is concentrated on the perimeter of the region (see Fig. 8, bottom right, red).

To conclude this section, let us consider a special case of the simplest factorised solution for $\lambda = 0$ and $R = const$.

**Theorem 7.** *Let $\ell > 2$, $\lambda = 0$ and $R = const$ then the factorised solution $u = \Theta_{\lambda=0}(\theta)$ of equation (1.12) corresponds to the solution $\psi_{n,\ell}(r,\phi,t)$ of the Schrödinger equation*

$$i\hbar \partial_t \psi_{n,\ell}(r,\phi,t) = -\frac{\hbar^2}{2m} \Delta_r \psi_{n,\ell}(r,\phi,t) + U_{n,\ell}(r) \psi_{n,\ell}(r,\phi,t), \qquad (3.18)$$

$$\psi_{n,\ell}(r,\phi,t) = \sqrt{N_{n,\ell}} \left( \frac{\ell+1}{n} \right)^{\ell/2n} \frac{\sigma_{\langle r \rangle}^{\ell/2}}{r^{\ell/2}} \exp\left( -\frac{\ell+1}{2n} \frac{\sigma_{\langle r \rangle}^n}{r^n} + i \frac{\rho_T \sigma_{\langle r \rangle}}{2} \phi - i \frac{E}{\hbar} t \right), \qquad (3.19)$$

$$N_{n,\ell}^{-1} = \frac{2\pi \sigma_{\langle r \rangle}^2}{n} \left( \frac{\ell+1}{n} \right)^{\frac{2}{n}} \Gamma\left( \frac{\ell-2}{n} \right), \quad \ell = 3, 4, \ldots \qquad (3.20)$$

*where the potential $U_{n,\ell}(r)$ and the quantum potential $Q_{n,\ell}(r)$ take the form*

$$Q_{n,\ell}(r) = -\frac{\hbar^2}{8mr^2}\left[ \ell^2 - \frac{2(\ell+1)(\ell+n)\sigma_r^n}{r^n} + \frac{(\ell+1)^2 \sigma_r^{2n}}{r^{2n}} \right], \qquad (3.21)$$

$$U_{n,\ell}(r) = -\frac{\hbar^2}{8mr^2}\left[ \rho_T^2 \sigma_{\langle r \rangle}^2 - \ell^2 + \frac{2(\ell+1)(\ell+n)\sigma_{\langle r \rangle}^n}{r^n} - \frac{(\ell+1)^2 \sigma_{\langle r \rangle}^{2n}}{r^{2n}} \right] + E, \qquad (3.22)$$

*and satisfy the Hamilton-Jacobi equation (i.10) with $\langle \vec{v} \rangle = \sigma_{\langle r \rangle} \sigma_{\langle v \rangle} \frac{\vec{e}_\phi}{r}$. The constant $c_1$ in the definition (1.20) of the function $\Theta_0(\theta)$ is taken to be equal to $c_1 = -\rho_T \sigma_{\langle r \rangle}$.*

The proof of Theorem 7 is given in Appendix B.

**Remark 7.** It follows from a comparison of expressions (3.17) and (3.21) that $E = 0$. Despite the singularity at the point $r = 0$, the limit $\lim_{r \to 0} \psi = 0$ since the exponential function tends to zero faster than the power function. Thus, expression (3.19) for the wave function can be defined at the singular point as $\psi(0,\phi,t) = 0$. Note that the asymptotic behaviour of the probability density $f_{n,\ell} \sim r^{-\ell}$ when $r \to \infty$, so the condition $\ell > 2$ is necessary for the convergence of the normalisation integral $2\pi \int_0^\infty f_{n,\ell} r dr$. Theorem 7 gives an explicit expression $N_{n,\ell}$ only when $\ell = 3, 4, 5,...$ although the normalisation integral exists for any finite $\ell > 2$.

Consider the form of the potential (3.22) in which there is a term with the coefficient $\rho_T^2 \sigma_{\langle r \rangle}^2$. A similar coefficient $\sigma_{\langle r \rangle} \sigma_{\langle v \rangle} = \dfrac{\hbar}{2m} \rho_T \sigma_{\langle r \rangle}$ is present in the probability flux $\langle \vec{v} \rangle$. Depending on the values of the coefficient $c_1 = \rho_T \sigma_{\langle r \rangle}$, three variants of the potential energy are possible $U_{n,\ell}(r)$

$$\begin{cases} \rho_T^2 \sigma_{\langle r \rangle}^2 - \ell^2 < 0, \\ \rho_T^2 \sigma_{\langle r \rangle}^2 - \ell^2 = 0, \\ \rho_T^2 \sigma_{\langle r \rangle}^2 - \ell^2 > 0, \end{cases} \Rightarrow \begin{cases} \sigma_{\langle r \rangle} \sigma_{\langle p \rangle} < (\hbar/2)\ell, \\ \sigma_{\langle r \rangle} \sigma_{\langle p \rangle} = (\hbar/2)\ell, \\ \sigma_{\langle r \rangle} \sigma_{\langle p \rangle} > (\hbar/2)\ell, \end{cases} \Rightarrow U_{n,\ell}(r) \sim \begin{cases} +1/r^2 \\ -1/r^{n+2}, \text{ for } r \to +\infty, \\ -1/r^2 \end{cases} \qquad (3.23)$$

where $\sigma_{\langle p \rangle} \overset{\text{def}}{=} m\sigma_{\langle v \rangle}$ is the characteristic length in terms of momentum. The variants of (3.23) differ in the behaviour of the potential (3.22) over large distances $r$ as well as in the presence of different numbers of zeros $r_j : U_{n,\ell}(r_j) = 0$, $j = 1, 2$

$$r_j = \sigma_{\langle r \rangle} \sqrt[n]{\ell+1} \begin{cases} \dfrac{1}{\sqrt[n]{2(\ell+n)}}, & \sigma_{\langle r \rangle} \sigma_{\langle p \rangle} < (\hbar/2)\ell, \\ \sqrt[n]{\dfrac{-\ell - n \pm \sqrt{D_{n,\ell}}}{\rho_T^2 \sigma_{\langle r \rangle}^2 - \ell^2}}, & \sigma_{\langle r \rangle} \sigma_{\langle p \rangle} = (\hbar/2)\ell, \quad D_{n,\ell} = n^2 + 2n\ell + \rho_T^2 \sigma_{\langle r \rangle}^2 > 0, \\ \sqrt[n]{\dfrac{\sqrt{D_{n,\ell}} - \ell - n}{\rho_T^2 \sigma_{\langle r \rangle}^2 - \ell^2}}, & \sigma_{\langle r \rangle} \sigma_{\langle p \rangle} > (\hbar/2)\ell. \end{cases} \qquad (3.24)$$

Fig. 12 shows the radial distributions of the potentials $U_{n,\ell}$ on the left, and the corresponding (by colour) probability density distributions $f_{n,\ell} = |\psi_{n,\ell}|^2$ in the centre, and the vector field of the probability flux $\langle \vec{v} \rangle$ on the right. All graphs in Fig. 12 are plotted for the values $n = 4$ and $\ell = 6$. The values of the characteristic coordinate dimensions $\sigma_{\langle r \rangle}^{(k)}$, $k = 1...3$ are taken based on the three conditions (3.23)-(3.24). For all values of $\sigma_{\langle r \rangle}^{(k)}$, $k = 1...3$ the potential $U_{4,6}$ has a pole at the origin (infinite potential barrier) so all probability densities $f_{4,6}$ have maxima at certain radii outside the central region. The maxima of the densities $f_{4,6}$ are located in finite potential wells. For $\sigma_{\langle r \rangle}^{(1)}$ (in Fig. 12, left and centre), the potential $U_{4,6}$ has two zeros $r_1$ and $r_2$ (3.24). The maximum probability (the area under the function $f_{4,6}$) is localised within $r_1 < r < r_2$ (red curve in Fig. 12, centre). The potentials for the values $\sigma_{\langle r \rangle}^{(2)}$ (blue graph) and $\sigma_{\langle r \rangle}^{(3)}$ (green graph) each have a single zero $r_1$ (3.24), but different orders of asymptotics at $r \to \infty$ (3.23) – $\sim -1/r^6$ and $\sim -1/r^2$ respectively.

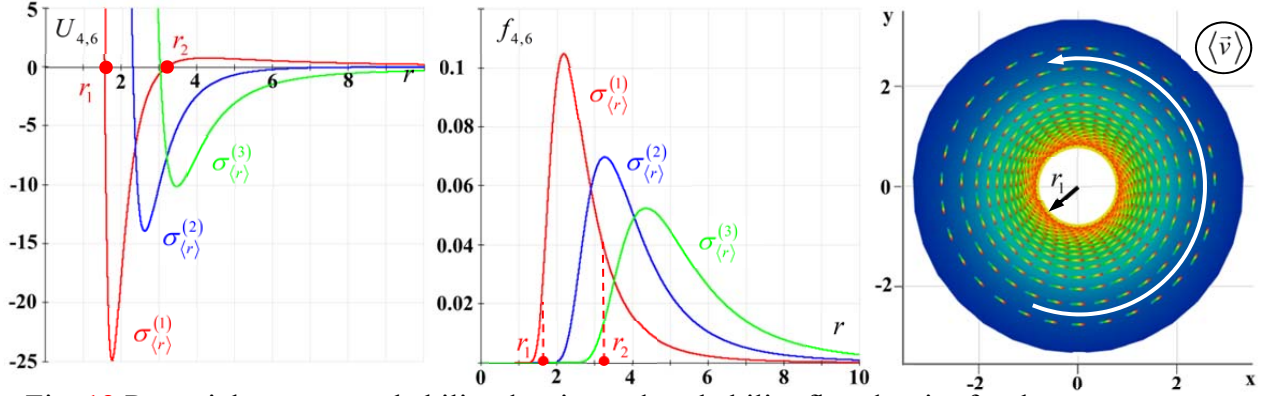

Fig. 12 Potential energy, probability density and probability flux density for the quantum system (3.18)-(3.22)

The vector field of the probability flux (Fig. 12, right) has a vortex structure $\langle \vec{v} \rangle = \sigma_{\langle r \rangle} \sigma_{\langle v \rangle} \frac{\vec{e}_\phi}{r}$. The white arrow on the right in Fig. 12 indicates the characteristic direction of the flux, which is anticlockwise. The velocity $\langle v \rangle$ of the probability flow is maximum in the central region (red in Fig. 12, right) and minimum (blue in Fig. 12, right) at a distance from it. A comparison of the graphs $f_{4,6}$ and $\langle v \rangle$ in Fig. 12 shows that in the vicinity of the pole ($r=0$) the probability density $f_{4,6}$ is negligibly small. For this reason, the circular region has been cropped out of the right-hand side of Fig. 12 of radius $r_1$ (the first zero (3.24) of the potential $U_{4,6}$ at $\sigma^{(1)}_{\langle r \rangle}$).

Note that the field $\langle \vec{v} \rangle$ according to (1.3) is potential, however $\mathrm{curl}_r \langle \vec{v} \rangle \neq \vec{0}$. The problem is that the potential $\Phi$ (the phase of the wave function $\varphi = \Phi/2$) is not a smooth function $\Phi(x,y) = -c_1 \phi$, i.e. there is a discontinuity at $\phi = 0$ and $\phi = 2\pi$. This peculiarity of the wave function's phase leads to the presence of a vortex in Fig. 12 (right). A consequence of this vortex is the natural fulfilment of the Bohr-Sommerfeld quantisation rule for momentum $\langle \vec{p} \rangle = m \langle \vec{v} \rangle$

$$\oint_{\gamma_0} \langle \vec{p} \rangle d\vec{l} = \frac{h}{2}|c_1|, \quad h = 2\pi\hbar, \tag{3.25}$$

where the integration is performed over the closed contour $\gamma_0$, containing the origin (the velocity pole $r=0$). Since $|c_1| > 0$ it can act as a quantum number $s$, for example, $|c_1| = 2s$. In this case, expression (3.25) coincides exactly with the Bohr-Sommerfeld quantisation rule.

**Remark 8.** In the general case, instead of the potential expansion (1.3) for representing the field $\langle \vec{v} \rangle$, one can use the full Helmholtz decomposition (i.7), containing the vortex component $\gamma \vec{A}$. In this case, instead of the Schrödinger equation (3.18), the electromagnetic equation (i.9) applies. From a physical point of view, there is only one quantum system with a vortex field $\langle \vec{v} \rangle$. Describing it using the Schrödinger equation (3.18) or (i.9) is equivalent, and is related to the gauge invariance of the Schrödinger equation itself. This issue is discussed in detail in [64], which also shows that for the field $\langle \vec{v} \rangle$ there exists positive calibration-invariant the Weyl-Stratonovich function of the electromagnetic quantum system. Note that the solution (3.19) is a special case of the solution for the so-called $\Psi$-model of micro- and macro-systems [61].

Note that the value $c_1$ is an arbitrary constant when constructing the solution for the angular part (1.20). In the formulation of Theorem 7, the value $c_1$ is formally redefined in terms of another constant $\sigma_{\langle r \rangle}$, that is $c_1 = -\rho_T \sigma_{\langle r \rangle}$. Looking at the inequalities on the right-hand sides of systems (3.23)-(3.24), one is reminded of a well-known inequality in quantum mechanics – the Heisenberg uncertainty principle $\sigma_r \sigma_p \geq \hbar/2$. A natural question arises regarding the relationship between these inequalities. Recall that in §1, the quantity $\sigma_{\langle v \rangle}$ was introduced as a characteristic length in momentum space (1.11) for the size of the ellipticity region ($\rho < \rho_T$) of equation (1.8) for the phase of the wave function. The quantity $\sigma_{\langle r \rangle}$ is a certain characteristic length in coordinate space, appearing in the expression for the potential $U_{n,\ell}$ and the velocity $\langle \vec{v} \rangle$. Nevertheless, there is a direct link between the quantities $\sigma_{\langle r \rangle}$ and $\sigma_r$. Indeed, given the probability density function $f_{n,\ell}$, one can calculate $\sigma_r$ (see Appendix B)

$$\sigma_r^2 = \langle r^2 \rangle - \langle r \rangle^2,$$

$$\sigma_r = \sigma_{\langle r \rangle} \left( \frac{\ell+1}{n} \right)^{\frac{1}{n}} \frac{1}{\Gamma\left(\frac{\ell-2}{n}\right)} \sqrt{\Gamma\left(\frac{\ell-4}{n}\right) \Gamma\left(\frac{\ell-2}{n}\right) - \Gamma^2\left(\frac{\ell-3}{n}\right)}, \ \ell = 5, 6, \ldots \quad (3.26)$$

A similar procedure can be carried out for the quantity $\sigma_{\langle v \rangle}$ by moving from the coordinate wave function (3.19) to the momentum function, and calculating $\sigma_p$. The quantity $\sigma_{\langle v \rangle}$ associated with the ellipticity region (1.11) is a parameter in the initial distribution $F_{n,\ell}$ (i.18) and allows conditions to be imposed on it.

**Conclusion**
Within the Wigner-Vlasov formalism, the exact solutions for coordinate space obtained in this work can be extended to phase space. Indeed, knowledge of the wave function $\psi(\vec{r}, t) = \sqrt{f(\vec{r})} \exp[i\varphi(\vec{r}, t)]$ allows one to find the Wigner function $W(\vec{r}, \vec{p}) = \frac{1}{(2\pi\hbar)^2} \int \psi^*(\vec{r}_-, t) \psi(\vec{r}_+, t) \exp(-\vec{p} \cdot \vec{s}/\hbar) d^2 s$, where $\vec{r}_\pm = \vec{r} \pm \vec{s}/2$. The function $W$ satisfies the Moyal equation or the second Vlasov equation (i.2) with the Vlasov-Moyal approximation (i.15). Note that when deriving the chain of Vlasov equations, no condition was imposed regarding the non-positivity of the functions $f_n$. Therefore, the presence of negative values in the quasi-probability density function $W$ does not contradict the chain of Vlasov equations.

**Appendix A**
*Proof of Theorem 1*

It follows from the theory of differential equations that the characteristic function is a solution to the following equation

$$\frac{d\theta}{d\rho} = \lambda_{1,2}(\rho, \theta) = \frac{a_{12} \pm \sqrt{\Delta_{n,\ell}}}{a_{11}} \Rightarrow \pm d\theta = \sqrt{\Delta_{n,\ell}(\rho)} \frac{d\rho}{\rho}, \qquad (A.1)$$

where, for equation (1.12), $a_{11} = 1$, $a_{12} = 0$ and (1.13) is taken into account. For the hyperbolic equation $\rho > \rho_T$ and $\Delta_{n,\ell}(\rho) > 0$. Therefore, the integration of expression (A.1) takes the form

$$\pm\theta + const = \int \sqrt{\Delta_{n,\ell}(\rho)} \frac{d\rho}{\rho} = \sqrt{\ell+1} \int \sqrt{\frac{\rho^n}{\rho_T^n} - 1} \frac{d\rho}{\rho} = \sqrt{\ell+1} \int \sqrt{\bar{\rho}^n - 1} \frac{d\bar{\rho}}{\bar{\rho}} = \frac{2}{n}\sqrt{\ell+1} \int \frac{t^2 dt}{1+t^2} =$$

$$= \frac{2}{n}\sqrt{\ell+1}(t - \operatorname{arctg} t) = \frac{2}{n}\sqrt{\ell+1}\left(\sqrt{\frac{\Delta_{n,\ell}}{\ell+1}} - \operatorname{arctg}\sqrt{\frac{\Delta_{n,\ell}}{\ell+1}}\right), \qquad (A.2)$$

where $\bar{\rho} = \rho/\rho_T > 1$, $t = \sqrt{\bar{\rho}^n - 1}$. From expression (A.2) it follows that (1.14) holds in the hyperbolic region $\rho > \rho_T$ of equation (1.12). In the elliptic region $\rho < \rho_T$, the determinant is $\Delta_{n,\ell}(\rho) < 0$ and equation (A.1) has an imaginary right-hand side $\pm\sqrt{\Delta_{n,\ell}} = \pm i\sqrt{-\Delta_{n,\ell}}$. The integration of such an equation is analogous to the procedure in (A.2)

$$\pm\theta + const = \int \sqrt{-\Delta_{n,\ell}(\rho)} \frac{d\rho}{\rho} = \sqrt{\ell+1} \int \sqrt{1 - \frac{\rho^n}{\rho_T^n}} \frac{d\rho}{\rho} = \sqrt{\ell+1} \int \sqrt{1 - \bar{\rho}^n} \frac{d\bar{\rho}}{\bar{\rho}} = -\frac{2}{n}\sqrt{\ell+1}\int \frac{t^2 dt}{1-t^2} =$$

$$= \frac{2}{n}\sqrt{\ell+1}(t - \operatorname{arcth} t) = \frac{2}{n}\sqrt{\ell+1}\left(\sqrt{\frac{-\Delta_{n,\ell}}{\ell+1}} - \operatorname{arcth}\sqrt{\frac{-\Delta_{n,\ell}}{\ell+1}}\right), \qquad (A.3)$$

where $t = \sqrt{1 - \bar{\rho}^n}$. Expression (A.3) proves the validity of (1.15). When $\rho = \rho_T$ (the parabolic region of the equation), both characteristics (A.2) and (A.3) coincide. Let us reduce equation (1.12) to canonical form. Using the representation (A.2) for the hyperbolic equation, we obtain

$$\partial_\rho \chi_{n,\ell}^{(h,\pm)} = \frac{2}{n}\left(1 - \frac{\rho_T^n}{\rho^n}\right) \partial_\rho \sqrt{\Delta_{n,\ell}} = \frac{\rho^n - \rho_T^n}{\rho^n} \frac{(\ell+1)\rho^{n-1}}{\rho_T^n \sqrt{\Delta_{n,\ell}}} = \left(\frac{\rho^n}{\rho_T^n} - 1\right) \frac{(\ell+1)}{\rho\sqrt{\Delta_{n,\ell}}},$$

$$\partial_\rho \chi_{n,\ell}^{(h,\pm)} = \frac{\sqrt{\Delta_{n,\ell}(\rho)}}{\rho} = \sqrt{\bar{\Delta}_{n,\ell}(\rho)}, \qquad (A.4)$$

$$\partial_\rho^2 \chi_{n,\ell}^{(h,\pm)} = \frac{n(\ell+1)\frac{\rho^n}{\rho_T^n} - 2\Delta_{n,\ell}}{2\rho^2 \sqrt{\Delta_{n,\ell}}} = \frac{(n-2)\Delta_{n,\ell} + n(\ell+1)}{2\rho^2 \sqrt{\Delta_{n,\ell}}}. \qquad (A.5)$$

The partial derivatives of equation (1.12) take the form

$$u_\rho = \Omega_{\chi^{(+)}} \chi_\rho^{(+)} + \Omega_{\chi^{(-)}} \chi_\rho^{(-)},$$

$$u_{\rho\rho} = \Omega_{\chi^{(+)}\chi^{(+)}}\left[\chi_\rho^{(+)}\right]^2 + 2\Omega_{\chi^{(+)}\chi^{(-)}}\chi_\rho^{(+)}\chi_\rho^{(-)} + \Omega_{\chi^{(-)}\chi^{(-)}}\left[\chi_\rho^{(-)}\right]^2 + \Omega_{\chi^{(+)}}\chi_{\rho\rho}^{(+)} + \Omega_{\chi^{(-)}}\chi_{\rho\rho}^{(-)}, \quad (A.6)$$

$$u_{\theta\theta} = \Omega_{\chi^{(+)}\chi^{(+)}} - 2\Omega_{\chi^{(+)}\chi^{(-)}} + \Omega_{\chi^{(-)}\chi^{(-)}}.$$

Substituting expressions (A.4)-(A.6) into equation (1.12), we obtain

$$0 = \left[\Omega_{\chi^{(+)}\chi^{(+)}} + 2\Omega_{\chi^{(+)}\chi^{(-)}} + \Omega_{\chi^{(-)}\chi^{(-)}}\right]\frac{\Delta_{n,\ell}}{\rho^2} + \left[\Omega_{\chi^{(+)}} + \Omega_{\chi^{(-)}}\right]\frac{n(\ell+1)+(n-2)\Delta_{n,\ell}}{2\rho^2\sqrt{\Delta_{n,\ell}}} -$$

$$-\frac{\Delta_{n,\ell}}{\rho}\left[\Omega_{\chi^{(+)}} + \Omega_{\chi^{(-)}}\right]\frac{\sqrt{\Delta_{n,\ell}}}{\rho} - \frac{\Delta_{n,\ell}}{\rho^2}\left[\Omega_{\chi^{(+)}\chi^{(+)}} - 2\Omega_{\chi^{(+)}\chi^{(-)}} + \Omega_{\chi^{(-)}\chi^{(-)}}\right] =$$

$$= 4\frac{\Delta_{n,\ell}}{\rho^2}\Omega_{\chi^{(+)}\chi^{(-)}} + \frac{1}{\rho^2}\left[\Omega_{\chi^{(+)}} + \Omega_{\chi^{(-)}}\right]\left[\frac{n(\ell+1)+(n-2)\Delta_{n,\ell}}{2\sqrt{\Delta_{n,\ell}}} - \Delta_{n,\ell}\sqrt{\Delta_{n,\ell}}\right] =$$

$$= 4\frac{\Delta_{n,\ell}}{\rho^2}\Omega_{\chi^{(+)}\chi^{(-)}} + \frac{n(\ell+1)+(n-2)\Delta_{n,\ell}-2\Delta_{n,\ell}^2}{\rho^2 2\sqrt{\Delta_{n,\ell}}}\left[\Omega_{\chi^{(+)}} + \Omega_{\chi^{(-)}}\right],$$

$$\Omega_{\chi^{(+)}\chi^{(-)}} + \frac{n(\ell+1)+(n-2)\Delta_{n,\ell}-2\Delta_{n,\ell}^2}{8\Delta_{n,\ell}^{3/2}}\left[\Omega_{\chi^{(+)}} + \Omega_{\chi^{(-)}}\right] = 0. \quad (A.7)$$

Equation (A.7) coincides with the second equation in (1.16). Taking (A.3) into account, we perform similar transformations for the elliptic equation

$$\partial_\rho \chi_{n,\ell}^{(e,\pm)} = \frac{\sqrt{-\Delta_{n,\ell}}}{\rho}, \quad \partial_\rho^2 \chi_{n,\ell}^{(e,\pm)} = -\frac{(n-2)\Delta_{n,\ell}+n(\ell+1)}{2\rho^2\sqrt{-\Delta_{n,\ell}}}, \quad (A.8)$$

Substituting (A.8) and (A.6) into equation (1.12), we obtain

$$0 = -\left[\Omega_{\chi^{(+)}\chi^{(+)}} + 2\Omega_{\chi^{(+)}\chi^{(-)}} + \Omega_{\chi^{(-)}\chi^{(-)}}\right]\frac{\Delta_{n,\ell}}{\rho^2} - \left[\Omega_{\chi^{(+)}} + \Omega_{\chi^{(-)}}\right]\frac{(n-2)\Delta_{n,\ell}+n(\ell+1)}{2\rho^2\sqrt{-\Delta_{n,\ell}}} -$$

$$-\frac{\Delta_{n,\ell}}{\rho}\left[\Omega_{\chi^{(+)}} + \Omega_{\chi^{(-)}}\right]\frac{\sqrt{-\Delta_{n,\ell}}}{\rho} - \frac{\Delta_{n,\ell}}{\rho^2}\left[\Omega_{\chi^{(+)}\chi^{(+)}} - 2\Omega_{\chi^{(+)}\chi^{(-)}} + \Omega_{\chi^{(-)}\chi^{(-)}}\right] =$$

$$= -\left[\Omega_{\chi^{(+)}\chi^{(+)}} + \Omega_{\chi^{(-)}\chi^{(-)}}\right]\frac{2\Delta_{n,\ell}}{\rho^2} - \frac{(n-2)\Delta_{n,\ell}+n(\ell+1)+2\Delta_{n,\ell}^2}{2\sqrt{-\Delta_{n,\ell}}\,\rho^2}\left[\Omega_{\chi^{(+)}} + \Omega_{\chi^{(-)}}\right],$$

$$\Omega_{\chi^{(+)}\chi^{(+)}} + \Omega_{\chi^{(-)}\chi^{(-)}} - \frac{(n-2)\Delta_{n,\ell}+n(\ell+1)+2\Delta_{n,\ell}^2}{4(-\Delta_{n,\ell})^{3/2}}\left[\Omega_{\chi^{(+)}} + \Omega_{\chi^{(-)}}\right] = 0. \quad (A.9)$$

Equation (A.9) coincides with the first equation in (1.16). Theorem 1 is proved.

*Proof of Theorem 2*

Let the variable $\rho$ be expressed in terms of the variable $\varsigma$, then equation (1.19) for the function $R$ takes the form

$$R'_\rho = \bar{R}'_\varsigma \varsigma'_\rho, \quad R''_{\rho\rho} = \bar{R}'_\varsigma \varsigma''_{\rho\rho} + \bar{R}''_{\varsigma\varsigma}\varsigma'_\rho \varsigma'_\rho,$$

$$\overline{R}'_{\varsigma\varsigma}\varsigma'_\rho\varsigma'_\rho + \left(\varsigma''_{\rho\rho} + \frac{g_{n,\ell}}{\rho}\varsigma'_\rho\right)\overline{R}'_\varsigma - \frac{g_{n,\ell}\lambda^2}{\rho^2}\overline{R} = 0. \qquad (A.10)$$

We require that the condition

$$\varsigma''_{\rho\rho} + \frac{g_{n,\ell}}{\rho}\varsigma'_\rho = 0 \Rightarrow \ln\overline{\varsigma} = -\int g_{n,\ell}\frac{d\rho}{\rho} = (\ell+1)\int(\overline{\rho}^n - 1)\frac{d\overline{\rho}}{\overline{\rho}} = (\ell+1)\left(\frac{\overline{\rho}^n}{n} - \ln\overline{\rho}\right) + const,$$

$$\overline{\varsigma}(\rho) = c_0 \frac{e^{\frac{\ell+1}{n}\overline{\rho}^n}}{\overline{\rho}^{\ell+1}} = c_0\left(\frac{\rho_T}{\rho}\right)^{\ell+1}\exp\left(\frac{\ell+1}{n}\frac{\rho^n}{\rho_T^n}\right), \qquad (A.11)$$

where the substitution $\varsigma'_\rho = \overline{\varsigma}$ and $\overline{\rho} = \rho/\rho_T$ has been made. Using expression (A.11), we find the variable $\varsigma$

$$\varsigma_{n,\ell}(\rho) = c_0\rho_T\int e^{\frac{\ell+1}{n}\overline{\rho}^n}\frac{d\overline{\rho}}{\overline{\rho}^{\ell+1}}. \qquad (A.12)$$

When condition (A.12) is satisfied, equation (A.10) takes the form

$$0 = \overline{R}''_{\varsigma\varsigma} - \frac{g_{n,\ell}\lambda^2}{\rho^2\overline{\varsigma}^2}\overline{R} = \overline{R}''_{\varsigma\varsigma} + \frac{\lambda^2(\ell+1)}{c_0^2\rho_T^2}\overline{\rho}^{2\ell}(\overline{\rho}^n - 1)e^{-2\frac{\ell+1}{n}\overline{\rho}^n}\overline{R}, \qquad (A.13)$$

which proves the validity of expressions (1.21)-(1.22). Let us evaluate the integral (A.12) for two cases: $\ell = kn$ and $\ell \neq kn$, $k = 0,1,2...$ Without loss of generality, let us begin with the case $\ell = kn$. For convenience of transformation, let us introduce the notation

$$J_{n,k}(x) \stackrel{def}{=} \int \frac{e^{\beta_{n,k}x^n}}{x^{kn+1}}dx \Rightarrow \varsigma_{n,\ell}(\rho) = c_0\rho_T J_{n,k}(\overline{\rho}). \qquad (A.14)$$

Let us compute the derivative of the integral exponential function (1.25) and take into account expression (A.14)

$$\mathrm{Ei}(x)' = \frac{1}{x} + \sum_{k=1}^{+\infty}\frac{x^{k-1}}{k!} = \frac{1}{x}\left(1 + \sum_{k=1}^{+\infty}\frac{x^k}{k!}\right) = \frac{1}{x}\sum_{k=0}^{+\infty}\frac{x^k}{k!} = \frac{e^x}{x},$$

$$\mathrm{Ei}(\beta_{n,k}x^n)' = \frac{ne^{\beta_{n,k}x^n}}{x} \Rightarrow J_{n,0}(x) = \frac{1}{n}\mathrm{Ei}(\beta_{n,0}x^n). \qquad (A.15)$$

Note that the following relation holds

$$\left(\frac{e^{\beta_{n,k}x^n}}{x^{nk}}\right)' = \frac{e^{\beta_{n,k}x^n}(x^n\beta_{n,k}n - nk)}{x^{nk+1}} \Rightarrow \frac{e^{\beta_{n,k}x^n}}{x^{nk+1}} = \frac{\beta_{n,k}}{k}\frac{e^{\beta_{n,k}x^n}}{x^{(k-1)n+1}} - \frac{1}{nk}\left(\frac{e^{\beta_{n,k}x^n}}{x^{nk}}\right)'. \qquad (A.16)$$

Let us integrate expression (A.16) and take into account the notation in (A.14)

$$J_{n,k}(x) = \frac{\beta_{n,k}}{k} \int \frac{e^{\beta_{n,k} x^n} dx}{x^{(k-1)n+1}} - \frac{1}{kn} \frac{e^{\beta_{n,k} x^n}}{x^{kn}}. \qquad (A.17)$$

We transform the integral in expression (A.17) into the form $J_{n,k-1}$ by performing a substitution $x = \bar{x} \sqrt[n]{\beta_{n,k-1}/\beta_{n,k}}$

$$\int \frac{e^{\beta_{n,k} x^n} dx}{x^{(k-1)n+1}} = \left(\frac{\beta_{n,k}}{\beta_{n,k-1}}\right)^{k-1} \int \frac{e^{\beta_{n,k-1} \bar{x}^n} d\bar{x}}{\bar{x}^{(k-1)n+1}} = \left(\frac{\beta_{n,k}}{\beta_{n,k-1}}\right)^{k-1} J_{n,k-1}\left(x\sqrt[n]{\beta_{n,k}/\beta_{n,k-1}}\right). \qquad (A.18)$$

Substituting (A.18) into expression (A.17), we obtain

$$J_{n,k}(x) = \frac{\beta_{n,k}^k}{k \beta_{n,k-1}^{k-1}} J_{n,k-1}\left(x\sqrt[n]{\beta_{n,k}/\beta_{n,k-1}}\right) - \frac{1}{kn} \frac{e^{\beta_{n,k} x^n}}{x^{kn}}. \qquad (A.19)$$

Expressions (A.15), (A.19) and (A.14) prove the validity of expressions (1.23) and (1.24) for $\ell = kn$. Let us evaluate the integral (A.12) for $\ell \neq kn$. In this case, the direct expansion of the exponential into a series and its integration term by term yields the expression

$$\int e^{\frac{\ell+1}{n} \bar{\rho}^n} \frac{d\bar{\rho}}{\bar{\rho}^{\ell+1}} = \sum_{k=0}^{+\infty} \left(\frac{\ell+1}{n}\right)^k \frac{1}{k!} \int \bar{\rho}^{nk-\ell-1} d\bar{\rho} = \sum_{k=0}^{+\infty} \left(\frac{\ell+1}{n}\right)^k \frac{\bar{\rho}^{nk-\ell}}{(nk-\ell)k!} + const, \qquad (A.20)$$

which corresponds to expression (1.23) for $\ell \neq kn$. Theorem 2 is proved.

*Proof of Theorem 3*

Let us make a change of variables in the hyperbolic region of the equation

$$\mu_{n,\ell}^{(\pm)} = \frac{1}{2}\left[\chi_{n,\ell}^{(h,+)} \pm \chi_{n,\ell}^{(h,-)}\right], \quad \Omega\left[\chi_{n,\ell}^{(h,+)}, \chi_{n,\ell}^{(h,-)}\right] = \tilde{\Omega}\left[\mu_{n,\ell}^{(+)}, \mu_{n,\ell}^{(-)}\right], \qquad (A.21)$$

then equation (1.16) takes the form

$$\Omega_{\chi_{n,\ell}^{(h,+)}} = \frac{1}{2}\tilde{\Omega}_{\mu_{n,\ell}^{(+)}} + \frac{1}{2}\tilde{\Omega}_{\mu_{n,\ell}^{(-)}}, \quad \Omega_{\chi_{n,\ell}^{(h,-)}} = \frac{1}{2}\tilde{\Omega}_{\mu_{n,\ell}^{(+)}} - \frac{1}{2}\tilde{\Omega}_{\mu_{n,\ell}^{(-)}},$$

$$\Omega_{\chi_{n,\ell}^{(h,+)}\chi_{n,\ell}^{(h,-)}} = \frac{1}{4}\tilde{\Omega}_{\mu_{n,\ell}^{(+)}\mu_{n,\ell}^{(+)}} - \frac{1}{4}\tilde{\Omega}_{\mu_{n,\ell}^{(+)}\mu_{n,\ell}^{(-)}} + \frac{1}{4}\tilde{\Omega}_{\mu_{n,\ell}^{(-)}\mu_{n,\ell}^{(+)}} - \frac{1}{4}\tilde{\Omega}_{\mu_{n,\ell}^{(-)}\mu_{n,\ell}^{(-)}} = \frac{1}{4}\left[\tilde{\Omega}_{\mu_{n,\ell}^{(+)}\mu_{n,\ell}^{(+)}} - \tilde{\Omega}_{\mu_{n,\ell}^{(-)}\mu_{n,\ell}^{(-)}}\right],$$

$$\tilde{\Omega}_{\mu_{n,\ell}^{(+)}\mu_{n,\ell}^{(+)}} - \tilde{\Omega}_{\mu_{n,\ell}^{(-)}\mu_{n,\ell}^{(-)}} + 4\kappa_{n,\ell}^{(h)} \tilde{\Omega}_{\mu_{n,\ell}^{(+)}} = 0. \qquad (A.22)$$

We shall seek the solution to equation (A.22) in the form $\tilde{\Omega} = \bar{\Omega}\left[\mu_{n,\ell}^{(+)}\right]$, then the equation for the function $\bar{\Omega}$ takes the form

$$\bar{\Omega}'' + 4\kappa_{n,\ell}^{(h)} \bar{\Omega}' = 0 \Rightarrow \Lambda' + 4\kappa_{n,\ell}^{(h)} \Lambda = 0, \quad \Lambda \stackrel{def}{=} \bar{\Omega}'. \qquad (A.23)$$

We shall find the solution to equation (A.23), taking into account the form of the coefficient $\kappa_{n,\ell}^{(h)}$ (1.17)

$$\frac{d\Lambda}{\Lambda} = -4\kappa_{n,\ell}^{(h)} d\mu_{n,\ell}^{(+)} \Rightarrow \Lambda = c_1 \exp\left[-4\int \kappa_{n,\ell}^{(h)} d\mu_{n,\ell}^{(+)}\right], \tag{A.24}$$

where $c_1$ is a constant. Let us evaluate the integral (A.24) using expressions (A.21) and (1.14). Let us introduce the substitution

$$\varepsilon_n = \frac{\Delta_{n,\ell}}{\ell+1} = \bar{\rho}^n - 1 \Rightarrow d\mu_{n,\ell}^{(+)} = \frac{2\sqrt{\ell+1}}{n} d\left(\sqrt{\varepsilon_n} - \mathrm{arctg}\sqrt{\varepsilon_n}\right) = \frac{\sqrt{\ell+1}}{n} \frac{\sqrt{\varepsilon_n}}{1+\varepsilon_n} d\varepsilon_n, \tag{A.25}$$

hence

$$\int \kappa_{n,\ell}^{(h)}(\rho) d\mu_{n,\ell}^{(+)} = \frac{\sqrt{\ell+1}}{8n} \int \frac{n(\ell+1)+(n-2)\Delta_{n,\ell}-2\Delta_{n,\ell}^2}{\Delta_{n,\ell}^{3/2}} \frac{\sqrt{\varepsilon_n}}{1+\varepsilon_n} d\varepsilon_n =$$

$$= \frac{\sqrt{\ell+1}}{8n} \int \frac{n(\ell+1)+(n-2)(\ell+1)\varepsilon_n - 2(\ell+1)^2 \varepsilon_n^2}{(\ell+1)^{3/2} \varepsilon_n^{3/2}} \frac{\sqrt{\varepsilon_n}}{1+\varepsilon_n} d\varepsilon_n =$$

$$= \frac{1}{8n} \int \frac{n+(n-2)\varepsilon_n - 2(\ell+1)\varepsilon_n^2}{\varepsilon_n(1+\varepsilon_n)} d\varepsilon_n = \frac{-2\varepsilon_n \ell - 2\varepsilon_n + n\ln\varepsilon_n + 2\ell\ln(1+\varepsilon_n)}{8n},$$

$$\int \kappa_{n,\ell}^{(h)} d\mu_n^{(+)} = -\frac{(\ell+1)\varepsilon_n}{4n} + \ln \varepsilon_n^{1/8} \left(1+\varepsilon_n\right)^{\ell/4n}. \tag{A.26}$$

Substituting the integral (A.26) into equation (A.24), we find the function $\Lambda$ and from this, according to (A.23), the function $\bar{\Omega}$

$$\frac{d\bar{\Omega}}{d\mu_n^{(+)}} = \Lambda = c_1 \exp\left[\frac{(\ell+1)\varepsilon_n}{n} - \ln \varepsilon_n^{1/2}\left(1+\varepsilon_n\right)^{\ell/n}\right], \tag{A.27}$$

hence

$$\bar{\Omega} = c_1 \frac{\sqrt{\ell+1}}{n} \int \frac{e^{\frac{(\ell+1)\varepsilon_n}{n}}}{\varepsilon_n^{1/2}\left(1+\varepsilon_n\right)^{\ell/n}} \frac{\sqrt{\varepsilon_n}}{1+\varepsilon_n} d\varepsilon_n = c_1 \frac{\sqrt{\ell+1}}{n} e^{-\frac{\ell+1}{n}} \int \frac{e^{\frac{\ell+1}{n}\bar{\varepsilon}_n}}{\bar{\varepsilon}_n^{\ell/n+1}} d\bar{\varepsilon}_n, \tag{A.28}$$

where (A.25) has been taken into account and the substitution $\bar{\varepsilon}_n = \varepsilon_n + 1$ has been made. The integral in expression (A.28) is similar to the integral (A.12). Let us consider the case where $\ell = kn$, $k = 0,1,2...$. By analogy with expressions (A.15)–(A.17), we obtain

$$I_{n,k}(\bar{\varepsilon}_n) \stackrel{\text{def}}{=} \int \frac{e^{\beta_{n,k}\bar{\varepsilon}_n}}{\bar{\varepsilon}_n^{k+1}} d\bar{\varepsilon}_n, \qquad \frac{d}{d\bar{\varepsilon}_n} \mathrm{Ei}(\beta_{n,0}\bar{\varepsilon}_n) = \frac{e^{\beta_{n,0}\bar{\varepsilon}_n}}{\bar{\varepsilon}_n} \Rightarrow I_{n,0}(\bar{\varepsilon}_n) = \mathrm{Ei}(\beta_{n,0}\bar{\varepsilon}_n), \tag{A.29}$$

$$\frac{d}{d\bar{\varepsilon}_n}\left(\frac{e^{\beta_{n,k}\bar{\varepsilon}_n}}{\bar{\varepsilon}_n^k}\right) = e^{\beta_{n,k}\bar{\varepsilon}_n} \frac{\bar{\varepsilon}_n \beta_{n,k} - k}{\bar{\varepsilon}_n^{k+1}} \Rightarrow \frac{e^{\beta_{n,k}\bar{\varepsilon}_n}}{\bar{\varepsilon}_n^{k+1}} = \frac{\beta_{n,k}}{k} \frac{e^{\beta_{n,k}\bar{\varepsilon}_n}}{\bar{\varepsilon}_n^k} - \frac{1}{k}\frac{d}{d\bar{\varepsilon}_n}\left(\frac{e^{\beta_{n,k}\bar{\varepsilon}_n}}{\bar{\varepsilon}_n^k}\right),$$

$$I_{n,k}(\bar{\varepsilon}_n) = \frac{\beta_{n,k}}{k} \int \frac{e^{\beta_{n,k}\bar{\varepsilon}_n} d\bar{\varepsilon}_n}{\bar{\varepsilon}_n^k} - \frac{1}{k}\frac{e^{\beta_{n,k}\bar{\varepsilon}_n}}{\bar{\varepsilon}_n^k}. \tag{A.30}$$

We transform the integral in expression (A.30)

$$\int \frac{e^{\beta_{n,k}\bar{\varepsilon}_n} d\bar{\varepsilon}_n}{\bar{\varepsilon}_n^k} = \left(\frac{\beta_{n,k}}{\beta_{n,k-1}}\right)^{k-1} \int \frac{e^{\beta_{n,k-1}\tilde{\varepsilon}_n} d\tilde{\varepsilon}_n}{\tilde{\varepsilon}_n^{k-1+1}} = \left(\frac{\beta_{n,k}}{\beta_{n,k-1}}\right)^{k-1} I_{n,k-1}\left(\frac{\beta_{n,k}}{\beta_{n,k-1}}\bar{\varepsilon}_n\right), \quad \bar{\varepsilon}_n = \frac{\beta_{n,k-1}}{\beta_{n,k}}\tilde{\varepsilon}_n. \tag{A.31}$$

Substituting (A.31) into expression (A.30), we obtain the recurrence relation

$$I_{n,k}(\bar{\varepsilon}_n) = \frac{\beta_{n,k}^k}{k\beta_{n,k-1}^{k-1}} I_{n,k-1}\left(\frac{\beta_{n,k}}{\beta_{n,k-1}}\bar{\varepsilon}_n\right) - \frac{1}{k}\frac{e^{\beta_{n,k}\bar{\varepsilon}_n}}{\bar{\varepsilon}_n^k}. \tag{A.32}$$

Expressions (A.29), (A.32) and (A.28) prove the validity of expressions (2.1)-(2.3) for $\ell = kn$. Let us evaluate the integral (A.28) for $\ell \neq kn$. In this case, the direct expansion of the exponential into a series and its integration term by term yields the expression

$$\int e^{\frac{\ell+1}{n}\bar{\varepsilon}_n} \frac{d\bar{\varepsilon}_n}{\bar{\varepsilon}_n^{\ell/n+1}} = \sum_{k=0}^{+\infty} \left(\frac{\ell+1}{n}\right)^k \frac{1}{k!} \int \bar{\varepsilon}_n^{k-\ell/n-1} d\bar{\varepsilon}_n = \sum_{k=0}^{+\infty} \left(\frac{\ell+1}{n}\right)^k \frac{\bar{\varepsilon}_n^{k-\ell/n}}{(k-\ell/n)k!} + c_2, \tag{A.33}$$

which corresponds to expression (2.1) when $\ell \neq kn$. Theorem 3 is proved.

**Appendix B**
*Proof of Theorem 4*

We shall seek a solution to equation (2.5) in accordance with expressions (1.26) and (1.28) in the form

$$\bar{R}(\bar{\rho}) = \bar{\rho}^{\upsilon} T(\tau), \quad \tau = \delta\bar{\rho}^s, \quad \bar{g}_{n,\ell}(\bar{\rho}) \stackrel{\text{def}}{=} \gamma_{n,\ell}(\tau) = (\ell+1)\left[1-\left(\frac{\tau}{\delta}\right)^{n/s}\right], \tag{B.1}$$

where $\delta, s, \upsilon$ are certain numbers to be determined. Substituting expression (B.1) into equation (2.5)

$$\bar{R}'_{\bar{\rho}} = \upsilon\bar{\rho}^{\upsilon-1} T + \delta s\bar{\rho}^{\upsilon+s-1} T'_\tau,$$
$$\bar{R}''_{\bar{\rho}\bar{\rho}} = \upsilon(\upsilon-1)\bar{\rho}^{\upsilon-2} T + \delta s\bar{\rho}^{\upsilon+s-2}(2\upsilon+s-1) T'_\tau + \delta^2 s^2 \bar{\rho}^{\upsilon+2s-2} T''_{\tau\tau}, \tag{B.2}$$

$$\delta^2 s^2 \frac{\bar{\rho}^{2s}}{\bar{\rho}^2} T''_{\tau\tau} + \delta s \frac{\bar{\rho}^s}{\bar{\rho}^2}(2\upsilon+s-1+\gamma_{n,\ell}) T'_\tau + \frac{(\upsilon-\lambda^2)\gamma_{n,\ell}+\upsilon(\upsilon-1)}{\bar{\rho}^2} T = 0,$$

$$\tau T''_{\tau\tau} + \frac{1}{s}(2\upsilon+s-1+\gamma_{n,\ell}) T'_\tau + \frac{(\upsilon-\lambda^2)\gamma_{n,\ell}+\upsilon(\upsilon-1)}{s^2\tau} T = 0. \tag{B.3}$$

We impose a condition on the coefficient of the function $T$ in equation (B.3)

$$(\upsilon-\lambda^2)\gamma_{n,\ell}+\upsilon(\upsilon-1) = \upsilon^2+\ell\upsilon-\lambda^2(\ell+1)-(\upsilon-\lambda^2)(\ell+1)\left(\frac{\tau}{\delta}\right)^{n/s} = -(\upsilon-\lambda^2)(\ell+1)\frac{\tau}{\delta},$$

$$s = n, \quad \upsilon^2+\ell\upsilon-\lambda^2(\ell+1) = 0, \tag{B.4}$$

where the roots of equation (B.4) are the numbers $\upsilon_{\ell,\lambda}^{(\pm)}$ (2.7). Satisfying the conditions in (B.4) reduces equation (B.3) to the form

$$\tau T''_{\tau\tau} + \left[\frac{2v_{\ell,\lambda}^{(\pm)}}{n} + 1 - \frac{1}{n} + \frac{\ell+1}{n}\left(1 - \frac{\tau}{\delta}\right)\right]T'_\tau - \frac{\left(v_{\ell,\lambda}^{(\pm)} - \lambda^2\right)(\ell+1)}{\delta n^2}T = 0,$$

$$\tau T''_{\tau\tau} + \left(\frac{2v_{\ell,\lambda}^{(\pm)} + n + \ell}{n} - \frac{\ell+1}{\delta n}\tau\right)T'_\tau - \frac{\left(v_{\ell,\lambda}^{(\pm)} - \lambda^2\right)(\ell+1)}{\delta n^2}T = 0. \tag{B.5}$$

Let us determine the last parameter $\delta = (\ell+1)/n$. As a result, equation (B.5) will become (2.6). The resulting equation (2.6) is the Kummer equation and a solution (2.8) is known for it. Theorem 4 is proved.

*Proof of Lemma 1*

The direct solution of the equation $a_{\ell,\lambda}^{(+)} = -k$ leads to the expression

$$\sqrt{\ell^2 + 4\lambda^2\ell + 4\lambda^2} = \ell + 2(\lambda^2 - kn), \tag{B.6}$$

$$\ell kn = (\lambda^2 - kn)^2 - \lambda^2, \tag{B.7}$$

where $kn \leq \ell/2 + \lambda^2$ is taken into account when squaring equation (B.6). Substituting the value $\ell$ from equation (B.7) into the inequality yields the condition

$$k^2 n^2 \leq \lambda^2(\lambda^2 - 1). \tag{B.8}$$

Note that for the periodic boundary conditions (1.20), we have $\lambda \geq 1$. Thus, expression (2.12) holds. We shall prove that $\bar{\alpha} \neq 0$ by contradiction. Suppose $\bar{\alpha} = 0$, then

$$\bar{\alpha} = b_{n,\ell,\lambda}^{(+)} - 1 = \frac{2v_{\ell,\lambda}^{(+)} + \ell}{n} = 0 \Rightarrow 2v_{\ell,\lambda}^{(+)} + \ell = \sqrt{\ell^2 + 4\lambda^2(\ell+1)} = 0, \tag{B.9}$$

It follows from expressions (B.9) and (B.6) that

$$\ell + 2(\lambda^2 - kn) = 0. \tag{B.10}$$

Consider the special case of the equality (B.8)

$$\lambda^4 - \lambda^2 - k^2 n^2 = 0 \Rightarrow 2\lambda^2 = 1 + \sqrt{1 + 4k^2 n^2}, \tag{B.11}$$

where $\lambda \geq 1$ by the condition of the lemma. Substituting (B.11) into expression (B.10) yields a violation of the lemma's condition $\ell > -1$, that is

$$\ell = -\left(2kn + 1 + \sqrt{1 + 4k^2 n^2}\right) > -1 \Rightarrow \sqrt{1 + 4k^2 n^2} < -2kn. \tag{B.12}$$

Inequality (B.12) is false. The factor $c_0$ follows directly from the connection relation

$$M(a,b,z) = \frac{\Gamma(1-a)\Gamma(b)}{\Gamma(b-a)} L_{-a}^{(b-1)}(z) \Rightarrow M(-k,1+\bar{\alpha},z) = \frac{\Gamma(1+k)\Gamma(1+\bar{\alpha})}{\Gamma(1+\bar{\alpha}+k)} L_k^{(\bar{\alpha})}(z). \qquad (B.13)$$

Lemma 1 is proved.

*Proof of Theorem 5*

Let us write the Legendre transformation (1.7) in polar coordinates and use the factorised representation of the solution $u \sim R(\rho)\Theta(\theta)$

$$x = u'_\rho \cos\theta - u'_\theta \frac{\sin\theta}{\rho} = R'\Theta\cos\theta - R\Theta'\frac{\sin\theta}{\rho} = \frac{R\Theta}{\rho}\left(\rho\frac{R'}{R}\cos\theta - \frac{\Theta'}{\Theta}\sin\theta\right), \qquad (B.14)$$

$$y = u'_\rho \sin\theta + u'_\theta \frac{\cos\theta}{\rho} = R'\Theta\sin\theta + R\Theta'\frac{\cos\theta}{\rho} = \frac{R\Theta}{\rho}\left(\rho\frac{R'}{R}\sin\theta + \frac{\Theta'}{\Theta}\cos\theta\right), \qquad (B.15)$$

$$\Phi(x,y) = \rho u'_\rho - u(\rho,\theta) = R\Theta\left(\rho\frac{R'}{R} - 1\right). \qquad (B.16)$$

Let us compute the derivative $R'$ for the solution restricted to the origin

$$\bar{R}'_{\bar{\rho}} = \upsilon\bar{\rho}^{\upsilon-1}M(a,b,\tau) + (\ell+1)\bar{\rho}^{\upsilon+n-1}M'_\tau(a,b,\tau) = \bar{\rho}^{\upsilon-1}M(a,b,\tau)\left[\upsilon + (\ell+1)\bar{\rho}^n\frac{M'_\tau(a,b,\tau)}{M(a,b,\tau)}\right],$$

$$R'_\rho = \frac{\bar{\rho}^{\upsilon-1}}{\bar{\rho}_T}M(a,b,\tau)\left[\upsilon + n\tau\mathcal{M}(a,b,\tau)\right] \Rightarrow \rho\frac{R'}{R} = \upsilon + n\tau\mathcal{M}(a,b,\tau). \qquad (B.17)$$

Substituting (B.17) into (B.14)-(B.16) yields the validity of expressions (3.1)-(3.3). We find the expression for the Jacobian of the Legendre transformation using formula (1.7). Let us calculate the partial derivatives for the factorised solution

$$\omega_{\xi\xi} = \frac{R\Theta}{\rho^2}\left[\rho\frac{R'}{R}\left(\sin^2\theta - g\cos^2\theta - \Upsilon\sin 2\theta\right) - \lambda^2\sin^2\theta + g\lambda^2\cos^2\theta + \Upsilon\sin 2\theta\right],$$

$$\omega_{\xi\eta} = \frac{R\Theta}{\rho^2}\left[\rho\frac{R'}{R}\left(\Upsilon\cos 2\theta - \frac{1+g}{2}\sin 2\theta\right) + \lambda^2\frac{1+g}{2}\sin 2\theta - \Upsilon\cos 2\theta\right], \qquad (B.18)$$

$$\omega_{\eta\eta} = \frac{R\Theta}{\rho^2}\left[\rho\frac{R'}{R}\left(\cos^2\theta - g\sin^2\theta + \Upsilon\sin 2\theta\right) - \lambda^2\cos^2\theta + g\lambda^2\sin^2\theta - \Upsilon\sin 2\theta\right],$$

where it is noted that the functions $R$ and $\Theta$ satisfy the equations (1.19), i.e. $\rho^2 R''/R = g(\lambda^2 - \rho R'/R)$ and $\Theta''/\Theta = -\lambda^2$. Substituting the expressions (B.18) into (1.7) and performing intermediate calculations, we obtain

$$\frac{\rho^4}{u^2}J^{-1} = A\left(\rho\frac{R'}{R}\right)^2 + B\rho\frac{R'}{R} + C, \qquad (B.19)$$

where

$$A = \left(\cos^2\theta + \Upsilon\sin 2\theta - g\sin^2\theta\right)\left(\sin^2\theta - g\cos^2\theta - \Upsilon\sin 2\theta\right) - \frac{1}{4}\left(2\Upsilon\cos 2\theta - (g+1)\sin 2\theta\right)^2, \qquad (B.20)$$

$$B = \left[\lambda^2\left(g\sin^2\theta - \cos^2\theta\right) - \Upsilon\sin 2\theta\right]\left(-g\cos^2\theta + \sin^2\theta - \Upsilon\sin 2\theta\right) +$$
$$+\left(\cos^2\theta + \Upsilon\sin 2\theta - g\sin^2\theta\right)\left[\lambda^2\left(g\cos^2\theta - \sin^2\theta\right) + \Upsilon\sin 2\theta\right] - \quad (B.21)$$
$$-\frac{1}{2}\left[\lambda^2(g+1)\sin 2\theta - 2\Upsilon\cos 2\theta\right]\left[2\Upsilon\cos 2\theta - (g+1)\sin 2\theta\right],$$

$$C = \left[\lambda^2\left(g\sin^2\theta - \cos^2\theta\right) - \Upsilon\sin 2\theta\right]\left[\lambda^2\left(g\cos^2\theta - \sin^2\theta\right) + \Upsilon\sin 2\theta\right] -$$
$$-\frac{1}{4}\left[\lambda^2(g+1)\sin 2\theta - 2\Upsilon\cos 2\theta\right]^2. \quad (B.22)$$

Let us simplify the expressions for the coefficients (B.20)-(B.22).

$$A = \Upsilon g\left(\sin^2\theta - \cos^2\theta + \cos 2\theta\right)\sin 2\theta - g\left(\cos^2\theta + \sin^2\theta\right)^2 + g^2\left(\sin^2\theta\cos^2\theta - \sin^2\theta\cos^2\theta\right) +$$
$$+\cos^2\theta\sin^2\theta - \sin^2\theta\cos^2\theta + \Upsilon\sin 2\theta\left(\sin^2\theta - \cos^2\theta + \cos 2\theta\right) - \Upsilon^2\left(\sin^2 2\theta + \cos^2 2\theta\right),$$
$$A = -\left(g + \Upsilon^2\right). \quad (B.23)$$

$$B = -\lambda^2 g^2 \sin^2\theta\cos^2\theta + \lambda^2 g\cos^4\theta + \lambda^2 g\sin^4\theta - \lambda^2\sin^2\theta\cos^2\theta - \lambda^2\Upsilon g\sin^2\theta\sin 2\theta +$$
$$+\lambda^2\Upsilon\sin 2\theta\cos^2\theta + \Upsilon g\sin 2\theta\cos^2\theta - \Upsilon\sin 2\theta\sin^2\theta + \Upsilon^2\sin 2\theta\sin 2\theta + \lambda^2 g\cos^4\theta -$$
$$-\lambda^2\cos^2\theta\sin^2\theta + \lambda^2\Upsilon g\sin 2\theta\cos^2\theta - \lambda^2\Upsilon\sin 2\theta\sin^2\theta - \lambda^2 g^2\sin^2\theta\cos^2\theta + \lambda^2 g\sin^4\theta +$$
$$+\Upsilon\sin 2\theta\cos^2\theta + \Upsilon^2\sin^2 2\theta - \Upsilon g\sin 2\theta\sin^2\theta - \lambda^2\Upsilon(g+1)\sin 2\theta\cos 2\theta + 2\Upsilon^2\cos^2 2\theta +$$
$$+2\lambda^2 g^2\sin^2\theta\cos^2\theta + 4\lambda^2 g\sin^2\theta\cos^2\theta + 2\lambda^2\sin^2\theta\cos^2\theta - \Upsilon(g+1)\cos 2\theta\sin 2\theta =$$
$$= \lambda^2 g^2\left(-2\sin^2\theta\cos^2\theta + 2\sin^2\theta\cos^2\theta\right) + 2\lambda^2 g\left(\cos^4\theta + \sin^4\theta + 2\sin^2\theta\cos^2\theta\right) +$$
$$+\lambda^2\left(-2\sin^2\theta\cos^2\theta + 2\sin^2\theta\cos^2\theta\right) + \lambda^2\Upsilon g\left(+\cos^2\theta - \sin^2\theta - \cos 2\theta\right)\sin 2\theta +$$
$$+\lambda^2\Upsilon\sin 2\theta\left(\cos^2\theta - \sin^2\theta - \cos 2\theta\right) + \Upsilon g\sin 2\theta\left(\cos^2\theta - \sin^2\theta - \cos 2\theta\right) +$$
$$+\Upsilon\sin 2\theta\left(\cos^2\theta - \sin^2\theta - \cos 2\theta\right) + 2\Upsilon^2\left(\sin^2 2\theta + \cos^2 2\theta\right),$$
$$B = 2\left(\lambda^2 g + \Upsilon^2\right). \quad (B.24)$$

$$C = \lambda^4 g^2\cos^2\theta\sin^2\theta - \lambda^4 g^2\sin^2\theta\cos^2\theta - \lambda^4 g\cos^2\theta\cos^2\theta - \lambda^4 g\sin^2\theta\sin^2\theta -$$
$$-2\lambda^4 g\sin^2\theta\cos^2\theta + \lambda^4\sin^2\theta\cos^2\theta - \lambda^4\sin^2\theta\cos^2\theta + \lambda^2\Upsilon g\sin 2\theta\sin^2\theta -$$
$$-\lambda^2\Upsilon g\sin 2\theta\cos^2\theta + \lambda^2\Upsilon g\cos 2\theta\sin 2\theta - \lambda^2\Upsilon\sin 2\theta\cos^2\theta + \lambda^2\Upsilon\sin 2\theta\sin^2\theta +$$
$$+\lambda^2\Upsilon\cos 2\theta\sin 2\theta - \Upsilon^2\sin^2 2\theta - \Upsilon^2\cos^2 2\theta =$$
$$= -\lambda^4 g\left(\cos^4\theta + \sin^4\theta + 2\sin^2\theta\cos^2\theta\right) - \lambda^2\Upsilon g\sin 2\theta\left(\cos^2\theta - \sin^2\theta - \cos 2\theta\right) -$$
$$-\lambda^2\Upsilon\sin 2\theta\left(-\cos 2\theta + \cos^2\theta - \sin^2\theta\right) - \Upsilon^2\left(\sin^2 2\theta + \cos^2 2\theta\right),$$
$$C = -\left(\lambda^4 g + \Upsilon^2\right). \quad (B.25)$$

Substituting expressions (B.23)-(B.25) into (B.19) gives

$$-\frac{\rho^4}{u^2}J^{-1} = \left(\lambda^4 g + \Upsilon^2\right) - 2\rho\frac{R'}{R}\left(\lambda^2 g + \Upsilon^2\right) + \left(\rho\frac{R'}{R}\right)^2\left(g + \Upsilon^2\right), \quad (B.26)$$

which, taking (B.17) into account, reduces to (3.4). Let us consider the special case of (B.26) for $\rho = \rho_T$ and $\theta = \theta_e$

$$J^{-1}(\rho_T,\theta_e) = -\frac{u^2(\rho_T,\theta_e)}{\rho_T^4}\Upsilon^2(\theta_e)\left[\mathcal{R}^2(\rho_T) - 2\mathcal{R}(\rho_T) + 1\right] =$$
$$= -\frac{u^2(\rho_T,\theta_e)}{\rho_T^4}\Upsilon^2(\theta_0)\left[\mathcal{R}(\rho_T) - 1\right]^2 = -\frac{\mathcal{R}^2(\rho_T)}{\rho_T^4}\Theta'^2(\theta_e)\left[\mathcal{R}(\rho_T) - 1\right]^2 = 0, \quad (B.27)$$

where it is assumed that $\Theta'(\theta_e) = 0$. It follows from (B.27) that (3.5) holds. In the case where $\lambda = 1$, we have $\upsilon_{\ell,1}^{(+)} = 1$ and $a_{n,\ell,1}^{(+)} = 0$. Consequently, $M(0,b,\tau) = 1$ and $M'_\tau(0,b,\tau) = 0$, that is

$$\mathcal{R}_{n,\ell,1}(\rho) = \upsilon_{\ell,1}^{(+)} = 1 \ . \tag{B.28}$$

Substituting (B.28) into (B.16) and then into (B.17), we obtain that $\Phi \equiv 0$, whilst expression (B.26) takes the form

$$-\frac{\rho^4}{u^2}J^{-1} = g + \Upsilon^2 - 2(g + \Upsilon^2) + g + \Upsilon^2 \equiv 0, \tag{B.29}$$

that is the transform is impossible. Theorem 5 is proved.

*Proof of Lemma 2*

It follows directly from expressions (B.14)-(B.15) that

$$x = \bar{\Omega}'_\rho \cos\theta, \ y = \bar{\Omega}'_\rho \sin\theta, \ \bar{\Omega}'_\rho = \frac{d\bar{\Omega}}{d\mu_{n,\ell}^{(+)}}\frac{d\mu_{n,\ell}^{(+)}}{d\varepsilon_n}\frac{d\varepsilon_n}{d\rho}, \tag{B.30}$$

where it is noted that $\bar{\Omega}$ does not depend on the angle $\theta$. Using expressions (A.25), (A.27) and (2.3), we obtain

$$\bar{\Omega}'_\rho = \frac{c_1 e^{\frac{\ell+1}{n}\varepsilon_n}}{\varepsilon_n^{1/2}(1+\varepsilon_n)^{\ell/n}}\frac{\sqrt{\ell+1}}{n}\frac{\sqrt{\varepsilon_n}}{1+\varepsilon_n}\frac{n}{\rho_T}\bar{\rho}^{n-1} = c_1 e^{-\frac{\ell+1}{n}}\frac{\sqrt{\ell+1}}{\rho_T}\frac{e^{\frac{\ell+1}{n}\bar{\rho}^n}}{\bar{\rho}^{\ell+1}} = \bar{\varsigma}(\rho), \tag{B.31}$$

where $c_1 = e^{\frac{\ell+1}{n}}\frac{c_0\rho_T}{\sqrt{\ell+1}}$ is an arbitrary constant and expression (A.11) is taken into account. Substituting (B.31) into (B.30) proves the coordinate transformation (3.10). Note that it also follows from (B.31) and (A.11) that the function $\bar{\Omega}$ is a solution to equation (1.19) for the radial part $R(\rho)$ when $\lambda = 0$, since $\varsigma'_\rho = \bar{\varsigma}$ and $\varsigma''_{\rho\rho} + \frac{g_{n,\ell}}{\rho}\varsigma'_\rho = 0$. Consequently, $\bar{\Omega} = R(\rho)$ and taking into account (B.16) and (B.31), we obtain

$$\Phi(x,y) = \rho\bar{\Omega}'_\rho - \bar{\Omega} = \rho\bar{\varsigma}(\rho) - R(\rho). \tag{B.32}$$

Equation (B.32) coincides with the coordinate solution (3.10). Lemma 2 is proved.

*Proof of Theorem 6*

Let us write the expression for the quantum potential (i.10) in terms of the quantity $S = \ln f$

$$Q = \frac{\alpha}{2\beta}\left(\Delta_r S + \frac{1}{2}|\nabla_r S|^2\right), \tag{B.33}$$

$$S = \ln f_{n,\ell} = \ln F_{n,\ell}(z) = \ln(N_{n,\ell} c_{n,\ell}) + \ln z^\ell - c_{n,\ell} z^n, \quad c_{n,\ell}^{-1} = \sigma_{\langle v \rangle, n, \ell}^n 2^{n/2}, \tag{B.34}$$

$$Q = \frac{\alpha}{2\beta}\left(\Delta_r \ln z^\ell - c_{n,\ell} \Delta_r z^n + \frac{1}{2}|\nabla_r \ln z^\ell - c_{n,\ell} \nabla_r z^n|^2\right). \tag{B.35}$$

Let us transform the terms in expression (B.35)

$$|\nabla_r \ln z^\ell - c_{n,\ell} \nabla_r z^n|^2 = |\nabla_r \ln z^\ell|^2 - 2 c_{n,\ell} \nabla_r \ln z^\ell \cdot \nabla_r z^n + c_{n,\ell}^2 |\nabla_r z^n|^2 =$$

$$= \frac{\ell^2}{z^2}|\nabla_r z|^2 + c_{n,\ell}^2 n^2 z^{2n-2}|\nabla_r z|^2 - 2 c_{n,\ell} \ell \frac{nz^{n-1}}{z}|\nabla_r z|^2,$$

$$|\nabla_r \ln z^\ell - c_{n,\ell} \nabla_r z^n|^2 = \left(c_{n,\ell}^2 n^2 z^{2n} - 2 c_{n,\ell} \ell n z^n + \ell^2\right)\frac{|\nabla_r z|^2}{z^2}, \tag{B.36}$$

$$\Delta_r \ln z = \nabla_r \cdot \nabla_r \ln z = \nabla_r \cdot \frac{\nabla_r z}{z} = \frac{\Delta_r z}{z} - \frac{1}{z^2}|\nabla_r z|^2, \tag{B.37}$$

$$\Delta_r z^n = \nabla_r \cdot \nabla_r z^n = n \nabla_r \cdot z^{n-1} \nabla_r z = n \nabla_r z^{n-1} \cdot \nabla_r z + n z^{n-1} \Delta_r z,$$

$$\Delta_r z^n = n(n-1) z^{n-2}|\nabla_r z|^2 + n z^{n-1} \Delta_r z. \tag{B.38}$$

Substitute (B.36)–(B.38) into (B.35)

$$\frac{2\beta}{\alpha} Q = \left(\ell - c_{n,\ell} n z^n\right)\frac{\Delta_r z}{z} + \left[\frac{1}{2}\left(c_{n,\ell}^2 n^2 z^{2n} - 2 c_{n,\ell} \ell n z^n + \ell^2\right) - \ell - c_{n,\ell} n(n-1) z^n\right]\frac{|\nabla_r z|^2}{z^2},$$

$$\frac{2\beta}{\alpha} Q = \left(\ell - c_{n,\ell} n z^n\right)\frac{\Delta_r z}{z} + \frac{1}{2}\left[c_{n,\ell}^2 n^2 z^{2n} - 2 c_{n,\ell} n(\ell+n-1) z^n + \ell(\ell-2)\right]\frac{|\nabla_r z|^2}{z^2}. \tag{B.39}$$

Let us calculate the values of $\nabla_r z$ and $\Delta_r z$, where $z = \langle v \rangle = |\alpha|\rho = |\alpha|\sqrt{\xi^2 + \eta^2}$. Using the coordinate transformation (3.1), we obtain

$$\begin{cases} x = r\cos\phi = X(\rho,\theta) \\ y = r\sin\phi = Y(\rho,\theta) \end{cases} \Rightarrow \begin{cases} x_r = \cos\phi = X_\rho \rho_r + X_\theta \theta_r \\ y_r = \sin\phi = Y_\rho \rho_r + Y_\theta \theta_r \end{cases}, \tag{B.40}$$

$$\cos\phi = \frac{X}{\sqrt{X^2 + Y^2}}, \quad \sin\phi = \frac{Y}{\sqrt{X^2 + Y^2}}. \tag{B.41}$$

The solution to system (B.40) is

$$\rho_r = \frac{Y_\theta \cos\phi - X_\theta \sin\phi}{Y_\theta X_\rho - X_\theta Y_\rho} = \frac{1}{|\alpha|}\frac{\partial z}{\partial r}, \quad \theta_r = \frac{X_\rho \sin\phi - Y_\rho \cos\phi}{Y_\theta X_\rho - X_\theta Y_\rho}. \tag{B.42}$$

Similarly, for the derivatives with respect to the azimuth angle $\phi$

$$\begin{cases} x_\phi = -r\sin\phi = X_\rho \rho_\phi + X_\theta \theta_\phi, \\ y_\phi = r\cos\phi = Y_\rho \rho_\phi + Y_\theta \theta_\phi, \end{cases} \tag{B.43}$$

$$\rho_\phi = r\frac{Y_\theta \sin\phi + X_\theta \cos\phi}{X_\theta Y_\rho - X_\rho Y_\theta} = \frac{1}{|\alpha|}\frac{\partial z}{\partial \phi}, \quad \theta_\phi = r\frac{X_\rho \cos\phi + Y_\rho \sin\phi}{X_\rho Y_\theta - X_\theta Y_\rho}. \tag{B.44}$$

Using equations (B.42), (B.44) and (B.41), we obtain

$$|\nabla_r z|^2 = \left|\frac{\partial z}{\partial r}\right|^2 + \frac{1}{r^2}\left|\frac{\partial z}{\partial \phi}\right|^2 = |\alpha|^2 \frac{(XY_\theta - YX_\theta)^2 + (YY_\theta + XX_\theta)^2}{(X^2+Y^2)(Y_\theta X_\rho - X_\theta Y_\rho)^2} = |\alpha|^2 \frac{Y_\theta^2 + X_\theta^2}{(Y_\theta X_\rho - X_\theta Y_\rho)^2},$$

$$\frac{|\nabla_r z|^2}{z^2} = \frac{X_\theta^2 + Y_\theta^2}{\rho^2 (X_\rho Y_\theta - X_\theta Y_\rho)^2}. \tag{B.45}$$

Thus, for expression (B.39), we need to find $\Delta_r z$

$$\Delta_r z = \frac{\partial^2 z}{\partial r^2} + \frac{1}{r}\frac{\partial z}{\partial r} + \frac{1}{r^2}\frac{\partial^2 z}{\partial \phi^2}. \tag{B.46}$$

Let us compute the second derivatives by differentiating systems (B.40) and (B.43)

$$\begin{cases} 0 = X_{\rho\rho}\rho_r^2 + 2X_{\rho\theta}\rho_r\theta_r + X_{\theta\theta}\theta_r^2 + X_\rho \rho_{rr} + X_\theta \theta_{rr}, \\ 0 = Y_{\rho\rho}\rho_r^2 + 2Y_{\rho\theta}\rho_r\theta_r + Y_{\theta\theta}\theta_r^2 + Y_\rho \rho_{rr} + Y_\theta \theta_{rr}, \end{cases} \tag{B.47}$$

$$\begin{cases} -X = X_{\rho\rho}\rho_\phi^2 + 2X_{\rho\theta}\rho_\phi\theta_\phi + X_{\theta\theta}\theta_\phi^2 + X_\rho \rho_{\phi\phi} + X_\theta \theta_{\phi\phi}, \\ -Y = Y_{\rho\rho}\rho_\phi^2 + 2Y_{\rho\theta}\rho_\phi\theta_\phi + Y_{\theta\theta}\theta_\phi^2 + Y_\rho \rho_{\phi\phi} + Y_\theta \theta_{\phi\phi}. \end{cases} \tag{B.48}$$

Since the values of $\rho_r, \theta_r$ and $\rho_\phi, \theta_\phi$ are known from the solutions to (B.42) and (B.44), we shall solve the systems (B.47)-(B.48) for $\rho_{rr}, \theta_{rr}$ and $\rho_{\phi\phi}, \theta_{\phi\phi}$. Note that to find (B.46), it suffices to know only $\rho_{rr}$ and $\rho_{\phi\phi}$ that is

$$-(X_\rho Y_\theta - X_\theta Y_\rho)^3 (X^2+Y^2)\rho_{rr} = (X_{\rho\rho}Y_\theta - X_\theta Y_{\rho\rho})(XY_\theta - X_\theta Y)^2 + \tag{B.49}$$
$$+(X_{\theta\theta}Y_\theta - X_\theta Y_{\theta\theta})(X_\rho Y - XY_\rho)^2 + 2(X_{\rho\theta}Y_\theta - X_\theta Y_{\rho\theta})(XY_\theta - X_\theta Y)(X_\rho Y - XY_\rho),$$

$$(X_\rho Y_\theta - X_\theta Y_\rho)^3 \rho_{\phi\phi} = (X_\theta Y - XY_\theta)(X_\rho Y_\theta - X_\theta Y_\rho)^2 - (X_{\rho\rho}Y_\theta - X_\theta Y_{\rho\rho})(XX_\theta + YY_\theta)^2 + \tag{B.50}$$
$$+2(X_{\rho\theta}Y_\theta - X_\theta Y_{\rho\theta})(XX_\theta + YY_\theta)(XX_\rho + YY_\rho) - (X_{\theta\theta}Y_\theta - X_\theta Y_{\theta\theta})(XX_\rho + YY_\rho)^2.$$

Substituting (B.49), (B.50) and (B.42) into (B.46), we obtain

$$\frac{\Delta_r z}{z} = \frac{1}{\rho}\left(\rho_{rr} + \frac{1}{r}\rho_r + \frac{1}{r^2}\rho_{\phi\phi}\right) =$$

$$= -\frac{(X_{\rho\rho}Y_\theta - X_\theta Y_{\rho\rho})(XY_\theta - X_\theta Y)^2}{\rho(X_\rho Y_\theta - X_\theta Y_\rho)^3(X^2+Y^2)} - \frac{(X_{\rho\rho}Y_\theta - X_\theta Y_{\rho\rho})(XX_\theta + YY_\theta)^2}{\rho(X_\rho Y_\theta - X_\theta Y_\rho)^3(X^2+Y^2)} -$$

$$-\frac{(X_{\theta\theta}Y_\theta - X_\theta Y_{\theta\theta})(X_\rho Y - XY_\rho)^2}{\rho(X_\rho Y_\theta - X_\theta Y_\rho)^3(X^2+Y^2)} - \frac{(X_{\theta\theta}Y_\theta - X_\theta Y_{\theta\theta})(XX_\rho + YY_\rho)^2}{\rho(X_\rho Y_\theta - X_\theta Y_\rho)^3(X^2+Y^2)} +$$

$$+\frac{(XY_\theta - X_\theta Y)(X_\rho Y_\theta - X_\theta Y_\rho)^2}{\rho(X_\rho Y_\theta - X_\theta Y_\rho)(X^2+Y^2)} + \frac{(X_\rho Y - XY_\rho)(X_\rho Y_\theta - X_\theta Y_\rho)^2}{\rho(X_\rho Y_\theta - X_\theta Y_\rho)^3(X^2+Y^2)} +$$

$$+\frac{2(X_{\rho\theta}Y_\theta - X_\theta Y_{\rho\theta})(XX_\theta + YY_\theta)(XX_\rho + YY_\rho)}{\rho(X_\rho Y_\theta - X_\theta Y_\rho)^3(X^2+Y^2)} - \frac{2(X_{\rho\theta}Y_\theta - X_\theta Y_{\rho\theta})(XY_\theta - X_\theta Y)(X_\rho Y - XY_\rho)}{\rho(X_\rho Y_\theta - X_\theta Y_\rho)^3(X^2+Y^2)}.$$

$$\rho(X_\rho Y_\theta - X_\theta Y_\rho)^3(X^2+Y^2)\frac{\Delta z}{z} = -(X_{\rho\rho}Y_\theta - X_\theta Y_{\rho\rho})\left[(XY_\theta - X_\theta Y)^2 + (XX_\theta + YY_\theta)^2\right] -$$

$$-(X_{\theta\theta}Y_\theta - X_\theta Y_{\theta\theta})\left[(X_\rho Y - XY_\rho)^2 + (XX_\rho + YY_\rho)^2\right] +$$

$$+2(X_{\rho\theta}Y_\theta - X_\theta Y_{\rho\theta})\left[(XX_\theta + YY_\theta)(XX_\rho + YY_\rho) - (XY_\theta - X_\theta Y)(X_\rho Y - XY_\rho)\right] =$$

$$= -(X_{\rho\rho}Y_\theta - X_\theta Y_{\rho\rho})(X_\theta^2 + Y_\theta^2)(X^2+Y^2) - (X_{\theta\theta}Y_\theta - X_\theta Y_{\theta\theta})(X_\rho^2 + Y_\rho^2)(X^2+Y^2) +$$

$$+2(X_{\rho\theta}Y_\theta - X_\theta Y_{\rho\theta})(X_\rho X_\theta + Y_\rho Y_\theta)(X^2+Y^2)$$

$$-\rho(X_\rho Y_\theta - X_\theta Y_\rho)^3 \frac{\Delta z}{z} = (X_{\rho\rho}Y_\theta - X_\theta Y_{\rho\rho})(X_\theta^2 + Y_\theta^2) + (X_{\theta\theta}Y_\theta - X_\theta Y_{\theta\theta})(X_\rho^2 + Y_\rho^2) -$$

$$-2(X_{\rho\theta}Y_\theta - X_\theta Y_{\rho\theta})(X_\rho X_\theta + Y_\rho Y_\theta).$$

(B.51)

Before substituting (B.45) and (B.51) into (B.39), it is necessary to find the partial derivatives of the functions $X$ and $Y$. From the factored solution of the Legendre transformation (3.1), it follows that

$$X = R'\Theta\cos\theta - \frac{R}{\rho}\Theta'\sin\theta,\ Y = R'\Theta\sin\theta + \frac{R}{\rho}\Theta'\cos\theta,\ \rho\frac{R'}{R} = \upsilon + n\tau\mathcal{M}(a,b,\tau) = \mathcal{R}, \quad \text{(B.52)}$$

Let us differentiate the expressions in (B.52) with respect to the variable $\rho$

$$X_\rho = \frac{R}{\rho^2}(\lambda^2 g\Theta\cos\theta + \Theta'\sin\theta) - \frac{R'}{\rho}(g\Theta\cos\theta + \Theta'\sin\theta),$$

$$Y_\rho = \frac{R}{\rho^2}(\lambda^2 g\Theta\sin\theta - \Theta'\cos\theta) + \frac{R'}{\rho}(\Theta'\cos\theta - g\Theta\sin\theta),$$

(B.53)

where $R'' = g(\lambda^2 R - \rho R')/\rho^2$ is taken into account. Redifferentiating (B.54) with respect to the variables $\rho$ and $\theta$ yields the expressions

$$X_{\rho\rho} = \left[-\lambda^2 g^2\Theta\cos\theta - \lambda^2 g(\Theta'\sin\theta + 2\Theta\cos\theta) + \lambda^2 \rho g'\Theta\cos\theta - 2\Theta'\sin\theta\right]\frac{R}{\rho^3} +$$

$$+\left[g^2\Theta\cos\theta + g\left[\Theta'\sin\theta + (1+\lambda^2)\Theta\cos\theta\right] - \rho g'\Theta\cos\theta + 2\Theta'\sin\theta\right]\frac{R'}{\rho^2},$$

(B.54)

$$X_{\rho\theta} = \frac{R}{\rho^2}\left[\left(\lambda^2 g+1\right)\Theta'\cos\theta-(g+1)\lambda^2\Theta\sin\theta\right]-\frac{R'}{\rho}\left[(g+1)\Theta'\cos\theta-\left(g+\lambda^2\right)\Theta\sin\theta\right], \quad (B.55)$$

$$Y_{\rho\rho} = \frac{R'}{\rho^2}\left(\lambda^2 g\Theta\sin\theta-\Theta'\cos\theta\right)-\frac{R'}{\rho^2}\left(\Theta'\cos\theta-g\Theta\sin\theta\right)-g\frac{R'}{\rho^2}\left(\Theta'\cos\theta-g\Theta\sin\theta\right)-$$
$$-g'\frac{R'}{\rho}\Theta\sin\theta-\frac{2R}{\rho^3}\left(\lambda^2 g\Theta\sin\theta-\Theta'\cos\theta\right)+\lambda^2 g'\frac{R}{\rho^2}\Theta\sin\theta+\lambda^2 g\frac{R}{\rho^3}\left(\Theta'\cos\theta-g\Theta\sin\theta\right),$$

$$Y_{\rho\rho} = \left[g^2\Theta\sin\theta+g\left[\left(\lambda^2+1\right)\Theta\sin\theta-\Theta'\cos\theta\right]-\rho g'\Theta\sin\theta-2\Theta'\cos\theta\right]\frac{R'}{\rho^2}+$$
$$+\left[-\lambda^2 g^2\Theta\sin\theta+\lambda^2 g\left(\Theta'\cos\theta-2\Theta\sin\theta\right)+\lambda^2\rho g'\Theta\sin\theta+2\Theta'\cos\theta\right]\frac{R}{\rho^3}, \quad (B.56)$$

$$Y_{\rho\theta} = \frac{R}{\rho^2}\left[\left(\lambda^2 g+1\right)\Theta'\sin\theta+\lambda^2(g+1)\Theta\cos\theta\right]-\frac{R'}{\rho}\left[\left(\lambda^2+g\right)\Theta\cos\theta+(1+g)\Theta'\sin\theta\right]. \quad (B.57)$$

Let us differentiate expression (B.52) with respect to the variable $\theta$

$$X_\theta = \left(R'-\frac{R}{\rho}\right)\Theta'\cos\theta+\left(\lambda^2\frac{R}{\rho}-R'\right)\Theta\sin\theta, \qquad (B.58)$$

$$Y_\theta = \left(R'-\frac{R}{\rho}\right)\Theta'\sin\theta+\left(R'-\lambda^2\frac{R}{\rho}\right)\Theta\cos\theta, \qquad (B.59)$$

$$X_{\theta\theta} = \left[2\lambda^2\frac{R}{\rho}-\left(\lambda^2+1\right)R'\right]\Theta\cos\theta+\left[\left(\lambda^2+1\right)\frac{R}{\rho}-2R'\right]\Theta'\sin\theta, \qquad (B.60)$$

$$Y_{\theta\theta} = \left[2\lambda^2\frac{R}{\rho}-\left(\lambda^2+1\right)R'\right]\Theta\sin\theta+\left[2R'-\left(\lambda^2+1\right)\frac{R}{\rho}\right]\Theta'\cos\theta. \qquad (B.61)$$

Using (B.53)-(B.61), let us calculate the relations appearing in expressions (B.45) and (B.51).

$$X_\rho^2+Y_\rho^2 = \frac{u^2}{\rho^4}\left[\Upsilon^2\left(\mathcal{R}-1\right)^2+g^2\left(\mathcal{R}-\lambda^2\right)^2\right], \qquad (B.62)$$

$$X_\theta^2+Y_\theta^2 = \frac{u^2}{\rho^2}\left[\Upsilon^2\left(\mathcal{R}-1\right)^2+\left(\mathcal{R}-\lambda^2\right)^2\right], \qquad (B.63)$$

$$X_\rho Y_\theta-X_\theta Y_\rho = -\frac{u^2}{\rho^3}\left[\Upsilon^2\left(\mathcal{R}-1\right)^2+g\left(\mathcal{R}-\lambda^2\right)^2\right], \qquad (B.64)$$

$$X_\rho X_\theta+Y_\rho Y_\theta = (1-g)\left(\mathcal{R}-\lambda^2\right)(\mathcal{R}-1)\frac{u^2}{\rho^3}\Upsilon, \qquad (B.65)$$

$$X_{\rho\rho}Y_\theta-X_\theta Y_{\rho\rho} = \left(g^2-\rho g'+g\right)\left(\mathcal{R}-\lambda^2\right)^2\frac{u^2}{\rho^4}+2\left(\mathcal{R}-1\right)^2\frac{u^2}{\rho^4}\Upsilon^2+$$
$$+g\left(\mathcal{R}-1\right)\left(\mathcal{R}-\lambda^2\right)\frac{u^2}{\rho^4}\left(\lambda^2+\Upsilon^2\right), \qquad (B.66)$$

$$X_{\theta\theta}Y_\theta-X_\theta Y_{\theta\theta} = -\left(\mathcal{R}-\lambda^2\right)^2\frac{u^2}{\rho^2}-\left(\mathcal{R}-1\right)^2\frac{u^2}{\rho^2}\Upsilon^2-\left(\lambda^2+\Upsilon^2\right)\left(\mathcal{R}-1\right)\left(\mathcal{R}-\lambda^2\right)\frac{u^2}{\rho^2}, \qquad (B.67)$$

$$X_{\rho\theta}Y_\theta-X_\theta Y_{\rho\theta} = \left[\lambda^2\left(\mathcal{R}-1\right)^2-g\left(\mathcal{R}-\lambda^2\right)^2+(g-1)\left(\mathcal{R}-\lambda^2\right)\left(\mathcal{R}-1\right)\right]\frac{u^2}{\rho^3}\Upsilon, \qquad (B.68)$$

where the representation (B.52) is taken into account $\mathcal{R}$. Using (B.63) and (B.64), we transform (B.45)

$$\frac{|\nabla_r z|^2}{z^2} = \frac{\rho^2}{u^2} \frac{\Upsilon^2(\mathcal{R}-1)^2 + (\mathcal{R}-\lambda^2)^2}{\left[\Upsilon^2(\mathcal{R}-1)^2 + g(\mathcal{R}-\lambda^2)^2\right]^2}. \tag{B.69}$$

For the convenience of further transformations, let us introduce the following notation

$$z_1 = \mathcal{R}-1, \quad z_2 = \mathcal{R}-\lambda^2, \quad z_3 = g, \quad z_4 = \Upsilon, \tag{B.70}$$

and also take into account the expression for the derivative $g'$ (1.13)

$$\rho g' = -n\bar{\rho}^{n-1}\frac{\rho}{\rho_T}(\ell+1) = -n(\ell+1)\bar{\rho}^n = n(g-1-\ell). \tag{B.71}$$

Using (B.66), (B.63), (B.68), (B.65), (B.67), (B.62), (B.71) and the notation (B.70), we obtain the following relations

$$(X_{\rho\rho}Y_\theta - X_\theta Y_{\rho\rho})(X_\theta^2 + Y_\theta^2) = \frac{u^4}{\rho^6}\left\{\left[z_3^2 + (1-n)z_3 + n(\ell+1)\right]z_2^2 + 2z_1^2z_4^2 + z_1z_2z_3(\lambda^2 + z_4^2)\right\}(z_1^2z_4^2 + z_2^2),$$

$$(X_{\rho\theta}Y_\theta - X_\theta Y_{\rho\theta})(X_\rho X_\theta + Y_\rho Y_\theta) = \frac{u^4}{\rho^6}z_1z_2(1-z_3)z_4^2\left[\lambda^2 z_1^2 - z_3 z_2^2 + (z_3-1)z_1z_2\right], \tag{B.72}$$

$$(X_{\theta\theta}Y_\theta - X_\theta Y_{\theta\theta})(X_\rho^2 + Y_\rho^2) = -\frac{u^4}{\rho^6}\left[z_2^2 + z_1^2z_4^2 + (\lambda^2 + z_4^2)z_1z_2\right](z_1^2z_4^2 + z_3^2z_2^2).$$

Equations (B.72) appear on the right-hand side of expression (B.51). The left-hand side of (B.51) is expressed in terms of (B.64). Consequently, (B.51) takes the form

$$\frac{u^2}{\rho^2}(z_4^2z_1^2 + z_3^2z_2^2)^3\frac{\Delta z}{z} = \left\{\left[z_3^2 + (1-n)z_3 + n(\ell+1)\right]z_2^2 + 2z_1^2z_4^2 + z_1z_2z_3(\lambda^2 + z_4^2)\right\}(z_1^2z_4^2 + z_2^2)$$

$$-\left[z_2^2 + z_1^2z_4^2 + (\lambda^2 + z_4^2)z_1z_2\right](z_1^2z_4^2 + z_3^2z_2^2) - 2z_1z_2(1-z_3)z_4^2\left[\lambda^2z_1^2 - z_3z_2^2 + (z_3-1)z_1z_2\right] =$$

$$= 2z_1^4z_4^4 + z_1^3z_2z_3z_4^4 - z_1^4z_4^4 - z_1^3z_2z_3z_4^4 + \left[z_3^2 + (1-n)z_3 + n(\ell+1)\right]z_1^2z_2^2z_4^2 + \lambda^2 z_1^3z_2z_3z_4^2 +$$

$$+2z_1^2z_2^2z_4^2 + z_1z_2^3z_3z_4^2 - z_1^2z_2^2z_4^2 - z_1^2z_2^2z_3^2z_4^2 - \lambda^2z_1^3z_2z_4^2 - z_1z_2^3z_3^2z_4^2 + \lambda^2z_1z_2^3z_3 - z_2^4z_3^2 -$$

$$-2z_1z_2(1-z_3)\left[\lambda^2z_1^2 - z_3z_2^2 + (z_3-1)z_1z_2\right]z_4^2 + z_2^4\left[z_3^2 + (1-n)z_3 + n(\ell+1)\right] - \lambda^2z_1z_2^3z_3^2.$$

$$\frac{u^2}{\rho^2}(z_4^2z_1^2 + z_3^2z_2^2)^3\frac{\Delta z}{z} = z_1^3(z_1 + z_2z_3 - z_2)z_4^4 +$$

$$+\left\{z_1z_2\left[2z_3^2 + (n+3)(1-z_3) + n\ell\right] + 3(z_2^2z_3 - \lambda^2z_1^2)(1-z_3)\right\}z_1z_2z_4^2 \tag{B.73}$$

$$+z_2^3\left\{\lambda^2z_1(1-z_3)z_3 + z_2\left[(1-n)z_3 + n(\ell+1)\right]\right\}.$$

Expressions (B.73) and (B.69) can be rewritten in a compact form using the notation (3.14)

$$\frac{\Delta z}{z} = \frac{\rho^2}{u^2} \frac{\sum_{k=0}^{2} A_{n,\ell,\lambda}^{(k)}(z_1, z_2, z_3) z_4^{2k}}{(z_1^2 z_4^2 + z_3 z_2^2)^3}, \qquad \frac{|\nabla_r z|^2}{z^2} = \frac{\rho^2}{u^2} \frac{z_1^2 z_4^2 + z_2^2}{(z_1^2 z_4^2 + z_3 z_2^2)^2}. \tag{B.74}$$

Let us simplify expression (B.34) for the coefficient $c_{n,\ell}^{-1}$. According to (1.11), we obtain

$$c_{n,\ell}^{-1} = |\alpha|^n \rho_T^n \left(\frac{n}{\ell+1}\right) \Rightarrow c_{n,\ell} n z^n = (\ell+1) \frac{\rho^n}{\rho_T^n} = (\ell+1) \bar{\rho}^n = \ell + 1 - g,$$

$$\ell - c_{n,\ell} n z^n = g - 1 = z_3 - 1, \tag{B.75}$$

$$c_{n,\ell}^2 n^2 z^{2n} - 2 c_{n,\ell} n (\ell + n - 1) z^n + \ell(\ell - 2) = (z_3 - 1)^2 + 2(z_3 - 1)(n - 1) - 2n\ell. \tag{B.76}$$

Substituting (B.74)-(B.76) into (B.39) and taking into account the notations (3.12)-(3.13), we obtain

$$\bar{Q} = \frac{\alpha}{2\beta} \frac{\rho^2}{u_{n,\ell,\lambda}^2} (\mathcal{A}_{n,\ell,\lambda} + \mathcal{B}_{n,\ell}). \tag{B.77}$$

Theorem 6 is proved.

***Proof of Theorem 7***

According to Remark 4, for $\lambda = 0$, the function $\Upsilon_0(\theta) = c_1/(c_1 \theta + c_2)$ may be used instead of $\Upsilon_\lambda(\theta)$. Since $R = const$, it follows from (B.52) that $\mathcal{R} = 0$. Thus, the transition (3.1) from the momentum $(\rho, \theta)$ to the coordinate $(x, y)$ representation takes the form

$$\Phi(x,y) = -\Theta_0(\theta), \quad \begin{pmatrix} x \\ y \end{pmatrix} = \frac{c_1}{\rho} \begin{pmatrix} -\sin\theta \\ \cos\theta \end{pmatrix}, \quad r = \frac{|c_1|}{\rho}, \quad \phi = \theta + \frac{\pi}{2}. \tag{B.78}$$

Let us simplify expressions (3.11)-(3.14) for the quantum potential

$$A_{n,\ell,0}^{(k)}(-1,0,g) = \begin{cases} 0, & k = 0, \\ 0, & k = 1, \\ 1, & k = 2, \end{cases} \Rightarrow \mathcal{A}(-1,0,g,c_1/\Theta_0) = \frac{g-1}{c_1^2} \Theta_0^2, \tag{B.79}$$

$$\mathcal{B}(-1,0,g,c_1/\Theta_0) = \frac{\Theta_0^2}{2c_1^2} \left[ g^2 + 2(n-2)g + 3 - 2n(\ell+1) \right], \tag{B.80}$$

$$\mathcal{C} = \left[ (\ell+1)^2 \bar{\rho}^{2n} - 2(\ell+1)(\ell+n) \bar{\rho}^n + \ell^2 \right] \frac{\Theta_0^2}{2c_1^2},$$

$$\bar{Q}(\rho,\theta) = \frac{\alpha \rho^2}{2\beta \Theta_0^2} \mathcal{C}(-1,0,g,c_1/\Theta_0) = \frac{\alpha}{4\beta r^2} \left[ \ell^2 - \frac{2(\ell+1)(\ell+n)|c_1|^n}{\rho_T^n r^n} + \frac{(\ell+1)^2 |c_1|^{2n}}{\rho_T^{2n} r^{2n}} \right], \tag{B.81}$$

which coincides with (3.21) when $|c_1| = \sigma_{(r)} \rho_T$. We obtain the expression for the potential $U$ from the Hamilton–Jacobi equation (3.16)

$$U(r,\phi) = \frac{\alpha}{4\beta r^2}\left\{ c_1^2 - \ell^2 + \frac{2(\ell+1)(\ell+n)\sigma_r^n}{r^n} - \frac{(\ell+1)^2 \sigma_r^{2n}}{r^{2n}} \right\} + E, \qquad (B.82)$$

which coincides with (3.22). Knowing the velocity $\langle v \rangle = |\alpha c_1|/r$, we obtain an expression for the probability density from formula (1.1) $f_{n,\ell}$

$$f_{n,\ell}(r) = N_{n,\ell}\left(\frac{\ell+1}{n}\right)^{\ell/n} \frac{\sigma_{\langle r \rangle}^\ell}{r^\ell} \exp\left(-\frac{\ell+1}{n}\frac{\sigma_{\langle r \rangle}^n}{r^n}\right), \qquad (B.83)$$

where the relations (1.11) are taken into account. The form of the wave function $\psi$ follows directly from the expression for its phase (B.78) and the square of the modulus (B.83). Let us calculate the normalisation integral

$$\int_{\mathbb{R}^2} f_{n,\ell}(r) d^2 r = 2\pi \sigma_{\langle r \rangle}^\ell N_{n,\ell} \left(\frac{\ell+1}{n}\right)^{\ell/n} \int_0^{+\infty} e^{-\frac{\ell+1}{n}\frac{\sigma_{\langle r \rangle}^n}{r^n}} \frac{dr}{r^{\ell-1}} = 2\pi \sigma_{\langle r \rangle}^{k+3} N_{n,k+3} \left(\frac{k+4}{n}\right)^{(k+3)/n} \int_0^{+\infty} e^{-\frac{k+4}{n}\sigma_{\langle r \rangle}^n \tau^n} \tau^k d\tau =$$

$$= 2\pi \sigma_{\langle r \rangle}^{k+3} \frac{N_{n,k+3}}{n} \left(\frac{k+4}{n}\right)^{(k+3)/n} \left(\frac{k+4}{n}\sigma_{\langle r \rangle}^n\right)^{-\frac{k+1}{n}} \Gamma\left(\frac{k+1}{n}\right),$$

$$1 = \int_{\mathbb{R}^2} f_{n,\ell}(r) d^2 r = 2\pi \sigma_{\langle r \rangle}^2 N_{n,k+3} \left(\frac{k+4}{n}\right)^{\frac{2}{n}} \frac{1}{n}\Gamma\left(\frac{k+1}{n}\right) = 2\pi \sigma_{\langle r \rangle}^2 \frac{N_{n,\ell}}{n}\left(\frac{\ell+1}{n}\right)^{\frac{2}{n}} \Gamma\left(\frac{\ell-2}{n}\right), \qquad (B.83)$$

where, during integration, the variables $\ell = k+3$ and $\tau = 1/r$ have been substituted. Expression (B.83) corresponds to (3.20). Theorem 7 is proved.

*Calculation of the standard deviation*

Let us derive an expression for the $s$-th moment of the distribution

$$\int_{\mathbb{R}^2} r^s f_{n,\ell}(r) d^2 r = 2\pi \sigma_{\langle r \rangle}^\ell N_{n,\ell}\left(\frac{\ell+1}{n}\right)^{\ell/n} \int_0^{+\infty} e^{-\frac{\ell+1}{n}\frac{\sigma_{\langle r \rangle}^n}{r^n}} \frac{dr}{r^{\ell-1-s}} =$$

$$= 2\pi \sigma_{\langle r \rangle}^{k+s+3} N_{n,\ell}\left(\frac{k+s+4}{n}\right)^{(k+s+3)/n} \int_0^{+\infty} e^{-\frac{k+s+4}{n}\sigma_{\langle r \rangle}^n \tau^n} \tau^k d\tau =$$

$$= 2\pi \sigma_{\langle r \rangle}^{k+s+3} N_{n,\ell}\left(\frac{k+s+4}{n}\right)^{(k+s+3)/n} \frac{1}{n}\left(\frac{k+s+4}{n}\sigma_{\langle r \rangle}^n\right)^{-\frac{k+1}{n}} \Gamma\left(\frac{k+1}{n}\right) =$$

$$= 2\pi \sigma_{\langle r \rangle}^{s+2} N_{n,\ell}\left(\frac{k+s+4}{n}\right)^{\frac{s+2}{n}} \frac{1}{n}\Gamma\left(\frac{k+1}{n}\right),$$

$$\int_{\mathbb{R}^2} r^s f_{n,\ell}(r) d^2 r = 2\pi \sigma_{\langle r \rangle}^{s+2} N_{n,\ell}\left(\frac{\ell+1}{n}\right)^{\frac{s+2}{n}} \frac{1}{n}\Gamma\left(\frac{\ell-s-2}{n}\right), \qquad (B.84)$$

where the variables $\ell = k + s + 3$ and $\tau = 1/r$ have been substituted. Using (B.84), we find $\langle r \rangle$ and $\langle r^2 \rangle$

$$\langle r \rangle = 2\pi\sigma_{\langle r \rangle}^3 N_{n,\ell} \left(\frac{\ell+1}{n}\right)^{\frac{3}{n}} \frac{1}{n} \Gamma\left(\frac{\ell-3}{n}\right), \quad \langle r^2 \rangle = 2\pi\sigma_{\langle r \rangle}^4 N_{n,\ell} \left(\frac{\ell+1}{n}\right)^{\frac{4}{n}} \frac{1}{n} \Gamma\left(\frac{\ell-4}{n}\right), \quad (B.85)$$

hence

$$\sigma_r^2 = \langle r^2 \rangle - \langle r \rangle^2 = \sigma_{\langle r \rangle}^2 \left(\frac{\ell+1}{n}\right)^{\frac{2}{n}} \frac{1}{\Gamma^2\left(\frac{\ell-2}{n}\right)} \left[\Gamma\left(\frac{\ell-4}{n}\right)\Gamma\left(\frac{\ell-2}{n}\right) - \Gamma^2\left(\frac{\ell-3}{n}\right)\right], \quad (B.86)$$

where (B.83) is taken into account.

## Acknowledgements

The study was conducted under a state commission from Lomonosov Moscow State University.